\documentclass[fleqn,usenatbib,useAMS]{mnras}

\usepackage{graphicx}	
\usepackage{amsmath}	
\usepackage{amssymb}	
\usepackage{multicol}        
\usepackage{bm}		
\usepackage{pdflscape}	

\usepackage[T1]{fontenc}
\usepackage{ae,aecompl}

\usepackage{txfonts}
\usepackage{array, makecell} 
\usepackage{multirow}

\title[]{Electron Energy Distributions in the Extended Gas Nebulae	
	associated with High-z AGN: Maxwell-Boltzmann vs. \textit{$\kappa$}-distributions }

\author[S. G.Morais]{S. G. Morais$^{1,2}$, A. Humphrey$^1$, M. Villar Martín$^3$, L. Binette$^{4,5}$, M. Silva %
	\thanks{Contact e-mail: \href{sandy.morais@astro.up.pt}{sandy.morais@astro.up.pt}}%
	\\
	$^{1}$ Instituto de Astrofísica e Ciências do Espaço, Universidade do Porto, CAUP, Rua das Estrelas, PT4150-762 Porto, Portugal \\
	$^{2}$ Departamento de Física e Astronomia, Faculdade de Ciências, Universidade do Porto, Rua do Campo Alegre 687, PT4169-007 Porto, Portugal \\
	$^3$  Centro de Astrobiología, (CAB, CSIC-INTA), Departamento de Astrofísica, Cra. de Ajalvir Km. 4, 28850, Torrejón de Ardoz, Madrid, Spain \\
	$^4$ Instituto de Astronomía, Universidad Nacional Autónoma de México, A.P. 70-264, 04510, Mexico, D.F., México \\
	$^5$ Département de physique, de génie physique et d’optique, Université Laval, Québec, QC G1V 0A6, Canada
} 

\pubyear{2021}

\begin{document}
	\label{firstpage}
	\pagerange{\pageref{firstpage}--\pageref{lastpage}}
	\maketitle
	
	\begin{abstract}
		Emission line observations together with photoionization models provide important information about the ionization mechanisms, densities, temperatures, and metallicities in AGN-ionized gas. Photoionization models usually assume Maxwell-Boltzmann (M-B) electron energy distributions (EED), but it has been suggested that using \textit{$\kappa$} distributions may be more appropriate and could potentially solve the discrepancies in temperatures and abundances found in HII regions and Planetary Nebulae (PNe). 
		We consider the impact of the presence of \textit{$\kappa$} distributions in photoionized nebulae associated with AGN and study how this might affect spectral modelling and abundance analyses for such regions. Using the photoionization code MAPPINGS 1e we compute models adopting M-B and \textit{$\kappa$} distributions of electron energies, and compare the behaviour of emission line ratios for different values of \textit{$\kappa$}, gas metallicity, density, ionization parameter and SED slope. We find that the choice of EED can have a large impact on some UV and optical emission lines emitted by photoionized nebulae associated with AGN, and that the impact of adopting a \textit{$\kappa$} distribution is strongly dependent on gas metallicity and ionization parameter. 	
		We compile a sample of line ratios for 143 type 2 AGN and compare our models against the observed line ratios. We find that for 98 objects \textit{$\kappa$} distributions provide a better fit to the observed line ratios than M-B distributions. In addition, we find that adopting \textit{$\kappa$}-distributed electron energies results in significant changes in the inferred gas metallicity and ionization parameter in a significant fraction of objects.	
		
	\end{abstract}

	\begin{keywords}
		galaxies: active, quasars: emission lines
	\end{keywords}

	\section{Introduction} \label{introduction}
	
	Active Galactic Nuclei (AGN) are compact objects found at the centre of galaxies. They are believed to be powered by the accretion of material onto supermassive black holes in the hearts of galaxies \citep[e.g.][]{Jones2004}, and are an important source of feedback potentially influencing the evolution of their host galaxies \citep[e.g.][]{Springel2005, Morganti2005, Croton2006, Moe2009, Cano2012}.
	Their spectra show strong emission lines from various chemical elements in a wide range of ionization states \citep[e.g.][]{Kriss1992, Netzer1997, Veron2000, Collins2005}. A detailed analysis of the emission line spectra provides information on the physical conditions of the emitting gas, including its density, temperature, chemical composition and ionization state \citep[e.g.][]{VM1999c, VM2001, Humphrey2008a}. Properties are usually estimated using photoionization models \citep[e.g.][]{Binette1996, Collins2009, Silva2018, Feltre2016}. Photoionization codes compute recombination rates, ionization rates, and fluxes of spectral lines, which depend on the shape of the electron velocity (or energy) distribution, usually assumed to be a Maxwell-Boltzmann (M-B) distribution \citep[e.g.][]{VM1997, Moy2002, Humphrey2008a, Matsuoka2009, Silva2018}.  
	
	The assumption that the velocity distribution of the free electrons in gaseous nebulae are described by a M-B distribution dates back to, at least, the 1930s, when the treatments of the physical state of Planetary Nebulae (PNe) already used this distribution \citep[e.g.][]{Bohm1947, Hebb1940}. This assumption is based on the electron thermalization timescale, the timescale a free electron requires to become thermalized and described by the M-B distribution \citep[e.g.][]{Bohm1947, Spitzer1962}. It is usually assumed that free electrons share their energy with the neighbouring medium so rapidly that they become thermalized before they excite any emission line. The thermalization timescale of energetic electrons is proportional to the cube of their velocity \citep{Spitzer1962}, hence plasmas with high energy electrons will take longer to reach equilibrium than plasmas with cooler electrons. Since the thermalization timescale depends on the frequency of collisions, electron energies in high density gas will equilibrate faster than in low density gas.
	
	The suggestion that the electron velocity distribution may be non-M-B also dates back to the 1930s, when \citet{Hagihara1939a, Hagihara1939b} proposed that the velocity distributions in PNe depart significantly from a M-B distribution.
	Later, empirical kappa (\textit{$\kappa$}) distributions were used to describe magnetospheric electron observations \citep[e.g.][]{Binsack1966, Olbert1968, Vasyliunas1968}. Since then, \textit{$\kappa$} electron energy distributions have been directly measured in several solar system plasmas. They are present in the solar wind \citep[e.g.][]{Gloeckler1992, Maksimovic1997}, in the outer heliosphere and the inner heliosleath \citep[e.g.][]{Decker2003, Heerikhuisen2008}, in planetary magnetospheres, including magnetosheath \citep[e.g.][]{Binsack1966, Gloeckler1987, Krimigis1983, Krimigis1986, Olbert1968}, and magnetospheres of planetary moons \citep[e.g.][]{Jurac2002, Moncuquet2002}. 
	
	It is clear that \textit{$\kappa$} distributions are prevalent in solar system plasmas, but not until recent years has the possibility of \textit{$\kappa$} distributions in extrasolar gaseous nebulae been significantly explored. \citet{Nicholls2012} used the \textit{$\kappa$} distribution in order to resolve longstanding discrepancies in the measurements of abundances and temperatures in HII regions and PNe, principally the discrepancies between electron temperatures derived from collisionally excited lines (CELs) and recombination lines (RLs), and discrepancies between chemical abundances inferred from RLs and CELs \citep[e.g.][]{Wyse1942}.
	
	Following the work of \citet{Nicholls2012}, a number of subsequent studies have addressed the possibility that \textit{$\kappa$} electron energy distributions are present in gaseous nebulae, in place of a M-B distribution \citep[e.g.][]{Binette2012, Dopita2013, Dopita2014, Nicholls2013, Storey2013, Storey2014a, Mendoza2014, Humphrey2014, Ferland2016, Zhang2016} with varied conclusions. While \citet{Nicholls2012, Nicholls2013, Binette2012, Humphrey2014} found that using \textit{$\kappa$} distributions solved some of the problems encountered when attempting to reproduce the temperatures observed in HII regions, PNe and AGN, \citet{Storey2013, Mendoza2014, Zhang2016, Ferland2016, Draine2018} found little or no evidence for the existence of \textit{$\kappa$} distributions in these objects. 
	
	Although \textit{$\kappa$} distributions were initially criticised for lacking a theoretical justification, \citet{Tsallis1995} showed how these distributions can appear using entropy considerations and it has also been shown that they can arise from Tsallis’s non-extensive statistical mechanics \citep[e.g.][]{Leubner2002, Livadiotis2009}. For a review on $\kappa$ distributions and on the mechanisms proposed to generate them in space plasmas see \citet{Pierrard2010}. 
	Supra-thermal electrons are present in the gas due to the photoionization process, but the question is how important are these electrons in comparison to the thermal ones. If the timescale at which energetic electrons enter the gas is of the same order or smaller than the thermalization timescale, then a high energy tail in the electron energy distribution could persist and these electrons may be able to affect the emitted spectrum, otherwise, the signature of a $\kappa$ distribution would disappear. 
	\citet{Nicholls2012} proposed several mechanisms capable of inducing $\kappa$ distributions, including accelerations induced by shocks, photoionization by a hard radiation source, photoionization of dust \citep{Dopita2000}, suprathermal atom or ion heating, X-ray ionization, and magnetic reconnection \citep[e.g.][]{Bradshaw2013}.
	
	Here, we consider the presence of \textit{$\kappa$} distributions in extrasolar AGN photoionized nebulae and we study how this might impact on spectral and abundance analyses for such regions. 
	The degree of ionization in the gas has a dependence on the form of the electron energy (or velocity) distribution, and thus if \textit{$\kappa$}-distributed electrons are present they may affect the observed spectra and this will affect the derivation of the physical conditions present in the nebulae. For example, if the free electrons follow a \textit{$\kappa$} distribution and the temperature is derived assuming that they follow a M-B distribution then the inferred temperature will be incorrect \citep[e.g.][]{Owocki1983}.
	
	In this work, we compare the emission line fluxes predicted by photoionization models using M-B distributions with predictions obtained using models that assume \textit{$\kappa$} distributions of electron energies, in an attempt to understand how they differ and what changes can be expected when using different distributions. We will be comparing the behaviour of emission line ratios using different electron energy distributions for several values of metallicity, density and ionization parameter. Choosing the appropriate photoionization model is crucial since we are relying on these models to study the properties of photoionized nebulae associated with AGNs.
	
	This paper is organized as follows. In Section \ref{sec:k dist} we describe the \textit{$\kappa$} distribution in more detail. In Section \ref{sec:models} the photoionization models used in this work are presented. The results are shown in Section \ref{sec:results}. In Section \ref{sec:data comparison} we compare our model calculations with observed line ratios and in Section \ref{sec:discussion} the results are discussed. Finally, we summarise our main results and final conclusions in Section \ref{sec:conclusion}.

	\section{The \textit{$\kappa$} distribution}\label{sec:k dist}
	
	\textit{$\kappa$} distributions describe stationary state systems outside of thermal equilibrium. In this system there are more particles with high energies than in a system in thermal equilibrium (i.e. a system described by the M-B distribution). 
	Following \citet{Vasyliunas1968} and \citet{Nicholls2012} the fraction $f(E)$ of electrons having energies between $E$ and $E + dE$ for the \textit{$\kappa$} distribution is given by 
	
	\begin{equation} \label{eq:k}
		\begin{split}
			f_{\small \kappa}(E) dE = \frac{2n_e}{\sqrt{\pi}} \, \frac{\Gamma(\kappa+1)}{\big(\kappa - 3/2 \big)^{\frac{3}{2}} \, \Gamma\big(\kappa - 1/2 \big)} \\
			\times \frac{\sqrt{E}}{(k_B T_{\kappa})^{3/2} \big(1+E / \big[\big(\kappa - 3/2 \big) k_B T_{\kappa}\big]\big)^{\kappa+1}} \, dE \,,
		\end{split}
	\end{equation}
	where $n_e$ is the electron density, $\Gamma$ is the gamma function, \textit{$\kappa$} is a parameter that characterizes the distribution and varies from $3/2 < \kappa \leq \infty$, $k_B$ is the Boltzmann constant, and $T_{\kappa}$ is the non-equilibrium temperature that characterizes the mean kinetic energy ($T_{\kappa}=<E>/(1.5k_B)$). 
	
	The \textit{$\kappa$} parameter measures the departure from an equilibrium distribution. As \textit{$\kappa$} increases the fraction of high energy electrons decreases. At the limit $\kappa \rightarrow \infty $, the energy \textit{$\kappa$} distribution reduces to the M-B distribution, 
	\begin{equation} \label{eq:MB_E}
		f_{\small{M-B}}(E) \, dE = 2n_e \sqrt{ \frac{E}{\pi} } \, \Bigg(\frac{1}{k_B T}\Bigg)^{3/2} \, exp\Bigg(-\frac{E}{k_B T}\Bigg) \, dE     \,, 
	\end{equation} 
	where $T$ is the kinetic temperature (electron equilibrium temperature) and the other quantities are the same as in the previous equation. 
	It has been shown that the physical meaning of the temperature for \textit{$\kappa$} distributions is the same as the kinetic temperature for a Maxwellian \citep[e.g.][]{Meyer-Vernet1995, Livadiotis2009}. 
	
\begin{figure}
		\centering
		\includegraphics[width=0.99\linewidth]{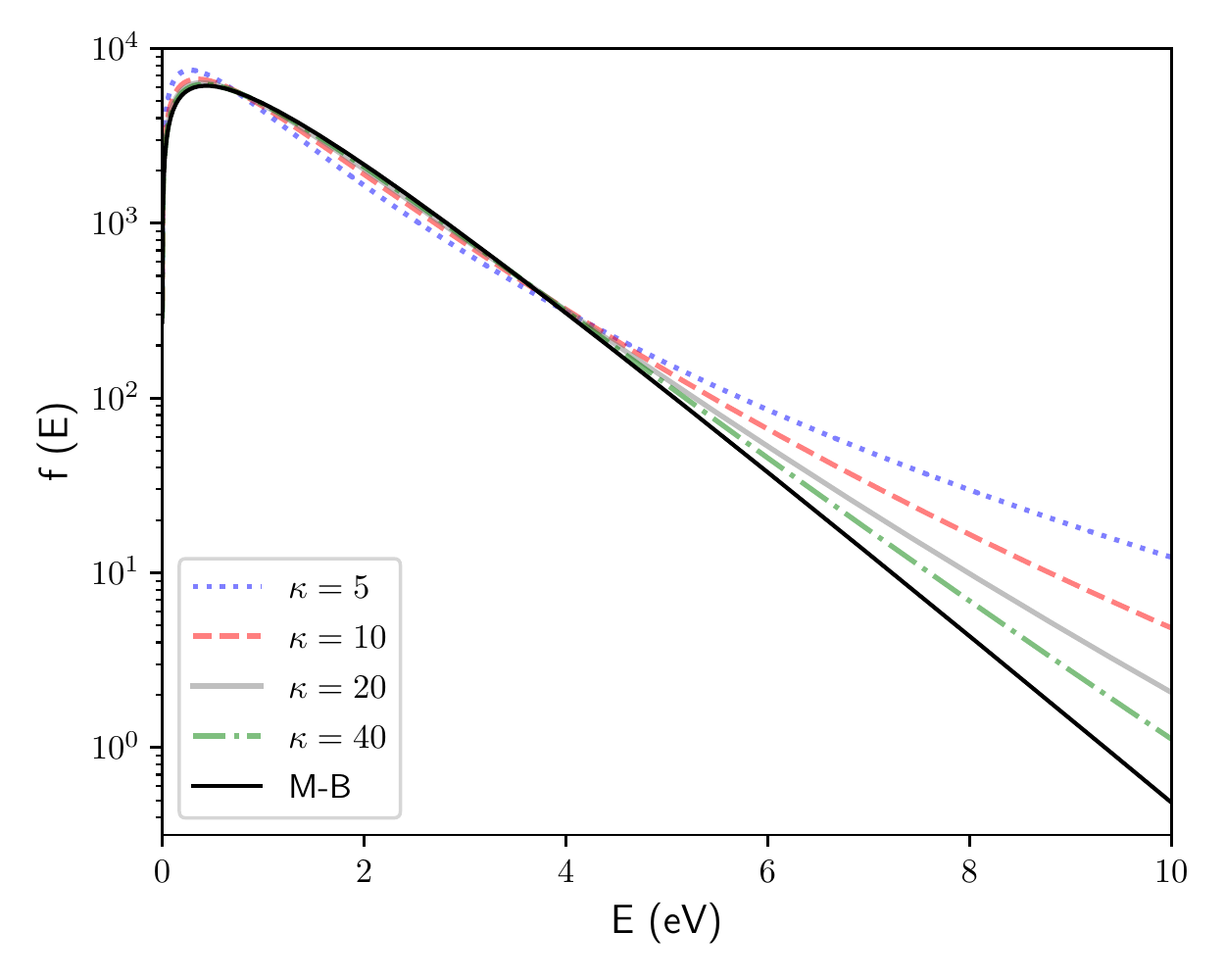}
		\vspace{-0.4cm}
		\caption{\textit{$\kappa$} electron energy distributions with \textit{$\kappa$}=5, \textit{$\kappa$}=10, \textit{$\kappa$}=20, \textit{$\kappa$}=40, and a Maxwellian.}\label{fig:kmb}
	\end{figure}

	Fig. \ref{fig:kmb} compares \textit{$\kappa$} electron energy distributions for several values of \textit{$\kappa$} (different level of deviations from equilibrium) with a M-B distribution, all at the same temperature.
	In comparison with a M-B distribution, \textit{$\kappa$} distributions reach their maximum at lower energies, and as \textit{$\kappa$} decreases the peak of the function moves to lower energies. In a \textit{$\kappa$} distribution there are more electrons both with high and low kinetic energies when compared with an M-B. At intermediate energies there is a depletion of electrons.

	\section{MAPPINGS 1e Models}\label{sec:models}
	
	We used the photoionization code MAPPINGS 1e \citep{Binette1985, Ferruit1997, Binette2012}, which includes the option of considering \textit{$\kappa$} distributions of electron energies instead of the equilibrium M-B distribution \citep[see][]{Binette2012}.
	
	To calculate the excitation and deexcitation rates of the collisionally excited lines for a \textit{$\kappa$} distribution, \citet{Binette2012} used the correction factor derived by \citet{Nicholls2012}. \citet{Binette2012} derived an expression for the temperature to compute recombination rates, and implemented the enhancement of collisional ionization using two sets of correction factors that depend on the ionization process. For more details on how the \textit{$\kappa$} distribution was implemented on MAPPINGS, 1e see \citet{Binette2012}.  
	
	The photoionization models consider an isochoric, plane-parallel slab of gas which is illuminated by a power-law spectrum of ionizing radiation ($S_{\nu} \propto \nu^{\alpha}$). The intensity of the radiation is given by the ionization parameter\footnote{$U=\frac{Q}{4\pi r^2 n_H c}$, where $Q$ is the ionizing photon luminosity, $r$ is the distance between the ionized cloud and the ionizing source, $c$ is the speed of light and $n_H$ is the hydrogen gas density.}, $U$, which indicates the intensity of the ionizing radiation as felt by the cloud. We adopted ionization-bounded termination and ended the calculation when the ionization fraction of hydrogen drops below 0.01. Our models did not consider depletion of metals unto dust. 
	
	Our goal is to investigate how line ratio predictions change depending on the adopted electron distribution function. In order to do this we compute a large grid of photoionization models using M-B and \textit{$\kappa$} electron energy distributions for identical parameters. In addition to models using the M-B distribution, we compute models where \textit{$\kappa$} has the values 5, 10, 20, and 40. \citet{Nicholls2012} suggested that for PNe and HII regions a value of \textit{$\kappa$} in the range 10$\lesssim$ \textit{$\kappa$} $\lesssim$ 20 would be enough to solve the discrepancies. However, we choose a wider range to explore all possibilities. 
	
	We compute sequences of models in which U varies over the range 10$^{-5}$ - 1.8. We investigate values of the power-law index $\alpha$=-2, -1.5, -1 which are consistent with values adopted in the literature \citep[e.g.][]{Robinson1987, VM1997, Zheng1997, Radovich1998, Groves2004}. 
	The models computed assuming $\alpha$=-1.5 and $\alpha$=-2 have a high energy cut-off for the ionizing continuum of 5$\times$10$^4$ eV, to ensure that the mean ionizing photon energy matches the value derived for radio-loud AGN \citep[e.g.][]{Mathews1987}. Models with an ionizing continuum with $\alpha$=-1 assume a cut-off energy of 1000 eV to ensure that the mean photon energy matches that of the $\alpha$=-1.5 power-law.
	
	For the hydrogen density we assume $n_H$ = 100, 10$^4$, 10$^6$ cm$^{-3}$ following the results from \citet{McCarthy1990, VM1999c}. 
	
	For the metallicity (defined as Z=O/H) we consider Z/Z$_{\odot}$=0.1, Z/Z$_{\odot}$=0.5, Z/Z$_{\odot}$=1.0, Z/Z$_{\odot}$=2.0 and Z/Z$_{\odot}$=3.0. The gas metallicity is expressed with respect to the solar abundance set of \citet{Asplund2006}.
	The chemical abundances have been scaled with all metals scaled proportionally to oxygen except nitrogen and carbon. Nitrogen and carbon are scaled quadratically with oxygen \citep[e.g.][]{Henry2000} when Z$\gtrsim$0.3Z$_{\odot}$ . Below this metallicity they are scaled linearly \citep[N/H $\propto$ O/H, e.g.][]{Henry2000, Vernet2001}.
	
	A summary of all the parameters used in the photoionization models is presented in Table \ref{table: Model parameters}. In total, the grid has (100 \textit{U}) x (5 \textit{Z}) x (3 n$_H$) x (5 \textit{$\kappa$}) = 7500 models.
	
	\begin{table}
		\caption{Parameters of the photoionization models. (1) Ionization parameter; (2) spectral index of the ionizing continuum; (3) metallicity; (4) hydrogen density in cm$^{-3}$; (5) electron energy distribution.}
		\label{table: Model parameters}     
		\centering                          
		\begin{tabular}{c c}       	
			Parameter   &   Value \\
			\hline \hline 
			\makecell{\bf{\textit{U}} \\ (1)}  &  10$^{-5}$ - 1.8 \\
			\hline 
			\makecell{$\alpha$ \\ (2)} &  -2, -1.5, -1 \\ 
			\hline 
			\makecell{\textit{Z/Z$_{\odot}$} \\ (3)} &  0.1, 0.5, 1.0, 2.0, 3.0  \\  
			\hline 
			\makecell{n$_H$ (cm$^{-3}$) \\  (4)}  &    100, 10$^4$, 10$^6$  \\
			\hline                                  
			\makecell{EED  \\  (5)}  &   M-B, \textit{$\kappa$}=5, 10, 20, 40  \\
			\hline                                 		
		\end{tabular}
	\end{table}

	\section{Results}\label{sec:results}
	
	In this section we present the results of the photoionization model calculations. We compare the results obtained using M-B and \textit{$\kappa$} electron energy distributions when different parameters are adopted. We hereafter refer to \textit{$\kappa$} / M-B of a given emission line as the quotient between the line flux obtained when a \textit{$\kappa$} distribution is used with respect to the same line flux assuming a M-B distribution. 
	In Figs. \ref{fig:evskmbiru1e-4} to \ref{fig:E_vs_k_MB_UV_U1_8} we show results from our grid of photoionization models, for emission lines in the IR, optical, and UV. 
	
	The figures are organized as follows: each emission line is shown for different model parameters, \textit{$\kappa$} and M-B electron energy distributions. Each panel in the figures corresponds to different assumptions concerning the gas metallicity, $Z$.

	\subsection{Infrared emission lines}\label{sec:IR} 
	
	We show our predictions for the following lines [ArIII]$\lambda$9 $\mu$m, [NeV]$\lambda$14 $\mu$m, [SIV]$\lambda$10.5 $\mu$m, [NeII]$\lambda$12.8 $\mu$m, [NeIII]$\lambda$15.6 $\mu$m, [NeV]$\lambda$24 $\mu$m, [OIV]$\lambda$25.9 $\mu$m, [OIII]$\lambda$52 $\mu$m, [NII]$\lambda$122 $\mu$m, [OI]$\lambda$63 $\mu$m, [OIII]$\lambda$88 $\mu$m, and [CII]$\lambda$158 $\mu$m.
	
	In general, our calculations show that the flux of infrared (IR) emission lines is not enhanced when assuming \textit{$\kappa$} distributions of electron energies in the photoionization models. On the contrary, the line fluxes are usually predicted to be weaker for \textit{$\kappa$} distributions. Our results are similar to those in \citet{Nicholls2012, Nicholls2013, Dopita2013} who showed that the effect of \textit{$\kappa$} on the IR lines was very weak. 
	To illustrate this, we show the ratio between predictions assuming \textit{$\kappa$} and M-B distributions of electron energies for a selection of IR emission lines typically observed in the narrow-line region of AGNs. 
	
\begin{figure*}
		\centering
		\includegraphics[width=0.99\linewidth]{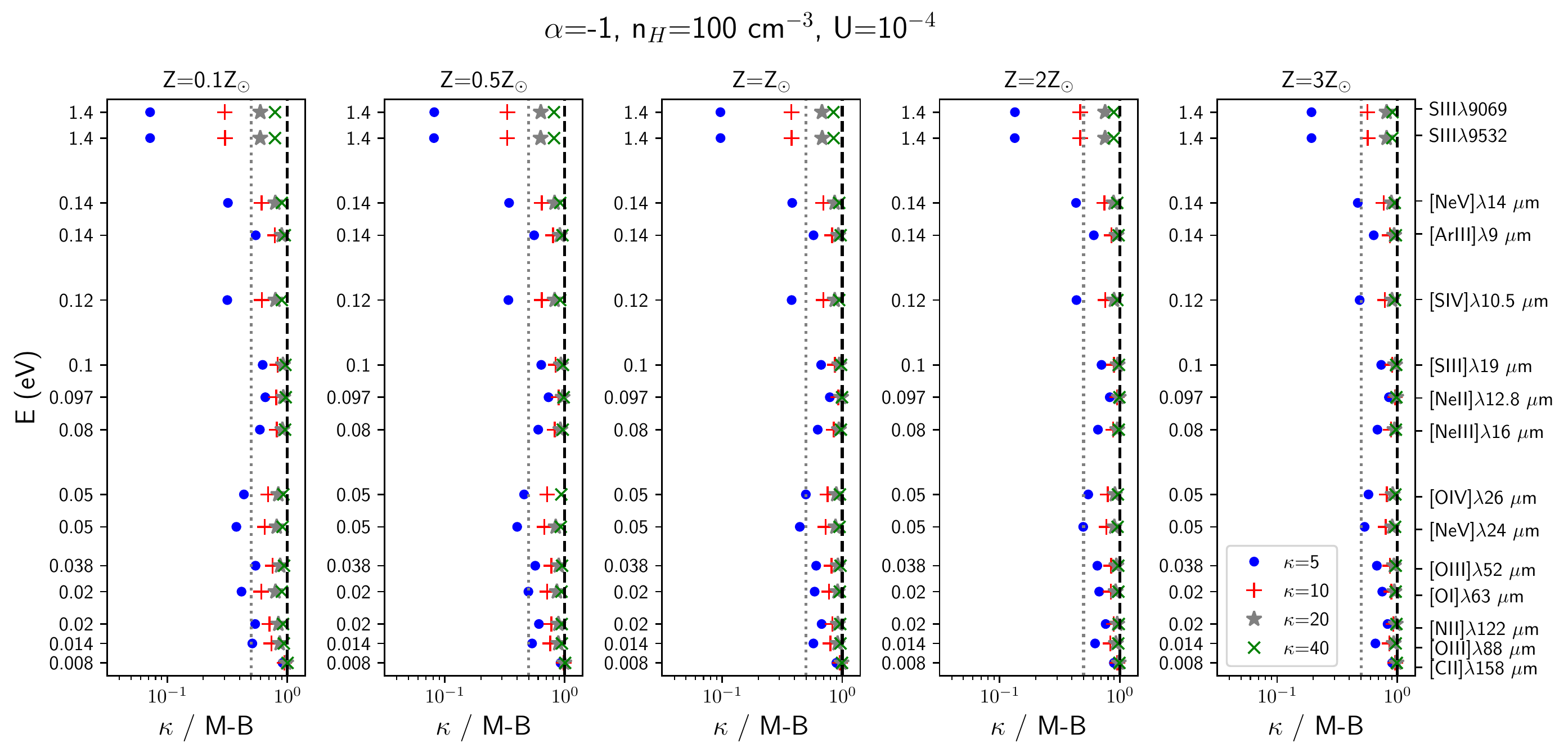}
		\vspace{-0.35cm} \caption{ Ratios of model fluxes for \textit{$\kappa$} and M-B electron energy distributions, for key nebular lines, arranged in order of increasing energy for $U$=1$\times$10$^{-4}$. Each panel corresponds to a different assumption about gas metallicity. Each marker corresponds to a different \textit{$\kappa$} distribution. The grey dotted vertical line shows the position where \textit{$\kappa$}/M-B = 0.5.}
		\label{fig:evskmbiru1e-4}
	\end{figure*}

	The IR line fluxes calculated for $U$=1$\times$10$^{-4}$ are shown in Fig. \ref{fig:evskmbiru1e-4}. In this figure we show the ratio between predictions assuming \textit{$\kappa$} and M-B distributions for all the lines as a function of the excitation energy of each line. The results cover different values of \textit{$\kappa$} and $Z$.    
	
	Similar results are observed across all emission lines: \textit{$\kappa$} distributions shift the line ratios towards lower fluxes. The shift becomes larger for low values of \textit{$\kappa$}, i.e., for larger deviations from equilibrium.  As the metallicity increases the predictions assuming \textit{$\kappa$} distributions approach the predictions from M-B distributions. 

\begin{figure*}
		\centering
		\includegraphics[width=0.99\linewidth]{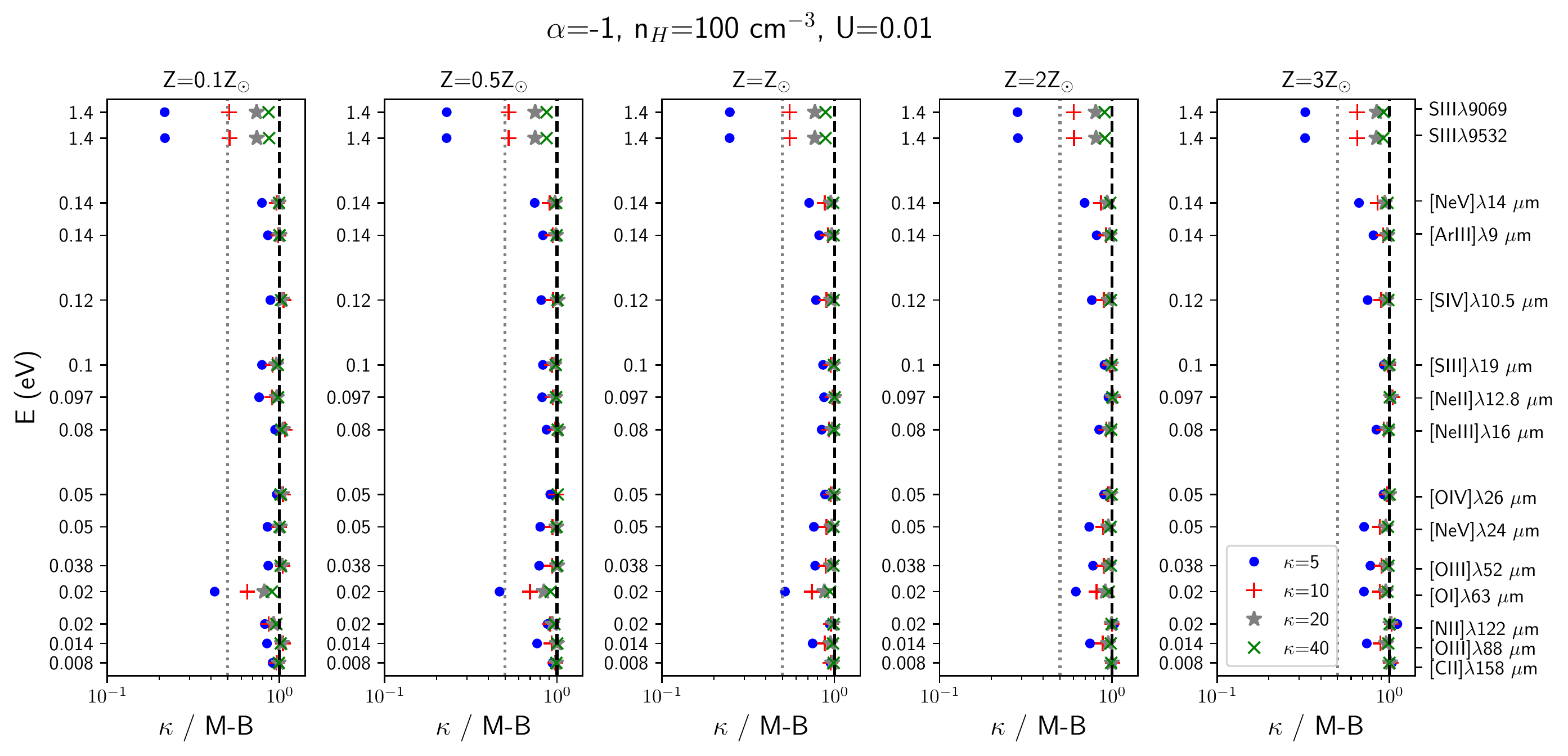}
		\vspace{-0.35cm}	\caption{Same as Fig. \ref{fig:evskmbiru1e-4} for $U$=0.01.}
		\label{fig:evskmbiru001}
	\end{figure*}

	Similar results are obtained for $U$=0.01, as shown in Fig. \ref{fig:evskmbiru001}. The lines are more intense when a Maxwellian is used. With such ionization parameter the differences between model predictions become smaller, with the ratio \textit{$\kappa$} / M-B of most lines being shifted closer to unity, when compared to the previous figure.     

\begin{figure*}
		\centering
		\includegraphics[width=0.99\linewidth]{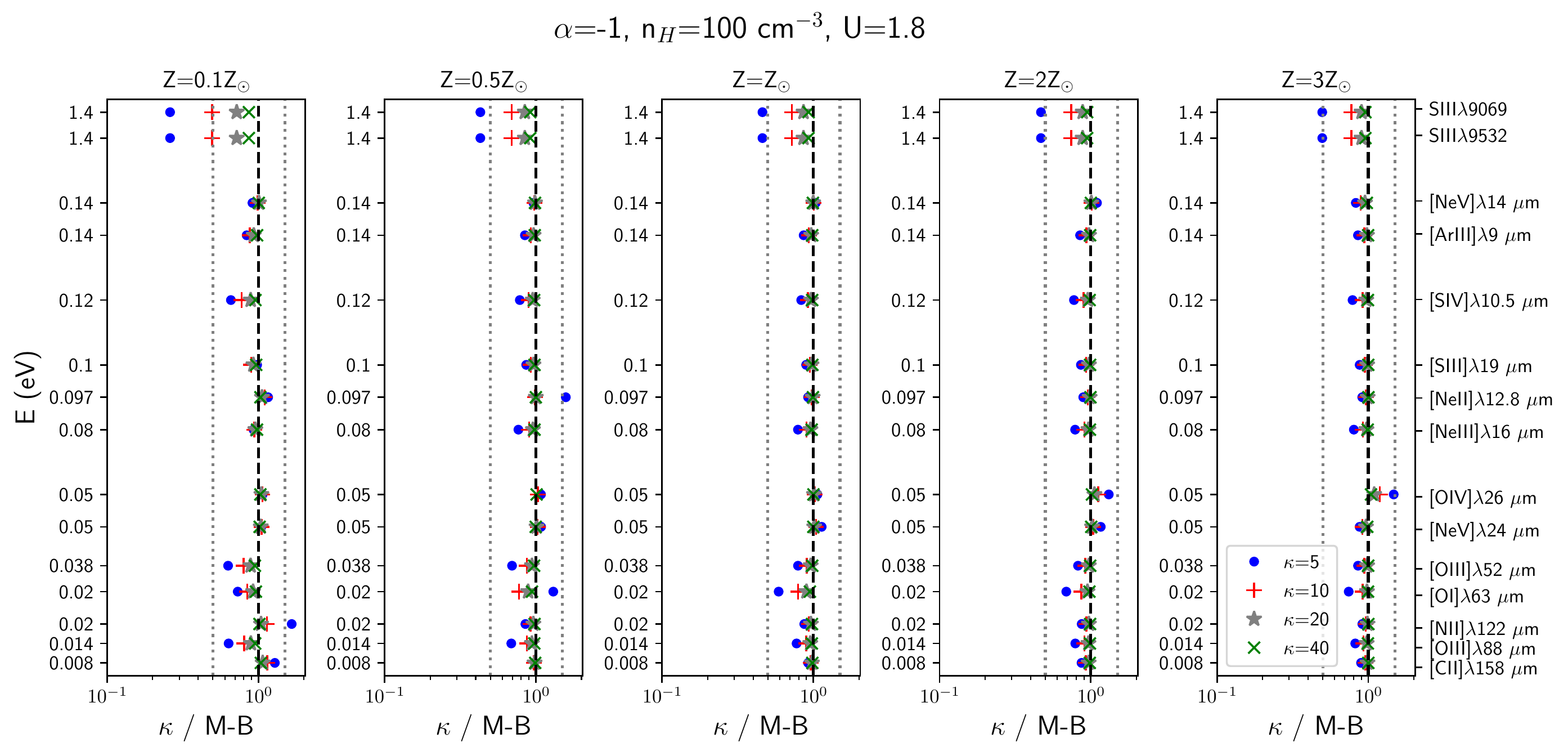}
		\vspace{-0.35cm}	\caption{ Same as Fig. \ref{fig:evskmbiru1e-4} for $U$=1.8. In this case another grey dotted vertical line was added to mark the position where \textit{$\kappa$} / M-B is 1.5. }
		\label{fig:evskmbiru18}
	\end{figure*}

	Finally, the results at $U$=1.8 are shown in Fig. \ref{fig:evskmbiru18} where most line ratios are predicted to be similar for both distributions, although there are some slight enhancements when assuming \textit{$\kappa$} distributions with low gas metallicity and \textit{$\kappa$}=5 (e.g. [NII]$\lambda$122 $\mu$m, [OI]$\lambda$63, [NeII]$\lambda$12.8 $\mu$m). [OIV]$\lambda$26 $\mu$m is also slightly enhanced but in this case only for higher metallicities.
	
	In summary, assuming \textit{$\kappa$} distributions of electron energies instead of M-B distributions moves the IR line fluxes to lower values, except for a small parameter space at low metallicities and with high ionization parameter. 
	Except within a small parameter space with high U (e.g. U=1.8), the predicted flux ratios are in the range 0.5 $\lesssim$ \textit{$\kappa$} / M-B \textless 1 for the majority of IR lines and model parameters.

	\subsection{Optical emission lines}\label{sec:optical} 
	
	The effect of \textit{$\kappa$} distributions on the emission spectra of \textit{optical} lines is complex. It depends on both metallicity and ionization parameter.
	
\begin{figure*}
		\centering
		\includegraphics[width=0.99\linewidth]{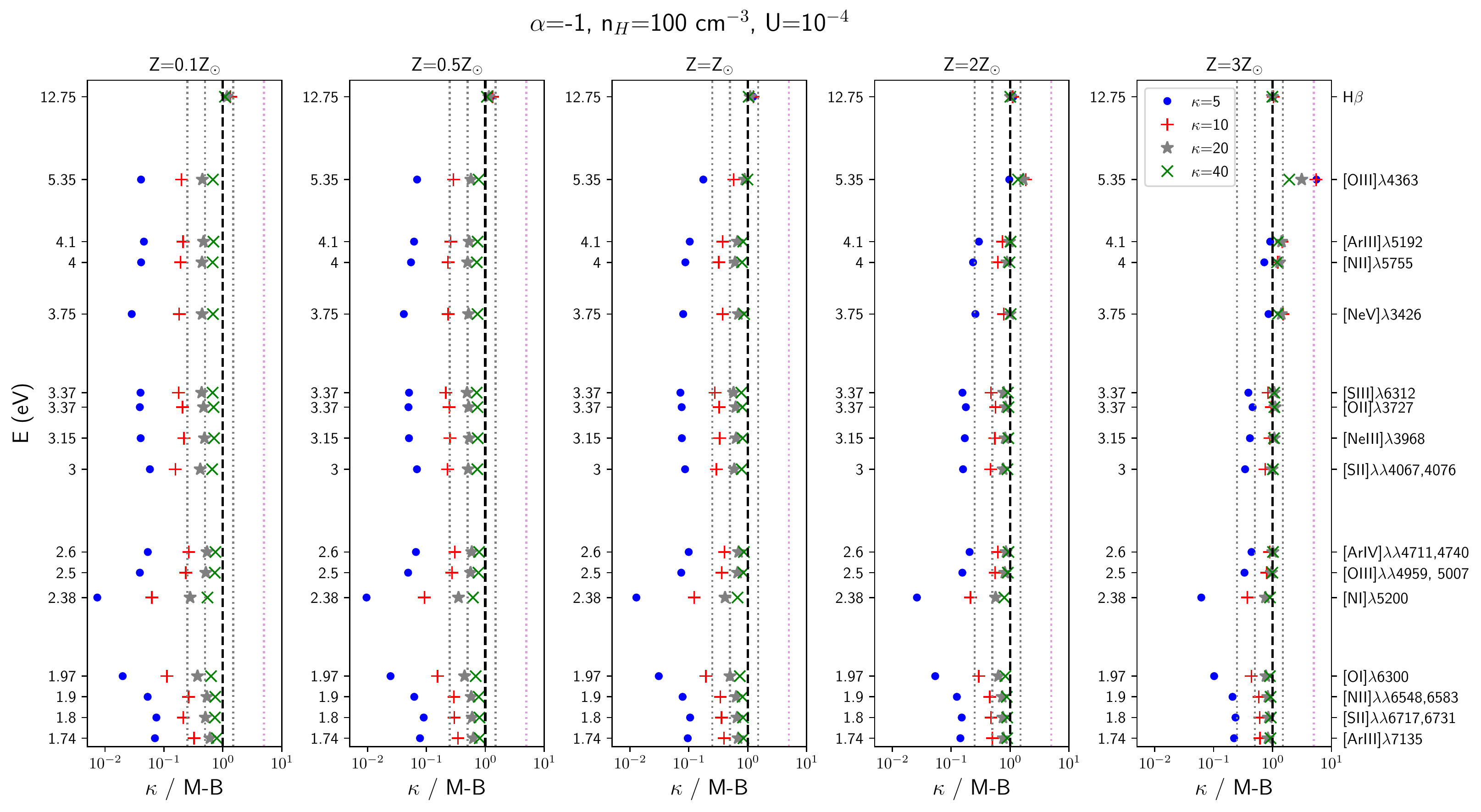}
		\vspace{-0.3cm} \caption{ Ratios of model flux predictions for \textit{$\kappa$} and M-B electron energy distributions, for key nebular lines, arranged in order of increasing energy for $U$=1$\times$10$^{-4}$. The grey dotted vertical lines mark the positions where \textit{$\kappa$} / M-B is 0.25, 0.5 and 1.5. The dashed line indicates where \textit{$\kappa$} / M-B is 5.}
		\label{fig:E_vs_k_MB_Optical_U1e-4}
	\end{figure*}

	Fig. \ref{fig:E_vs_k_MB_Optical_U1e-4} shows the predictions for $U$=1$\times$10$^{-4}$.
	At low metallicities, Z$\lesssim$Z$_{\odot}$, the predictions from models assuming M-B distributions exceed the flux predictions from \textit{$\kappa$} models (this is \textit{$\kappa$} / M-B \textless 1) for all the lines considered except H$\beta$, which is very similar for all distributions independently of the metallicity. As the metallicity increases, line fluxes predicted by \textit{$\kappa$} distributions increase slightly (\textit{$\kappa$} / M-B \textgreater 1), however the only flux that gets significantly higher when using \textit{$\kappa$} models is [OIII]$\lambda$4363.
	For this ionization parameter ($U$=1$\times$10$^{-4}$), a large deviation from equilibrium, i.e., a \textit{$\kappa$}=5 distribution, \textit{$\kappa$} / M-B becomes smaller than 1 for almost all lines considered, with the exception of [OIII]$\lambda$4363 at high metallicity (3Z$_{\odot}$).
	
\begin{figure*}
		\centering
		\includegraphics[width=0.99\linewidth]{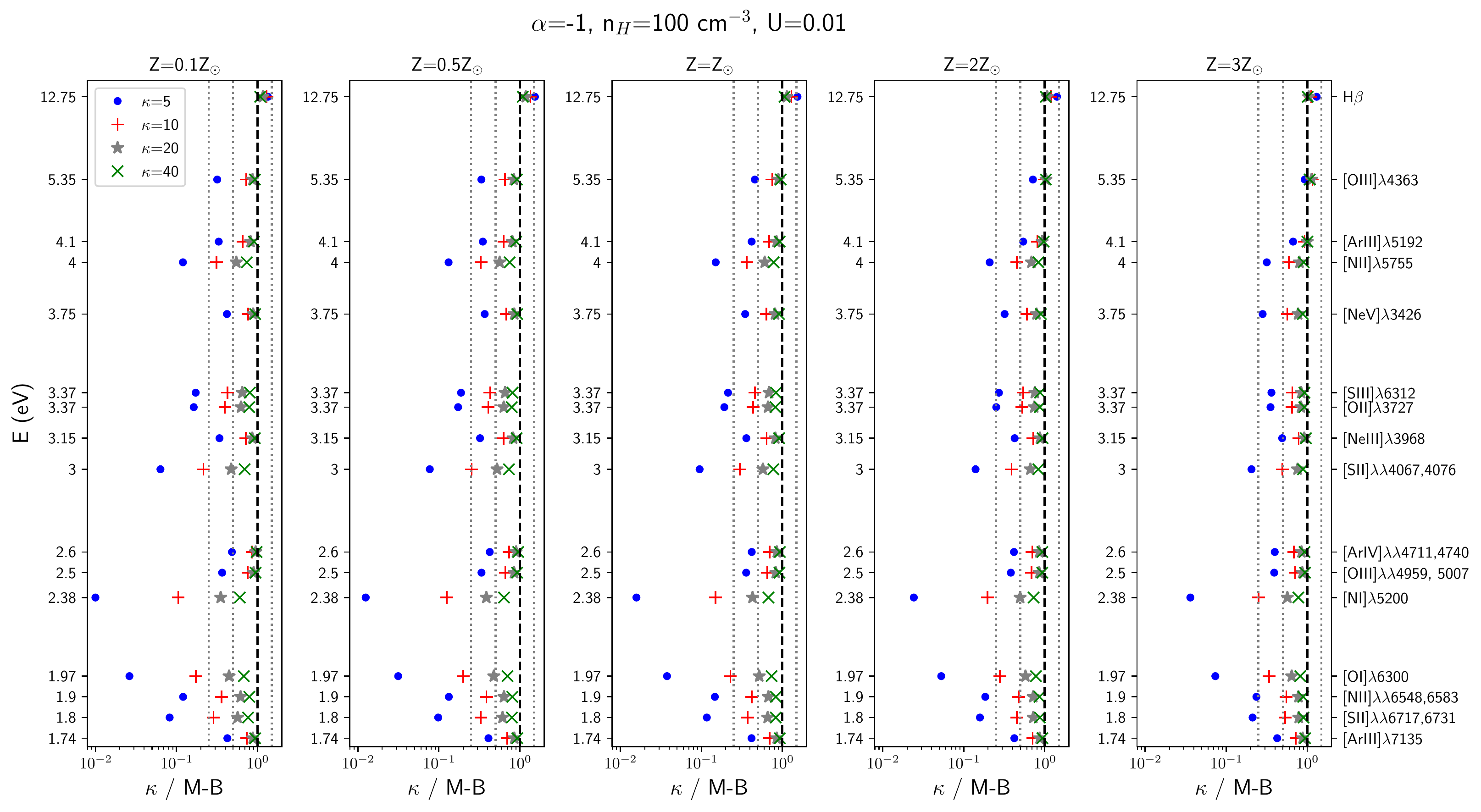}
		\vspace{-0.3cm} \caption{Same as Fig. \ref{fig:E_vs_k_MB_Optical_U1e-4} for $U$=0.01.}
		\label{fig:E_vs_k_MB_Optical_U0_01}
	\end{figure*}

	In Fig. \ref{fig:E_vs_k_MB_Optical_U0_01} the \textit{$\kappa$} / M-B predictions for $U$=0.01 are presented.
	In general, optical line fluxes are not enhanced when assuming \textit{$\kappa$} distributions, with the exception of H$\beta$. The maximum H$\beta$$_\textit{$\kappa$}$ / H$\beta$$_{M-B}$ is 1.5, reached for \textit{$\kappa$}=5 and Z=Z$_{\odot}$.
	Increasing the metallicity shifts most line fluxes predicted by \textit{$\kappa$} distributions closer to the fluxes predicted by a Maxwellian.
	
\begin{figure*}
		\centering
		\includegraphics[width=0.99\linewidth]{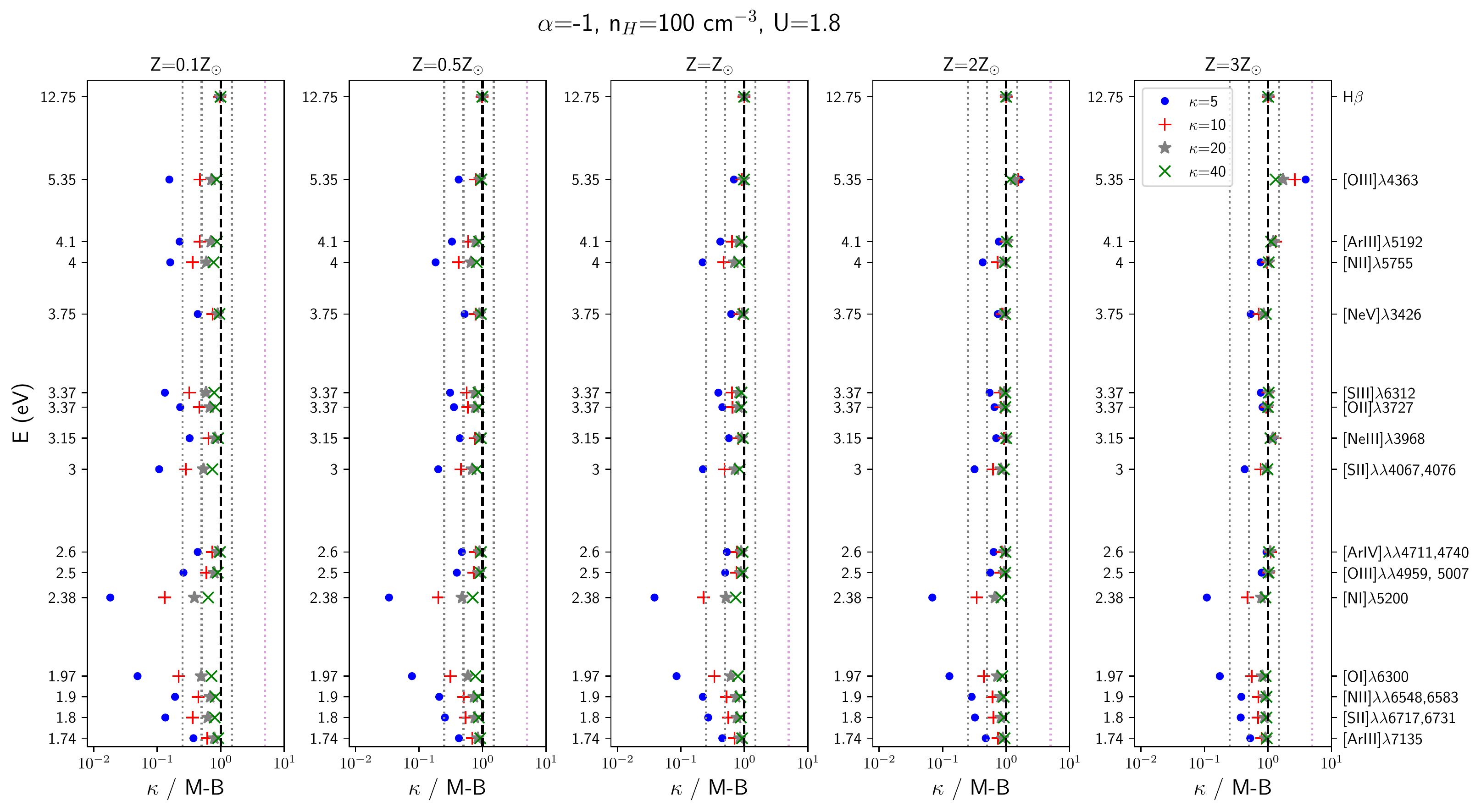}
		\vspace{-0.3cm} \caption{Same as Fig. \ref{fig:E_vs_k_MB_Optical_U1e-4} for $U$=1.8.}
		\label{fig:E_vs_k_MB_Optical_U1_8}
	\end{figure*}

	The \textit{$\kappa$} / M-B line ratios for $U$=1.8 are presented in Fig. \ref{fig:E_vs_k_MB_Optical_U1_8}.
	In line with the previous results, the predictions from M-B exceed or are very similar to the predictions from \textit{$\kappa$} models for most line ratios (this is, \textit{$\kappa$} / M-B $\lesssim$1). The only exception is [OIII]$\lambda$4363, for which the increase is \textgreater 1 when Z$\gtrsim$2Z$_{\odot}$. 
	
	The impact of \textit{$\kappa$} distributions depends on the metallicity: as the metallicity increases, \textit{$\kappa$} / M-B approaches 1.

	\subsection{UV emission lines}\label{sec:UV} 
	
	In general, when using \textit{$\kappa$} distributions the UV emission line fluxes are enhanced (\textit{$\kappa$} / M-B \textgreater 1) for a wide range of ionization and metallicity parameters. This is primarily due to the fact that the \textit{$\kappa$} distributions place relatively more electrons at energies capable of exciting the UV lines, compared to the M-B distribution. 
	
\begin{figure*}
		\centering
		\includegraphics[width=0.99\linewidth]{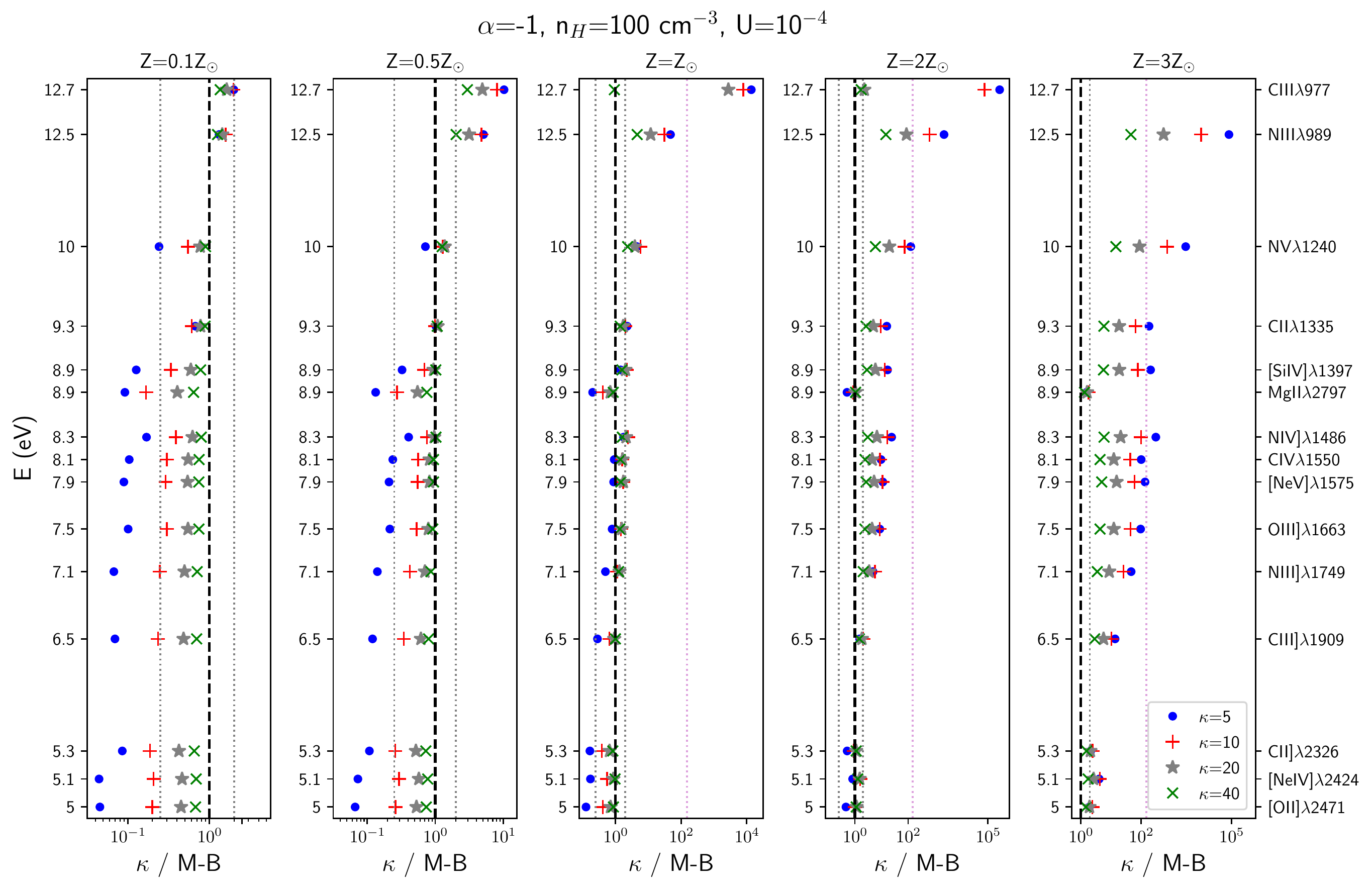}
		\vspace{-0.3cm} \caption{ Ratios of model fluxes for \textit{$\kappa$} and M-B electron energy distributions, for key nebular lines, arranged in order of increasing energy for $U$=1$\times$10$^{-4}$. The grey dotted vertical lines mark the positions where \textit{$\kappa$} / M-B is 0.25, and 2. The dashed line indicates where \textit{$\kappa$} / M-B is 150. Note that at Z=3Z$_{\odot}$, the ratio for the CIII$\lambda$977 line is not shown, as its flux in the M-B model is too low. }
		\label{fig:E_vs_k_MB_UV_U1e-4}
	\end{figure*}

	Fig. \ref{fig:E_vs_k_MB_UV_U1e-4} shows the predictions for $U$=1$\times$10$^{-4}$.
	In comparison with a M-B distribution, \textit{$\kappa$} distributions have high-energy tails, thus lines with high excitation energies will be enhanced when \textit{$\kappa$} distributions are assumed. This is particularly obvious for the UV lines with the highest excitation energies. The CIII$\lambda$977 \AA \, line which has an excitation energy of 12.3 eV, is strongly enhanced by the high-energy tail of \textit{$\kappa$} distributions (see Fig. \ref{fig:kmb}). Similar results are obtained for [NIII]$\lambda$991 \AA \, which has an excitation energy of 12.6 eV. 
	
	The CIII$\lambda$977 and NIII$\lambda$989 line fluxes are enhanced for \textit{$\kappa$} distributions for all metallicities considered. As the metallicity increases, \textit{$\kappa$} / M-B \textgreater 1 for an increasing number of lines.
	
\begin{figure*}
		\centering
		\includegraphics[width=0.99\linewidth]{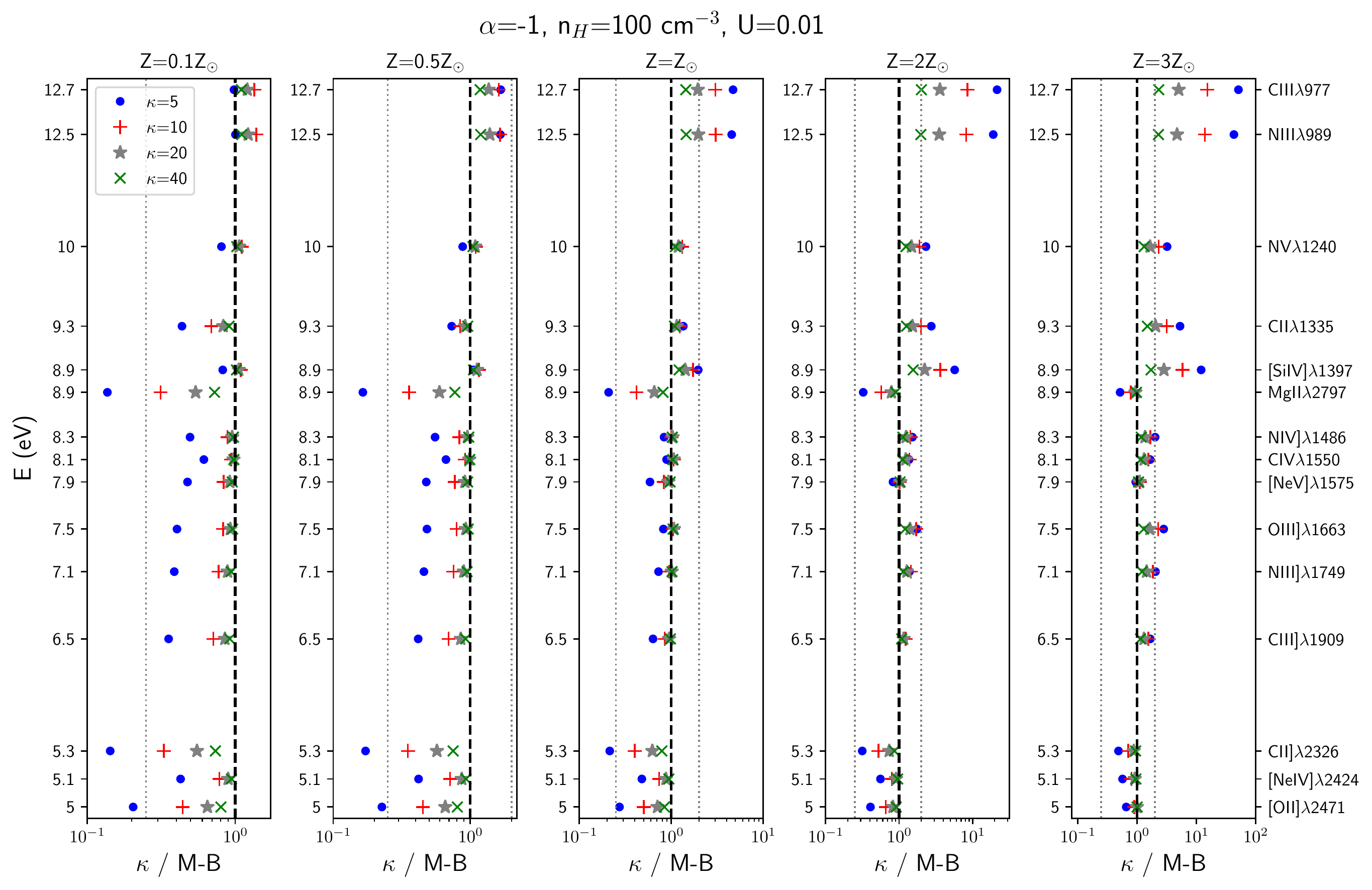}
		\vspace{-0.3cm} \caption{Same as Fig. \ref{fig:E_vs_k_MB_UV_U1e-4} for $U$=0.01.}
		\label{fig:E_vs_k_MB_UV_U0_01}
	\end{figure*}

\begin{figure*}
		\centering
		\includegraphics[width=0.99\linewidth]{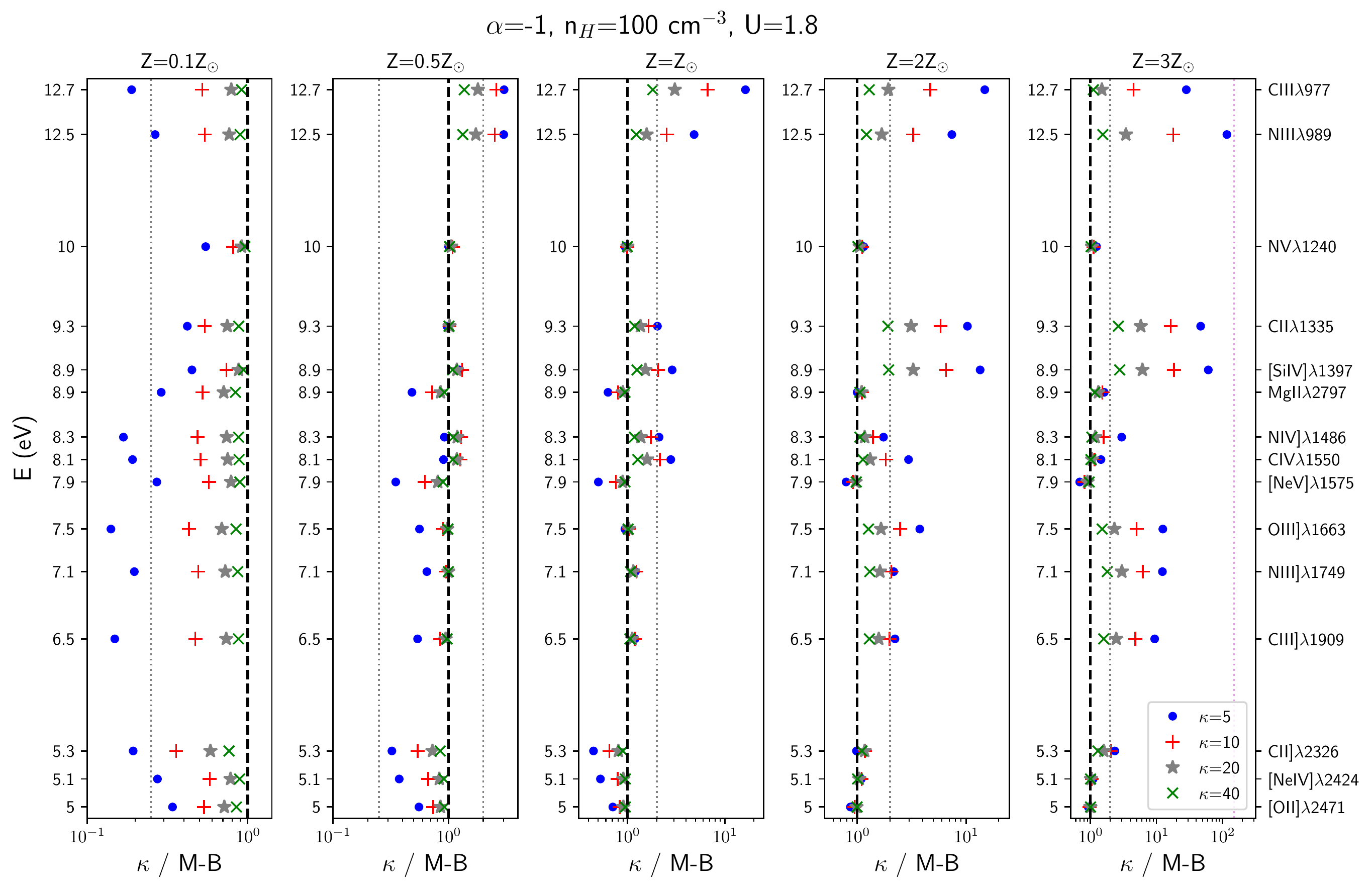}
		\vspace{-0.3cm} \caption{Same as Fig. \ref{fig:E_vs_k_MB_UV_U1e-4} for $U$=1.8.}
		\label{fig:E_vs_k_MB_UV_U1_8}
	\end{figure*}

	Fig. \ref{fig:E_vs_k_MB_UV_U0_01} shows the predictions for $U$=0.01.
	As the ionization parameter increases, the magnitude of the enhancement when assuming \textit{$\kappa$} distributions diminishes. This is clearly seen in the calculated fluxes of CIII$\lambda$977 and NIII$\lambda$989, which are still enhanced for all metallicities, but the enhancement is much smaller. 
	The other line fluxes also become less enhanced with increasing ionization parameter. They eventually reach at a turnover point and then the enhancement becomes gradually stronger as can be seen in Fig. \ref{fig:E_vs_k_MB_UV_U1_8}. This  turnover ionization parameter depends on the chosen electron energy distribution and on the emission line in question.  
	For constant ionization parameter, as the metallicity increases we see increases in  line fluxes when using \textit{$\kappa$} distributions. 
	
	Fig. \ref{fig:E_vs_k_MB_UV_U1_8} shows the predictions for $U$=1.8.
	At this value of U, the impact of \textit{$\kappa$} distributions is strongly dependent on the gas metallicity. For low metallicities no line fluxes show enhancements when \textit{$\kappa$} distributions are used, and as the metallicity increases UV lines become enhanced. Almost all ratios are enhanced (\textit{$\kappa$} / M-B \textgreater 1) for \textit{$\kappa$} distributions when Z=3Z$_{\odot}$.
	
	Although HI Ly$\alpha\lambda$1216 is often the most luminous of the UV emission lines, we omitted it from our analysis because it can be strongly affected by the presence of young stars \citep{VM2007a}, resonance scattering effects \citep{VM1996}, and contamination by OV] $\lambda\lambda$1213.8,1218.3 emission \citep{Humphrey2019b}. These effects make the line unreliable for quantitative analyses of the kind performed herein.

	\section{Comparison with observed line ratios} \label{sec:data comparison}
	
	It is interesting to investigate whether observed emission line ratios measured in the emission line regions of AGN are better reproduced using a M-B distribution or a \textit{$\kappa$} distribution. In the following figures, the predictions of the photoionization models described in Section \ref{sec:models} are compared with observed emission line ratios. The data was compiled from \citet{Solorzano2004}, \citet{Vernet2001}, \citet{Humphrey2008a} and \citet{Silva2018} and \citet{Silva2020}. It consists in 15 high-z radio galaxies (HzRGs), and two Seyfert galaxies, NGC 1068 \citep[][]{Kraemer1998}, NGC 3393 \citep{Diaz1988} and 126 Type 2 quasars (QSO2s). The original Type 2 quasar sample had 144 galaxies but we excluded objects where less than 3 lines were observed. 
	
\begin{figure*}
		\centering
		\includegraphics[width=0.99\linewidth]{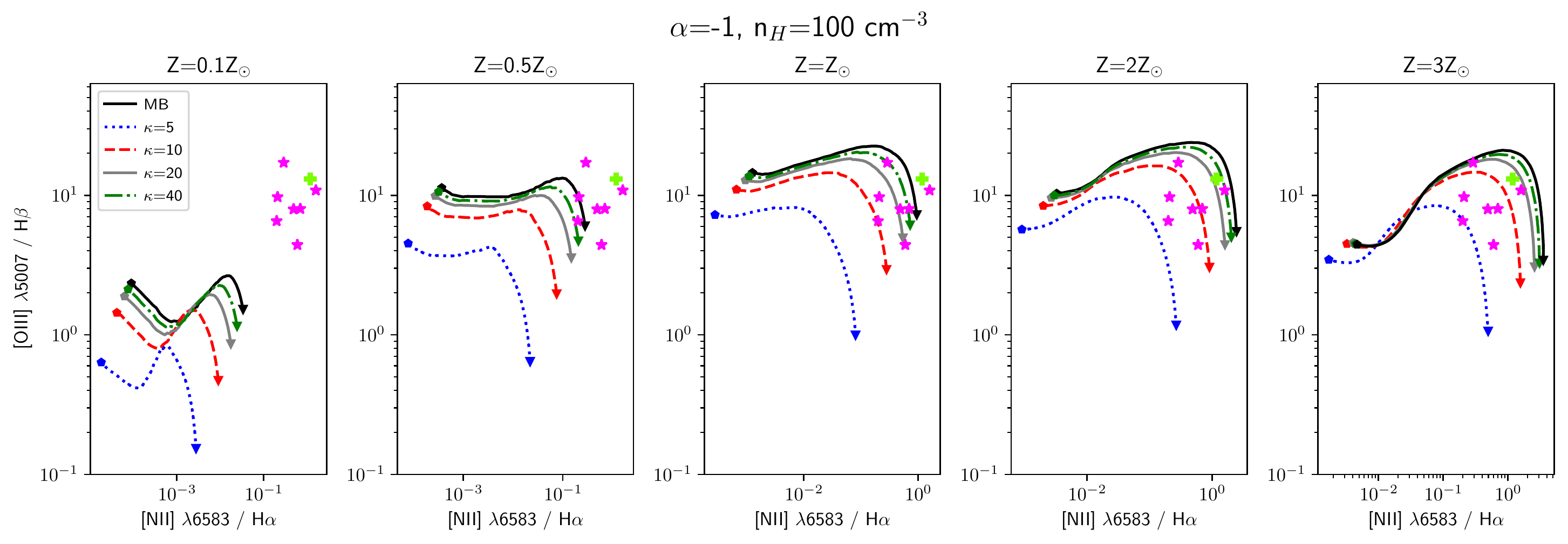}
		\vspace{-0.3 cm}\caption{Comparison of the observed emission line ratios from the selected samples: HzRGs (pink stars) and NGC 3393 (green plus) with photoionization models using a power law index $\alpha$=-1 and n$_H$=100 cm$^{-3}$. Curves with different colors represent models computed assuming different electron energy distributions. The ionization parameter sequences increase from the triangle (\textit{U}=1$\times$10$^{-3}$) to the pentagon (\textit{U}=1.8) symbols.}
		\label{fig:OIII_Hb_vs_NII_Ha_SecNC}
	\end{figure*}

\begin{figure*}
		\centering
		\includegraphics[width=0.99\linewidth]{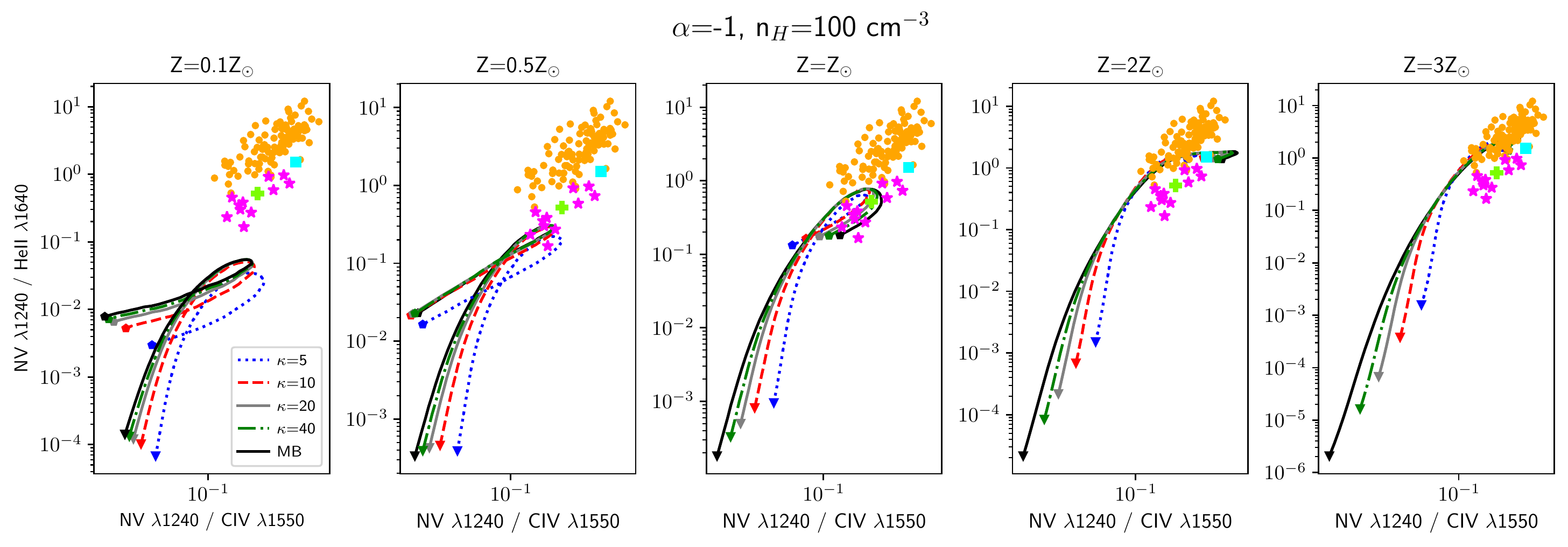}
		\vspace{-0.3 cm}\caption{Diagnostic diagram NV$\lambda$1240/HeII$\lambda$1640 versus NV$\lambda$1240/CIV$\lambda$1550. HzRGs are represented by pink stars, QSO2s by gold dots, NGC 1068 by a cyan square and NGC 3393 by a green plus.} 
		\label{fig:NV_HeII_vs_NV_CIV_alpha-1}
	\end{figure*}

\begin{figure*}
		\centering
		\includegraphics[width=0.99\linewidth]{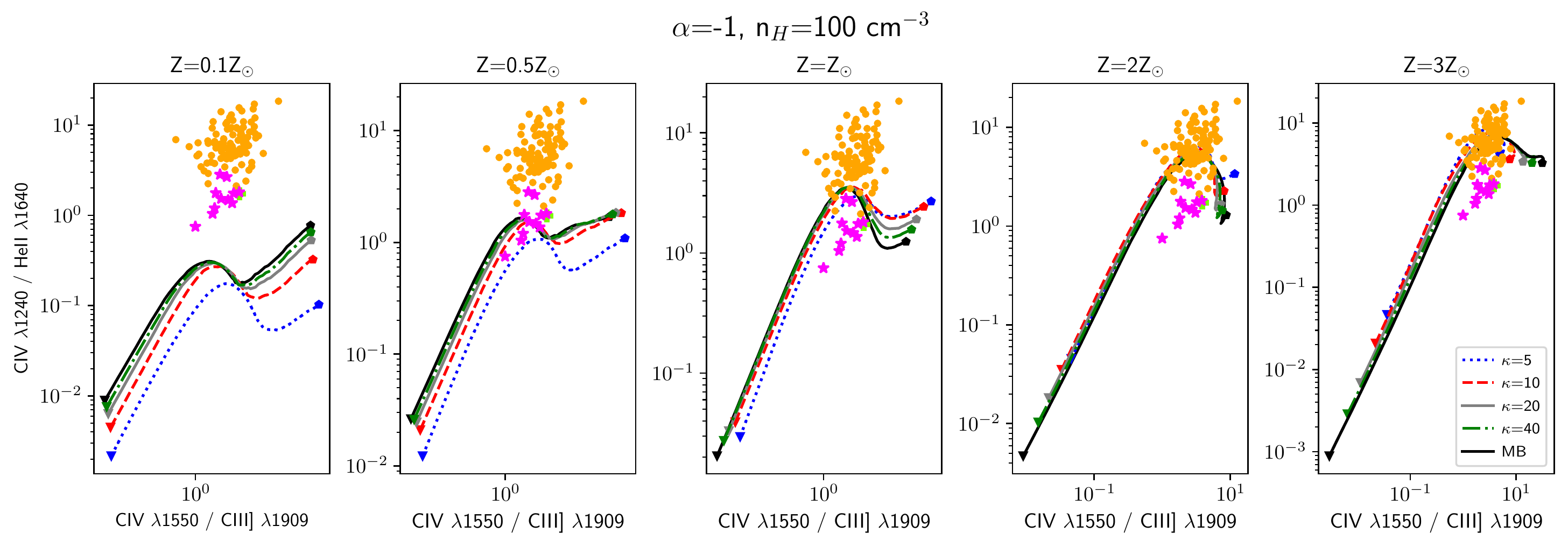}
		\vspace{-0.3 cm}\caption{Diagnostic diagram CIV$\lambda$1550/HeII$\lambda$1640 as a function of CIV$\lambda$1550/CIII]$\lambda$1909. Symbols and lines as in Fig. \ref{fig:OIII_Hb_vs_NII_Ha_SecNC}. }
		\label{fig:CIV_HeII_vs_CIV_CIII_SecNC_alpha-1}
	\end{figure*}

\begin{figure*}
		\centering
		\includegraphics[width=0.99\linewidth]{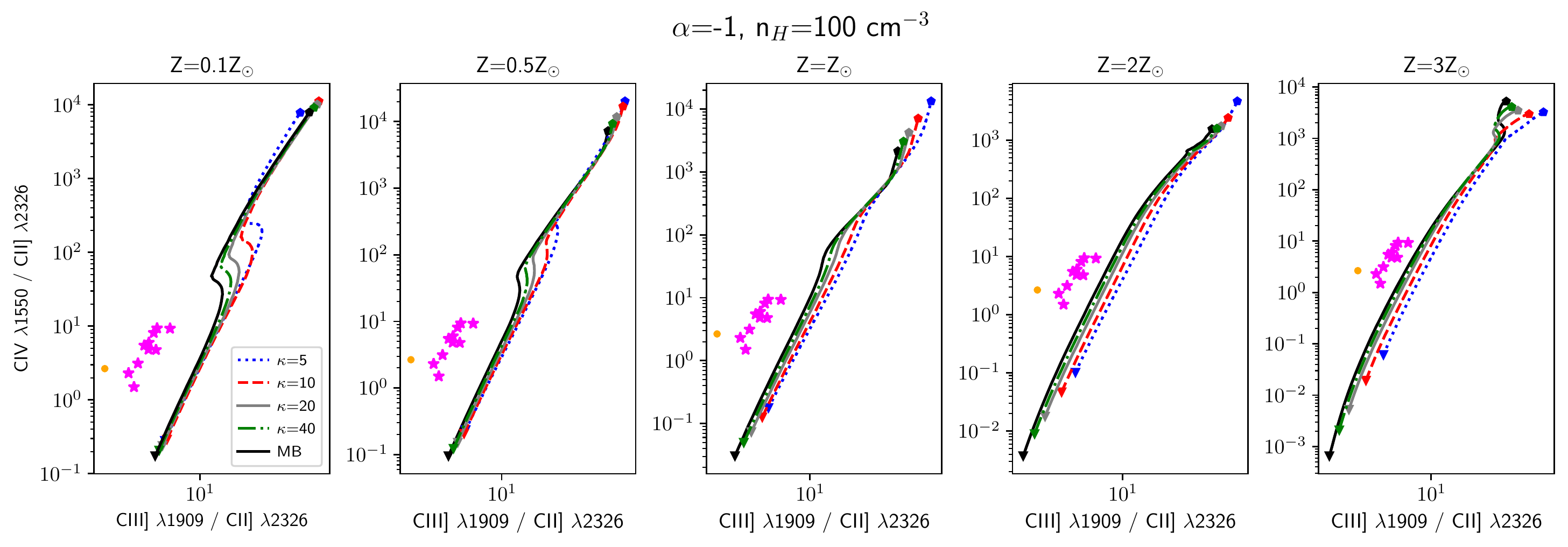}
		\vspace{-0.3 cm}\caption{Diagnostic diagram CIV$\lambda$1550/CII]$\lambda$2326 as a function of CIII]$\lambda$1909/CII]$\lambda$2326. Symbols and lines as in Fig. \ref{fig:OIII_Hb_vs_NII_Ha_SecNC}. Only one QSO2 has reported CII]$\lambda$2326 measurements. }
		\label{fig:civciivsciiicii_alpha-1}
	\end{figure*}

\begin{figure*}
		\centering
		\includegraphics[width=0.99\linewidth]{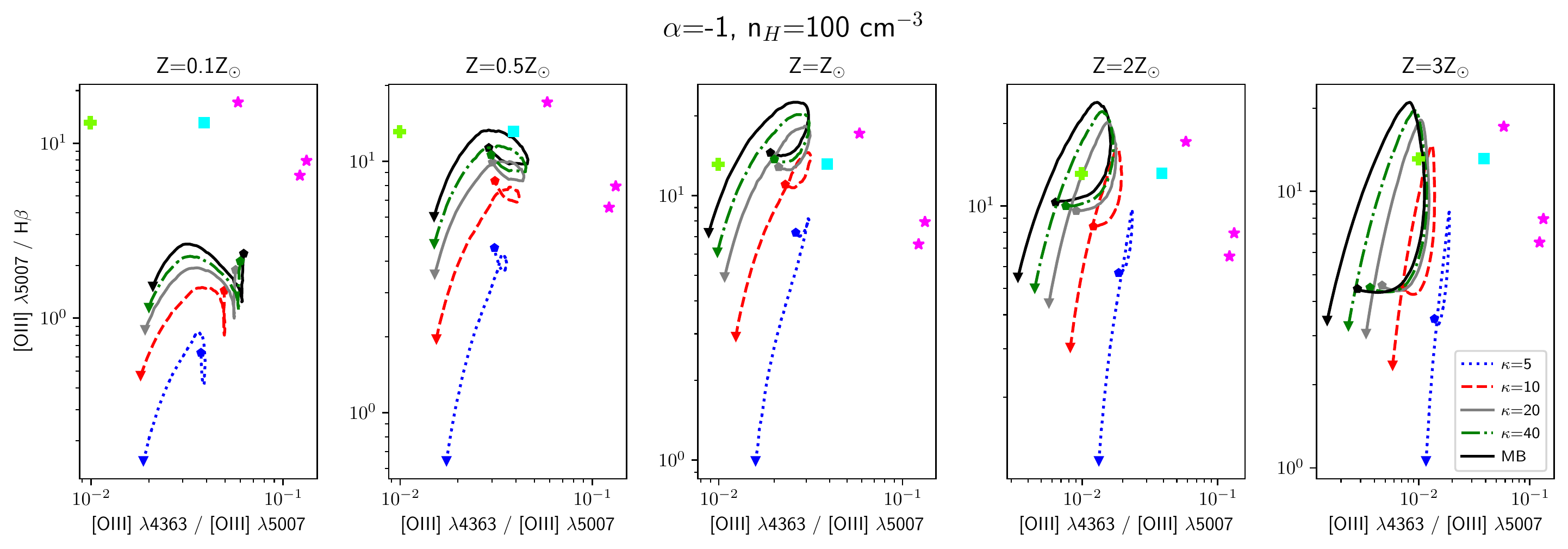}
		\vspace{-0.3 cm}\caption{Diagram [OIII]$\lambda$5007/H$\beta$ vs. [OIII]$\lambda$4363/[OIII]$\lambda$5007.}
		\label{fig:OIII_Hb_vs_OIII4363_OIII_SecNC_alpha-1}
	\end{figure*}

	The diagnostic diagrams are shown in Figs. \ref{fig:OIII_Hb_vs_NII_Ha_SecNC} - \ref{fig:OIII_Hb_vs_OIII4363_OIII_SecNC_alpha-1}. Different models are represented using different colors and patterns. The lines represent sequences of ionization parameter, from \textit{U}=1$\times$10$^{-3}$ to \textit{U}=1.8, computed assuming a different distribution of electron energies. To make visualization easier, in the figures we show \textit{U}=1$\times$10$^{-3}$ as the lowest ionization parameter because as we will see in Section \ref{sec:fit}, \textit{U}=1$\times$10$^{-4}$ is too low to reproduce the observed line ratios. For the same reason we will show figures for a power-law index of -1.   
	
	The black line represents models computed for a M-B distribution, the blue dashed line represents a \textit{$\kappa$} distribution with \textit{$\kappa$}=5, the red line represents a \textit{$\kappa$} distribution with \textit{$\kappa$}=10, magenta shows a \textit{$\kappa$} distribution with \textit{$\kappa$}=20 and a green line assumes a \textit{$\kappa$} distribution with \textit{$\kappa$}=40. Different objects are distinguished by different markers and colours.  A solid triangle indicates the lowest ionization parameter (\textit{U}=1$\times$10$^{-3}$) and a pentagon indicates the highest ionization parameter for each model (\textit{U}=1.8). 
	
	We start by showing predictions of the photoionization models in the standard [OIII]$\lambda$5007/H$\beta$ versus [NII]$\lambda$6584/H$\alpha$ BPT diagnostic diagram \citep{Baldwin1981}. 
	Only NGC 3393 and a few HzRGs have reported measurements for these lines. At low metallicity (Z$\lesssim$0.5Z$_{\odot}$), the models predict too low [OIII]$\lambda$5007/H$\beta$ and [NII]$\lambda$6584/H$\alpha$ compared to the observations. The fit improves with increasing metallicity. 
	When compared to the M-B distribution, \textit{$\kappa$} distributions shift the ratios [OIII]$\lambda$5007/H$\beta$ and [NII]$\lambda$6584/H$\alpha$ towards lower values for a given $U$. 
	
	In Fig. \ref{fig:NV_HeII_vs_NV_CIV_alpha-1} we show the diagnostic diagram NV$\lambda$1240/HeII$\lambda$1640 vs. NV$\lambda$1240/CIV$\lambda$1550, often used as an abundance indicator \citep[e.g.][]{Hamann1992, Hamann1993, VM1999b}. 
	The shape of the curves is similar for models computed assuming \textit{$\kappa$} and M-B distributions.
	Photoionization models assuming \textit{Z}=0.1\textit{Z}$_{\odot}$ predict NV$\lambda$1240/HeII$\lambda$1640 ratios that are too low to reproduce any of the observations, independently of the electron energy distribution assumed. As the metallicity increases the predicted NV$\lambda$1240/HeII$\lambda$1640 and NV$\lambda$1240/CIV$\lambda$1550 ratios increase. At high metallicity (\textit{Z}$\sim$3\textit{Z}$_{\odot}$) the models start to generate line ratios similar to those observed.
	For the same ionization parameter and Z \textgreater 0.5Z$_{\odot}$ \textit{$\kappa$} distributions predict higher NV$\lambda$1240/HeII$\lambda$1640 than a M-B. This is clear for the low end of the ionization parameter sequences.
	
	The diagnostic diagram showing CIV$\lambda$1550/HeII$\lambda$1640 as a function of CIV$\lambda$1550/CIII]$\lambda$1909 is presented in Fig. \ref{fig:CIV_HeII_vs_CIV_CIII_SecNC_alpha-1}. At low gas metallicity, and when compared to the M-B ditribution, \textit{$\kappa$} distributions shift the ratio CIV$\lambda$1550/HeII$\lambda$1640 to lower values. This situation reverses as the metallicity increases (and also for Z=Z$_{\odot}$ and high ionization parameter). 

	Fig. \ref{fig:civciivsciiicii_alpha-1} shows the predictions for the diagnostic diagram CIV$\lambda$1550/CII]$\lambda$2326 as a function of CIII]$\lambda$1909/CII]$\lambda$2326.      
	Compared to M-B distributions, \textit{$\kappa$} distributions predict higher CIII]$\lambda$1909/CII]$\lambda$2326 ratios. The data points lie far from the photoionization model predictions for all the parameter space considered. Compared to M-B distributions \textit{$\kappa$} distributions produce a slightly worse fit to the data.

	Finally, the temperature diagnostic diagram [OIII]$\lambda$5007/H$\beta$ vs. [OIII]$\lambda$4363/[OIII]$\lambda$5007 is presented in Fig. \ref{fig:OIII_Hb_vs_OIII4363_OIII_SecNC_alpha-1}. 
	Only the Seyfert galaxies and some HzRGs have observations for these lines. The predicted  [OIII]$\lambda$4363/[OIII]$\lambda$5007 ratios are too low to reproduce most of the observed line ratios. At high metallicities, \textit{$\kappa$}=5 distributions produce higher [OIII]$\lambda$4363/[OIII]$\lambda$5007 than the M-B distributions but the ratio is still low compared to the observations. This is the long standing T$_e$ problem of active galaxies \citep[e.g.,][]{Tadhunter1989, Binette1996, Richardson2014}.

	\subsection{Best fits} \label{sec:fit}
	
	A Python code was created in order to find the best fitting model for each galaxy, using a $\chi^2$ minimization. The code considers every possible line ratio from the data, and compares it with ratios calculated for the different photoionization models. The result is the best fit for each object. All possible line ratios from lines detected in each galaxy were used. All ratios are given equal weighting.
	
	The parameters of the best-fitting models are shown in Tables \ref{table: Best fit model parameters_Sy2}, \ref{table: Best fit model parameters_HzRGs},
	\ref{table: Best fit model parameters QSOs 1}, \ref{table: Best fit model parameters QSOs 2}, \ref{table: Best fit model parameters QSO2s 3}, \ref{table: Best fit model parameters QSO2s 4}, and \ref{table: Best fit model parameters QSO2s 5}. For each object, in the first row we present the best-fitting M-B model, and in the second row the best-fitting \textit{$\kappa$}-distribution model.
	
	To simplify NV, SiIV/OIV], SiII, CII, NIV], CIV,  HeII, OIII], NIII], CIII], [NeIV], CII], MgII, [ArIV], [NI], [OI], [ArIII] refer to NV$\lambda$1240, SiIV/OIV]$\lambda$1402, SiII$\lambda$1309, CII$\lambda$1335, NIV]$\lambda$1486, CIV$\lambda$1550, HeII$\lambda$1640, OIII]$\lambda$1663, NIII]$\lambda$1750, CIII]$\lambda$1909, [NeIV]$\lambda$2424, CII]$\lambda$2326, MgII$\lambda$2800, [ArIV]$\lambda$4740, [NI]$\lambda$5200, [OI]$\lambda$6300, [ArIII]$\lambda$7135, respectively.
	
	\subsubsection{Seyfert 2}
	
	In table \ref{table: Best fit model parameters_Sy2} the lines used in the minimization for the Seyfert 2 galaxies are presented. 
	\begin{table*}
		\caption{Best fit model parameters for a M-B distribution (first row) and the best fitting \textit{$\kappa$} distribution (second row). (1) Object; (2) electron energy distribution; (3) spectral index of the ionizing continuum; (4) ionization parameter; (5) metallicity; (6) hydrogen density in cm$^{-3}$; (7) $\chi^2$; (8) lines used in the fitting.}
		\vspace{-0.2cm}
		\label{table: Best fit model parameters_Sy2}     
		\begin{center}                         
			\begin{tabular}{c c c c c c c c}        
				Object   & EED &  $\alpha$ & $U$ & $Z$ & $n_H$ & $\chi^2$ & EL\\
				\hline \hline 	
				
				\multirow{3}{*}{ NGC 1068 } 
				& \vspace{-0.1cm} \textbf{M-B} & -1 & 0.004 & 2$Z_{\odot}$ & 10$^4$ & 4.2 & \multirowcell{3}{ \scriptsize NV, SiIV/OIV], NIV], CIV, HeII, OIII], NIII], CIII]+SiIII]$\lambda$1892, [NeIV], [OII]$\lambda$2470, MgII, \\ \scriptsize [NeV]$\lambda$$\lambda$3346,3426, [OII]$\lambda$3727, [NeIII]$\lambda$$\lambda$3869,3967, [SII]$\lambda$4072, [OIII]$\lambda$4363, HeII$\lambda$4686, H$\beta$, \\
					\scriptsize [OIII]$\lambda$5007, [OI], [NII]$\lambda$$\lambda$6548,6584, H$\alpha$, [SII]$\lambda$$\lambda$6716,6731 } \\ 
				&   &  &  &  & &  \\
				
				& \textit{$\kappa$}=40 & -1 & 0.0106 & 2$Z_{\odot}$ & 10$^4$ & 4.6 \\
				
				\hline  
				\multirow{3}{*}{NGC 3393}
				& M-B & -1.5 & 0.009 & 3$Z_{\odot}$ & 100 & 8.1 & \multirowcell{3}{ \scriptsize NV, CIV, HeII, CIII], [NeIV], MgII, [OII]$\lambda$3727, [NeIII]$\lambda$3869, \\
					\scriptsize [NeIII]$\lambda$3969, [OIII]$\lambda$4363, HeII$\lambda$4686, [ArIV]$\lambda$4740, H$\beta$, [OIII]$\lambda$$\lambda$4959,5007, [NI], [NII]$\lambda$6548, \\
					\scriptsize H$\alpha$, [NII]$\lambda$6584, [SII]$\lambda$$\lambda$67176731, [ArIII] } \\ 
				
				&   &  &  &  & &  \\
				
				& \textit{$\kappa$}=10 & -1 & 0.007 & 2$Z_{\odot}$ & 100 & 6.3 \\ 
				
				\hline 					      
			\end{tabular}
		\end{center}
	\end{table*}
	
	In the case of NGC 1068 the best-fitting \textit{$\kappa$} distribution and the M-B predict similar parameters, $\alpha$=-1, n$_H$=10$^{4}$ and Z=2Z$_{\odot}$, with the exception of the ionization parameter, which is higher when a \textit{$\kappa$} distribution is assumed ($U_{M-B}$=0.004, $U_{\kappa}$=0.01).
	For the Seyfert galaxy NGC 3393 the situation is different. In this case the best-fitting \textit{$\kappa$} distribution and the M-B predict different parameters with the exception of the gas density which is similar.

	\subsubsection{HzRGs}
	
	Table \ref{table: Best fit model parameters_HzRGs} presents the parameters of the best-fitting photoionization models.   
	\begin{table*}
		\caption{Best fit model parameters for the HzRGs, assuming a M-B distribution (first row) and the best fitting \textit{$\kappa$} distribution (second row). (1) Object; (2) electron energy distribution; (3) spectral index of the ionizing continuum; (4) ionization parameter; (5) metallicity; (6) hydrogen density in cm$^{-3}$; (7) $\chi^2$; (8) lines used in the fitting.}
		\vspace{-0.2cm}
		\label{table: Best fit model parameters_HzRGs}     
		\begin{center}                         
			\begin{tabular}{c c c c c c c c}        
				Object   & EED &  $\alpha$ & $U$ & $Z$ & $n_H$ & $\chi^2$ & ELs \\
				\hline \hline 	
				\multirow{3}{*}{TXS 0211-122} & M-B & -1 & 0.003 & $Z_{\odot}$ & 100 & 24.2 & 
				\multirowcell{3}{ \scriptsize NV, CIV, SiIV+OIV], SiII$\lambda$1309, NIV], HeII, OIII], CIII], [NeIV], [NeV]$\lambda$3426, \\
					\scriptsize [OII]$\lambda$3727, [NeIII]$\lambda$3869, [OIII]$\lambda$4363, HeII$\lambda$4686, [OIII]$\lambda$4960, [OIII]$\lambda$5007, \\
					\scriptsize [OI], [NII]$\lambda$6548, H$\alpha$, [NII]$\lambda$6583, [SII]$\lambda$$\lambda$6716,6731, H$\beta$ } \\ 
				&   &  &  &  &  \\
				& \textit{$\kappa$}=10 & -1 & 0.004 & 2$Z_{\odot}$ & 100 & 19.96 \\

				\hline
				\multirow{3}{*}{MRC 0406-242} & M-B & -1 & 0.0028 & 2$Z_{\odot}$ & 100 & 2.2 & 
				\multirowcell{3}{ \scriptsize NV, SiIV+OIV], NIV], CIV, HeII$\lambda$1640, \\
					\scriptsize	OIII], CIII], CII], [NeIV],  [NeV]$\lambda$3426, [OII]$\lambda$3727, \\
					\scriptsize [NeIII]$\lambda$3869, HeII$\lambda$4686, [OIII]$\lambda$5007, H$\beta$ } \\
				&   &   &  &  &   \\	
				& \textit{$\kappa$}=40 & -1 & 0.0045 & 2$Z_{\odot}$ & 100 & 2.5 \\

				\hline	
				\multirow{3}{*}{PKS 0529-549} & M-B & -1 & 0.003 & $Z_{\odot}$ & 100 & 2.9 & 
				\multirowcell{3}{\scriptsize [NeV]$\lambda$3426, [OII]$\lambda$3727, [NeIII]$\lambda$3869, [OIII]$\lambda$4363, HeII$\lambda$4686, \\
					\scriptsize [OIII]$\lambda$4960, [OIII]$\lambda$5007, [OI], [NII]$\lambda$6548, Halpha, \\
					\scriptsize [NII]$\lambda$6583,	[SII]$\lambda$$\lambda$6716,6731, H$\beta$ }\\
				&   &   &  &  &   \\
				& \textit{$\kappa$}=10 & -1 & 0.006 & 2$Z_{\odot}$ & 100 & 2.6 \\

				\hline
				\multirow{3}{*}{TXS 0828+193} & M-B & -1 & 0.0045 & 0.5$Z_{\odot}$ & 10$^4$ & 15.17  &
				\multirowcell{3}{\scriptsize NV, SiIV+OIV], SiII$\lambda$1309, CII, NIV], CIV, [NeV]$\lambda$1575, \\
					\scriptsize [NeIV]$\lambda$1602, HeII$\lambda$1640, OIII], NIII], SiIII], CIII], CII], [NeIV], \\
					\scriptsize [NeV]$\lambda$3426, [OII]$\lambda$3727, [OIII]$\lambda$4363, HeII$\lambda$4686, [OIII]$\lambda$5007, H$\beta$ } \\
				&   &   &  &  &   \\
				& \textbf{\textit{$\kappa$}=5} & -1 & 0.005 & 0.5$Z_{\odot}$ & 100 & 11.48 \\

				\hline
				\multirow{3}{*}{MRC 1138-262} & M-B & -1.5 & 0.002 & 3$Z_{\odot}$ & 100 & 0.8 &
				\multirowcell{3}{\scriptsize [OII]$\lambda$3727, [NeIII]$\lambda$3869, HeII$\lambda$4686, [OIII]$\lambda$4960, \\
					\scriptsize [OIII]$\lambda$5007, [OI]$\lambda$6300, [NII]$\lambda$6548, H$\alpha$, \\
					\scriptsize [NII]$\lambda$6583, [SII]$\lambda$$\lambda$6716,6731, H$\beta$ } \\
				&   &   &  &  &   \\
				& \textit{$\kappa$}=10 & -1 & 0.0008 & $Z_{\odot}$ & 10$^4$ & 0.7 \\

				\hline    
				\multirow{3}{*}{4C-00.54} & M-B & -1.5 & 0.01 & $Z_{\odot}$ & 100 & 7.2 &
				\multirowcell{3}{\scriptsize NV, SiIV+OIV], CII, CIV, HeII$\lambda$1640, OIII], CIII], CII], [NeIV], \\
					\scriptsize [OII]$\lambda$3727, [OIII]$\lambda$4363, HeII$\lambda$4686, [OIII]$\lambda$4960, [OIII]$\lambda$5007, \\
					\scriptsize [OI], [NII]$\lambda$6548, H$\alpha$, [NII]$\lambda$6583, [SII]$\lambda$$\lambda$6716,6731, H$\beta$} \\ 
				&   &   &  &  &   \\
				& \textit{$\kappa$}=40 & -1.5 & 0.015 & 2$Z_{\odot}$ & 100 & 6.5 \\

				\hline	
				\multirow{3}{*}{USS 1558-003} & M-B & -1 & 0.0035 & $Z_{\odot}$ & 100 & 5.3 &
				\multirowcell{3}{\scriptsize NV, SiIV+OIV], NIV], CIV, HeII$\lambda$1640, OIII], CIII], CII], [NeV]$\lambda$3426, \\ \scriptsize  [OII]$\lambda$3727, [NeIII]$\lambda$3869, HeII$\lambda$4686, [OIII]$\lambda$4960, [OIII]$\lambda$5007, \\
					\scriptsize [OI], [NII]$\lambda$6548, H$\alpha$, [NII]$\lambda$6583, [SII]$\lambda$$\lambda$6716,6731, H$\beta$ } \\ 
				&   &   &  &  &   \\
				& \textit{$\kappa$}=10 & -1 & 0.005 & $Z_{\odot}$ & 100 & 3.4 \\

				\hline
				\multirow{3}{*}{4C+40.36} & M-B & -1 & 0.0017 & $Z_{\odot}$ & 10$^4$ & 15.7 &
				\multirowcell{3}{\scriptsize NV, SiIV+OIV], SiII, CII, NIV], CIV, HeII$\lambda$1640, OIII], NIII], CIII], CII], \\
					\scriptsize [NeIV], [OII]$\lambda$2471, [OII]$\lambda$3727, [NeIII]$\lambda$3869, HeII$\lambda$4686, \\
					\scriptsize [OIII]$\lambda$5007, [OI], [NII]$\lambda$6548, H$\alpha$, [NII]$\lambda$6583, [SII]$\lambda$$\lambda$6716,6731, H$\beta$ } \\ 
				&   &   &  &  &   \\
				& \textit{$\kappa$}=10 & -1 & 0.003 & 2$Z_{\odot}$ & 100 & 11.4 		\\

				\hline
				\multirow{3}{*}{4C+23.56} & M-B & -1.5 & 0.004 & 2$Z_{\odot}$ & 100 & 6.6 &
				\multirowcell{3}{\scriptsize NV, SiIV+OIV], NIV], CIV, HeII$\lambda$1640, OIII], CIII], CII], [NeIV], \\ 
					\scriptsize [NeV]$\lambda$3426, [OII]$\lambda$3727, [NeIII]$\lambda$3869, HeII$\lambda$4686, [OIII]$\lambda$4960, [OIII]$\lambda$5007, \\
					\scriptsize [OI], [NII]$\lambda$6548, H$\alpha$, [NII]$\lambda$6583, [SII]$\lambda$$\lambda$6716,6731, H$\beta$ } \\ 
				&   &   &  &  &   \\
				& \textit{$\kappa$}=40 & -1.5 & 0.0065 & 2$Z_{\odot}$ & 100 & 7.5 \\
				
				\hline
				\multirow{2}{*}{0850-206} & M-B & -1 & 0.004 & 0.5$Z_{\odot}$ & 100 & 4.93 &
				\multirowcell{2}{\scriptsize HeII$\lambda$1640, NIII], SiIII], CIII], CII], [NeIV], [OII]$\lambda$2471, \\
					\scriptsize MgII, [NeV]$\lambda$3346, [NeV]$\lambda$3426, [OII]$\lambda$3727 } \\
				& \textit{$\kappa$}=40 & -1 & 0.006 & 0.5$Z_{\odot}$ & 100 & 5.7 \\
				
				\hline
				\multirow{2}{*}{1303+091} & M-B & -1 & 0.0035 & $Z_{\odot}$ & 10$^4$ & 1.8 &
				\multirowcell{2}{\scriptsize CIV, HeII$\lambda$1640, OIII], SiIII], CIII],  CII], [NeIV], \\
					\scriptsize [OII]$\lambda$2471, MgII, [NeV]$\lambda$3346, [NeV]$\lambda$3426  } \\
				& \textit{$\kappa$}=40 & -1 & 0.0045 & $Z_{\odot}$ & 10$^4$ & 2.1 \\

				\hline 
				\multirow{2}{*}{4C+03.24} & M-B & -1 & 0.0065 & 2$Z_{\odot}$ & 10$^6$ & 3.8 &
				\multirowcell{2}{\scriptsize NV, CII, SiIV+OIV], CIV, HeII$\lambda$1640 } \\
				& \textbf{\textit{$\kappa$}=5} & -1 & 0.0065 & 3$Z_{\odot}$ & 10$^6$ & 1.5 \\

				\hline
				\multirow{2}{*}{0731+438} & M-B & -1 & 0.0045 & 0.5$Z_{\odot}$ & 10$^4$ & 10.2 &
				\multirowcell{2}{\scriptsize NV, SiII, CII, SiIV+OIV], NIV], CIV, HeII$\lambda$1640, OIII], \\	\scriptsize SiIII]$\lambda$$\lambda$1882,1892, CIII], CII], [NeIV]$\lambda$2422 } \\
				& \textbf{\textit{$\kappa$}=5} & -1 & 0.0045 & $Z_{\odot}$ & 100 & 5.98 \\

				\hline
				\multirow{2}{*}{4C+48.48} & M-B & -1 & 0.01 & 2$Z_{\odot}$ & 100 & 2.44 &
				\multirowcell{2}{\scriptsize NV, CII, SiIV+OIV], NIV], CIV, HeII$\lambda$1640, OIII], \\
					\scriptsize CIII], CII], [NeIV]$\lambda$2422 } \\ 
				& \textit{$\kappa$}=40 & -1 & 0.012 & 2$Z_{\odot}$ & 100 & 2.26 \\

				\hline 
				\multirow{3}{*}{MRC 0943-242} & M-B & -1 & 0.0035 & $Z_{\odot}$ & 100 & 2.2 &
				\multirowcell{3}{\scriptsize NV, CIV, HeII$\lambda$1640, CIII], CII], [NeIV], MgII, \\
					\scriptsize [NeV]$\lambda$3426, [OII]$\lambda$3727, [NeIII]$\lambda$3869, \\
					\scriptsize [OIII]$\lambda$4363, [OIII]$\lambda$4960, [OIII]$\lambda$5007 } \\ 
				&   &   &  &  &   \\
				& \textit{$\kappa$}=40 & -1 & 0.0065 & $Z_{\odot}$ & 100 & 2.4  \\				   
				
				\hline 					      
			\end{tabular}
		\end{center}
	\end{table*}

\begin{figure}
		\centering
		\includegraphics[width=\linewidth]{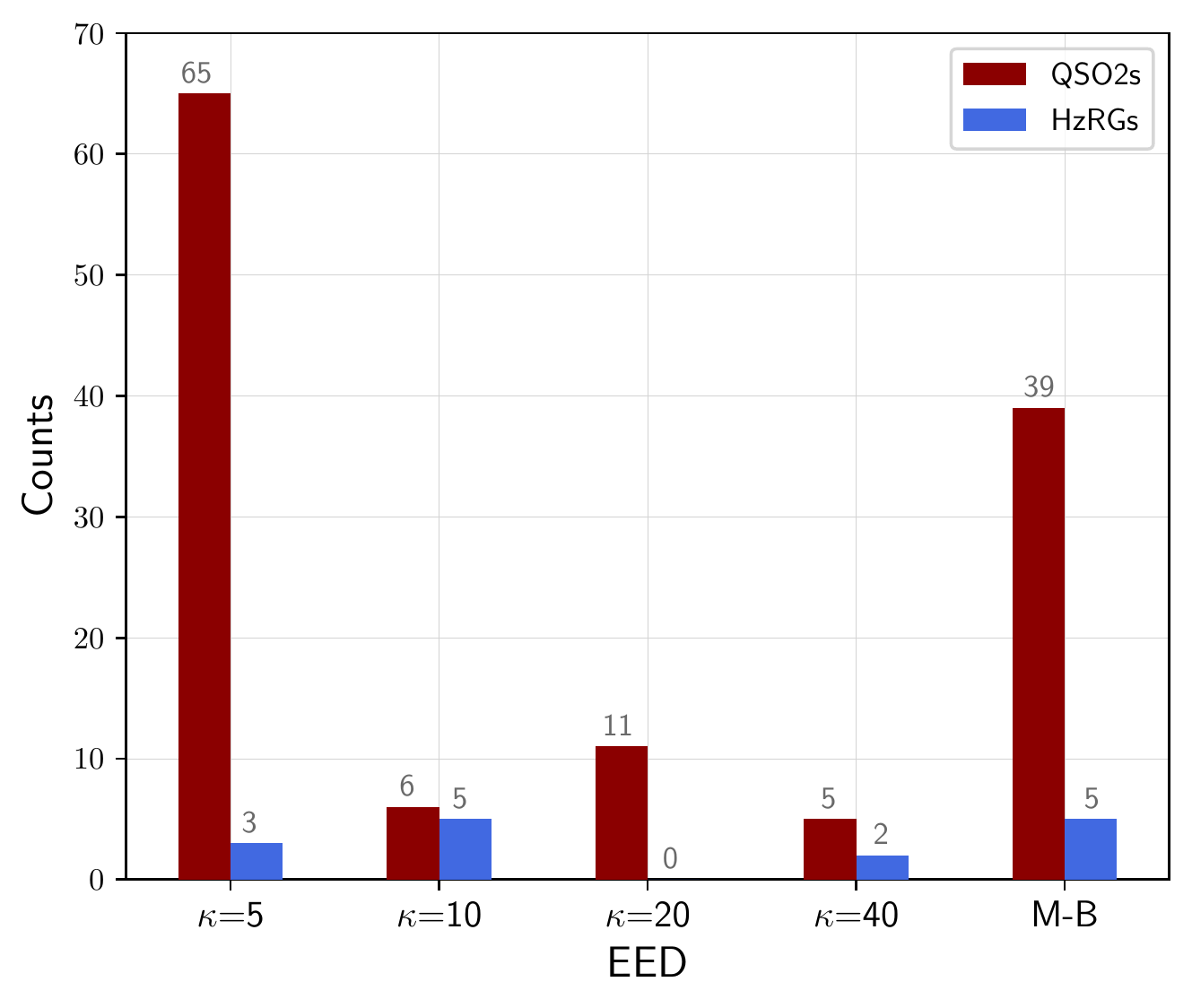}
		\vspace{-0.6 cm}\caption{ Number of objects (HzRGs and QSO2s) according to the best fitting electron energy distribution. } 	\label{fig:count_EED_SDSS_HzRGs}
	\end{figure}

	The best fitting distribution varies from object to object. Five HzRGs are best fitted by a M-B distribution, and the other ten objects are best fitted by \textit{$\kappa$} distributions, with varying \textit{$\kappa$} indices (see Fig. \ref{fig:count_EED_SDSS_HzRGs}).

\begin{figure*}
		\centering
		\includegraphics[width=0.96\linewidth]{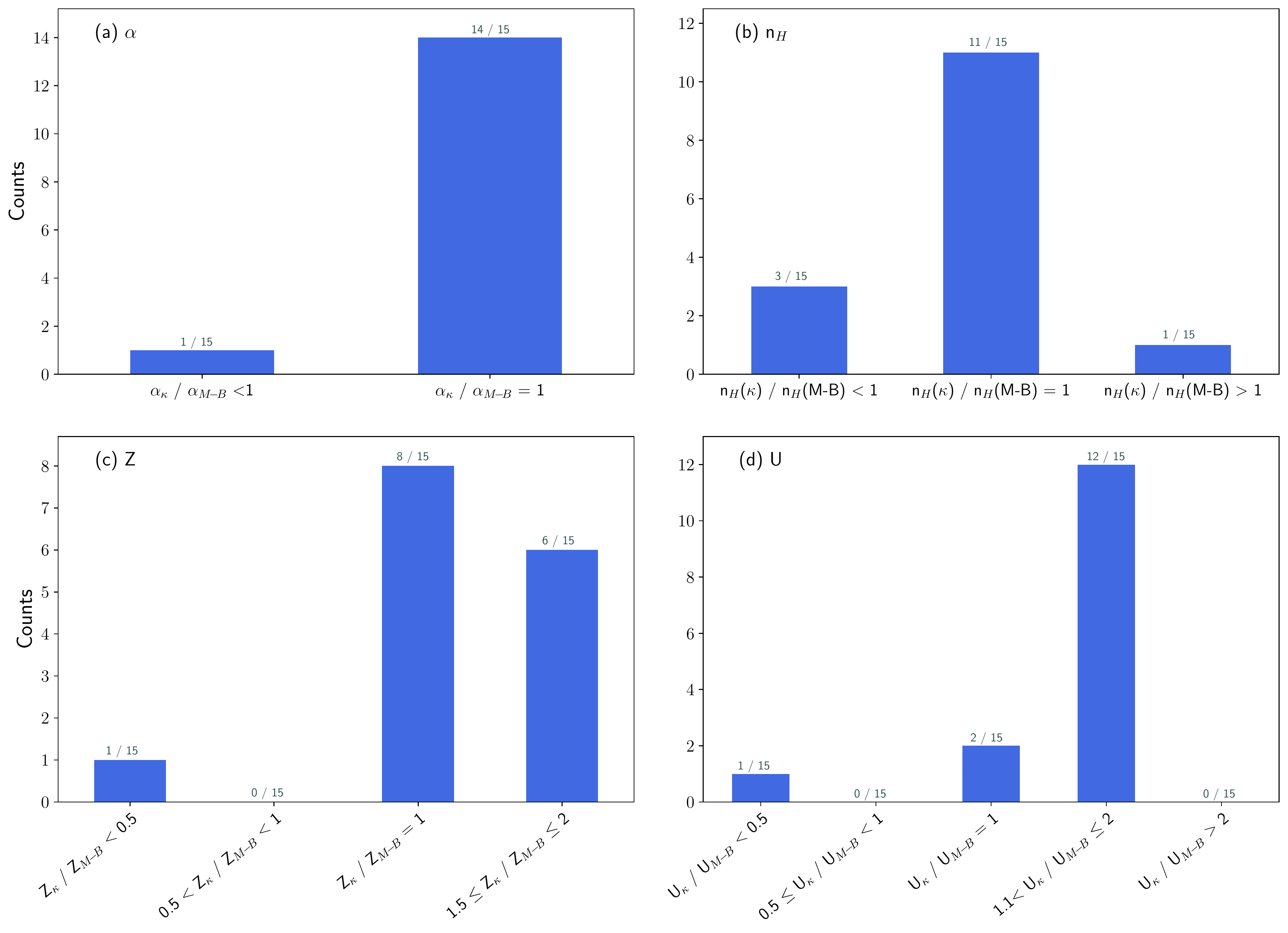}
		\vspace{-0.2 cm} \caption{HzRGs binned according to discrepancy in (a) $\alpha$, (b) n$_H$, (c) Z, and (d) U between \textit{$\kappa$} and M-B distributions.}
		\label{fig:hzrgs}
	\end{figure*}

	Fig. \ref{fig:hzrgs} shows the HzRGs binned according to discrepancy in $\alpha$ (panel (a)), n$_H$ (panel (b)), Z (panel (c)) and U (panel (d)).
	It shows how distant the \textit{$\kappa$} and M-B parameter predictions are and the fraction of objects in which these discrepancies are found. 
	
	In general, the HzRGs are best fitted by photoionization models with $\alpha$ = -1. The assumption of a \textit{$\kappa$} or M-B distribution has no impact on the inferred spectral index, $\alpha$, in most HzRGs (14/15 or 93\%; Fig. \ref{fig:hzrgs}, panel (a)). 
	
	In panel(b) of Fig. \ref{fig:hzrgs} we show how the choice of EED affects the predicted gas density. Assuming a \textit{$\kappa$} distribution instead of a M-B has no impact on the inferred density in 73\% of HzRGs. In these sources, \textit{$\kappa$} and the M-B models predict n$_H$=100 cm$^{-3}$.  	
	
	Panel (c) of Fig. \ref{fig:hzrgs} presents the results for the metallicity. In 93\% of HzRGs (14 sources), \textit{$\kappa$} models produce identical or higher $Z$ than the M-B models.
	The predicted metallicities are consistent between the \textit{$\kappa$} and M-B models (Z$_\textit{$\kappa$}$ / Z$_{M\!-\!B}$ = 1) in 53\% of objects. In 40\% of HzRGs, \textit{$\kappa$} models predict metallicities which are a factor of 1.5 to 2 times higher than M-B models. 
	Large discrepancies in the predicted metallicities (Z$_\textit{$\kappa$}$ / Z$_{M\!-\!B}$ \textless 0.5) occur only for one HzRG (7\%).	
	
	Fig. \ref{fig:hzrgs}, panel (d) presents the ratio U$_\textit{$\kappa$}$ / U$_{M\!-\!B}$.
	In 93\% of cases (14 sources), \textit{$\kappa$} models produce identical or higher ionization parameters than M-B models.  
	For 12 sources (80\%) \textit{$\kappa$} models predict an ionization parameter 1.1 to 2 times higher than the value predicted by the M-B model. Large discrepancies in the predicted ionization parameter (U$_\textit{$\kappa$}$ / U$_{M\!-\!B}$ \textless 0.5) occur only for one HzRG (7\%).

	
	\subsubsection{Type 2 quasars at z\textgreater2}
	
	The best fitting parameters for the QSO2s are presented in Tables \ref{table: Best fit model parameters QSOs 1}, \ref{table: Best fit model parameters QSOs 2}, \ref{table: Best fit model parameters QSO2s 3}, \ref{table: Best fit model parameters QSO2s 4}, and \ref{table: Best fit model parameters QSO2s 5}.

	\begin{table*}
		\caption{Best fit model parameters for the QSO2s. The first row shows the parameters when a M-B distribution is used and the second row shows the best fitting \textit{$\kappa$} distribution and its parameters. (1) Object; (2) electron energy distribution; (3) spectral index of the ionizing continuum; (4) ionization parameter; (5) metallicity; (6) hydrogen density in cm$^{-3}$; (7) $\chi^2$; (8) lines used in the fitting.}
		\vspace{-0.2cm}
		\label{table: Best fit model parameters QSOs 1}     
		\begin{center}                         
			\begin{tabular}{c c c c c c c c}        
				Object   & EED &  $\alpha$ & $U$ & $Z$ & $n_H$ & $\chi^2$ & ELs \\
				\hline \hline
				
				\multirow{2}{*}{SDSS J011506.65} & M-B & -1 & 0.0065 & $Z_{\odot}$ & 10$^4$ & 22.2 &
				\multirowcell{2}{\scriptsize NV, OI, SiIV+OIV], CIV, HeII, SiIII], CIII] } \\ 
				& \textbf{\textit{$\kappa$}=5} & -1 & 0.005 & 3$Z_{\odot}$ & 10$^4$ & 17.5  \\   
				
				\hline 
				\multirow{2}{*}{SDSS J022051.68} & M-B & -1 & 0.012 & $Z_{\odot}$ & 10$^4$ & 20.4 &
				\multirowcell{2}{\scriptsize NV, OI, SiIV+OIV], CIV, HeII, NIII], CIII] } \\
				& \textbf{\textit{$\kappa$}=5} & -1 & 0.0106 & 2$Z_{\odot}$ & 10$^4$ & 16.6 \\
				
				\hline 
				\multirow{2}{*}{SDSS J075119.09} & M-B & -1 & 0.0065 & $Z_{\odot}$ & 10$^4$ & 15.7 &
				\multirowcell{2}{\scriptsize NV, OI, SiIV+OIV], CIV, HeII, CIII] } \\ 
				& \textbf{\textit{$\kappa$}=5} & -1 & 0.005 & 3$Z_{\odot}$ & 10$^4$ & 10.9 \\					
				
				\hline
				\multirow{2}{*}{SDSS J081950.96} & M-B & -1 & 0.01 & 0.5$Z_{\odot}$ & 10$^4$ & 12.1 &
				\multirowcell{2}{\scriptsize OVI+CII, NV, OI, SiIV+OIV], CIV, HeII, CIII] } \\ 
				& \textbf{\textit{$\kappa$}=5} & -1 & 0.01 & $Z_{\odot}$ & 10$^4$ & 10.7 \\
				
				\hline 	
				\multirow{2}{*}{SDSS J084005.00} & M-B & -1 & 0.0058 & $Z_{\odot}$ & 10$^4$ & 10.7 & 
				\multirowcell{2}{\scriptsize NV, OI, SiIV+OIV], NIV], CIV, HeII, OIII],	SiIII]	CIII]} \\ 
				& \textbf{\textit{$\kappa$}=5} & -1 & 0.0045 & 3$Z_{\odot}$ & 10$^4$ & 8.6 \\
				
				\hline
				\multirow{2}{*}{SDSS J095118.93} & M-B & -1 & 0.0065 & $Z_{\odot}$ & 10$^4$ & 15.6 & 
				\multirowcell{2}{\scriptsize NV, OI, SiIV+OIV], CIV, [NeIV]1602, HeII, CIII], [NeIV]$\lambda$2422, MgII } \\
				& \textbf{\textit{$\kappa$}=5} & -1 & 0.0065 & $Z_{\odot}$ & 10$^4$ & 13.2 \\ 
				
				\hline
				\multirow{2}{*}{SDSS J100250.98} & M-B & -1 & 0.0136 & $Z_{\odot}$ & 10$^4$ & 23.1 & 
				\multirowcell{2}{\scriptsize OVI+CII, NV, OI, CIV, HeII, OIII], CIII] } \\ 
				& \textbf{\textit{$\kappa$}=5} & -1 & 0.0136 & 2$Z_{\odot}$ & 10$^4$ & 18.4 \\						
				
				\hline 
				\multirow{2}{*}{SDSS J100916.93} & M-B & -1 & 0.017 & $Z_{\odot}$ & 10$^4$ & 21.6 & 
				\multirowcell{2}{\scriptsize OVI+CII, NV, OI, SiIV+OIV], CIV, HeII, NIII], CIII] } \\
				& \textbf{\textit{$\kappa$}=5} & -1 & 0.017 & $Z_{\odot}$ & 10$^4$ & 19.6 \\ %
				
				\hline 
				\multirow{2}{*}{SDSS J125148.53} & M-B & -1 & 0.008 & $Z_{\odot}$ & 100 & 19.7 & 
				\multirowcell{2}{\scriptsize NV, OI, SiIV+OIV], CIV, HeII, OIII], CIII], MgII } \\ 
				& \textbf{\textit{$\kappa$}=5} & -1 & 0.008 & 2$Z_{\odot}$ & 100 & 13.7 \\
				
				\hline
				\multirow{2}{*}{SDSS J135531.46} & M-B & -1 & 0.009 & 0.5$Z_{\odot}$ & 10$^4$ & 16.6 & 
				\multirowcell{2}{\scriptsize OVI+CII, NV, OI, SiIV+OIV], CIV, [NeIV]$\lambda$1602, HeII, OIII], CIII], [NeIV]$\lambda$2422 } \\ 
				& \textbf{\textit{$\kappa$}=5} & -1 & 0.009 & $Z_{\odot}$ & 10$^4$ & 13.6 \\
				
				\hline
				\multirow{2}{*}{SDSS J150549.73} & M-B & -1 & 0.017 & $Z_{\odot}$ & 100 & 17.1 & 
				\multirowcell{2}{\scriptsize OVI+CII, NV, OI, SiIV+OIV], CIV, HeII, OIII], NIII], CIII] } \\  %
				& \textbf{\textit{$\kappa$}=5} & -1 & 0.015 & $Z_{\odot}$ & 10$^4$ & 14.03 \\	
				
				\hline
				\multirow{2}{*}{SDSS J160900.01} & M-B & -1 & 0.009 & $Z_{\odot}$ & 100 & 19.97  & 
				\multirowcell{2}{\scriptsize OVI+CII, NV, OI, SiIV+OIV], NIV], CIV, HeII, OIII], CIII], MgII } \\ 
				& \textbf{\textit{$\kappa$}=5} & -1 & 0.009 & $Z_{\odot}$ & 10$^4$ & 15.5 \\
				
				\hline
				\multirow{2}{*}{SDSS J161059.96} & M-B & -1 & 0.0058 & $Z_{\odot}$ & 100 & 19.2 & 
				\multirowcell{2}{\scriptsize NV, OI, SiIV+OIV], CIV, HeII, OIII], CIII], MgII } \\ 
				& \textbf{\textit{$\kappa$}=5} & -1 & 0.0065 & 2$Z_{\odot}$ & 100 & 12.7 \\	
				
				\hline
				\multirow{2}{*}{SDSS J162812.51} & M-B & -1 & 0.005 & $Z_{\odot}$ & 10$^4$ & 11.1 & 
				\multirowcell{2}{\scriptsize NV, OI, SiIV+OIV], NIV], CIV, HeII, OIII], NIII], CIII], MgII } \\ 
				& \textbf{\textit{$\kappa$}=5} & -1 & 0.0045 & 3$Z_{\odot}$ & 10$^4$ & 8.4 \\		
				
				\hline 
				\multirow{2}{*}{SDSS J013747.84} & M-B & -1 & 0.004 & 0.1$Z_{\odot}$ & 100 & 5.7 & 
				\multirowcell{2}{\scriptsize OVI+CII, NV, OI, CIV,} \\ 
				& \textbf{\textit{$\kappa$}=5} & -1 & 0.003 & 3$Z_{\odot}$ & 100 & 2.9 \\
				
				\hline
				\multirow{2}{*}{SDSS J014607.15} & M-B & -1 & 0.0106 & $Z_{\odot}$ & 10$^4$ & 46.1 & 
				\multirowcell{2}{\scriptsize OVI+CII, NV, OI, NIV], CIV, HeII, CIII] } \\ 
				& \textbf{\textit{$\kappa$}=5} & -1 & 0.0106 & 2$Z_{\odot}$ & 10$^4$ & 37.99 \\
				
				\hline						      
				\multirow{2}{*}{SDSS J015700.14} & M-B & -1 & 0.156 & 3$Z_{\odot}$ & 10$^6$ & 0.7 & 
				\multirowcell{2}{\scriptsize OVI+CII, NV, CIV, HeII, CIII] } \\ 
				& \textit{$\kappa$}=5 & -1 & 0.156 & 3$Z_{\odot}$ & 10$^6$ & 0.7 \\				 	
				
				\hline 
				\multirow{2}{*}{SDSS J020245.82} & M-B & -1 & 0.23 & 3$Z_{\odot}$ & 10$^4$ & 0.003 & 
				\multirowcell{2}{\scriptsize CIV, HeII, CIII] } \\ 
				& \textit{$\kappa$}=5 & -1 & 0.37 & 2$Z_{\odot}$ & 10$^6$ & 0.002 \\
				
				\hline
				\multirow{2}{*}{SDSS J020643.64} & M-B & -1 & 0.0058 & $Z_{\odot}$ & 10$^4$ & 23.2 &
				\multirowcell{2}{\scriptsize OVI+CII, NV, OI, CIV, HeII, OIII], SiIII], CIII] } \\ 
				& \textbf{\textit{$\kappa$}=5} & -1 & 0.0045 & 3$Z_{\odot}$ & 10$^4$ & 14.8 \\ 		
				
				\hline 
				\multirow{2}{*}{SDSS J091301.33} & M-B & -1.5 & 0.138 & 2$Z_{\odot}$ & 10$^4$ & 0.07 & 
				\multirowcell{2}{\scriptsize OVI+CII, NV, CIV, CIII] } \\ 
				& \textit{$\kappa$}=5 & -1 & 0.04 & 2$Z_{\odot}$ & 10$^6$ & 0.02 \\
				
				\hline
				\multirow{2}{*}{SDSS J091025.50} & M-B & -1 & 0.012 & $Z_{\odot}$ & 10$^4$ & 17.6 & 
				\multirowcell{2}{\scriptsize OVI+CII, NV, OI, SiIV+OIV], NIV], CIV, HeII, NIII], CIII] } \\
				& \textbf{\textit{$\kappa$}=5} & -1 & 0.012 & $Z_{\odot}$ & 10$^4$ & 14.6 \\			
				
				\hline		 
				\multirow{2}{*}{SDSS J023337.89} & M-B & -1 & 0.004 & 0.1$Z_{\odot}$ & 100 & 9.5 & 
				\multirowcell{2}{\scriptsize NV, OI, CIV } \\	
				& \textbf{\textit{$\kappa$}=5} & -1 & 0.004 & 3$Z_{\odot}$ & 100 & 5.9 \\			
				
				\hline	
				\multirow{2}{*}{SDSS J074725.50} & M-B & -1 & 0.036 & $Z_{\odot}$ & 100 & 0.24 & 
				\multirowcell{2}{\scriptsize OVI+CII, NV, CIV, HeII, OIII], CIII] } \\
				& \textit{$\kappa$}=5 & -1 & 0.036 & $Z_{\odot}$ & 100 & 0.1 \\					
				
				\hline       	               
				\multirow{2}{*}{SDSS J080428.80} & M-B & -1 & 0.008 & $Z_{\odot}$ & 100 & 22.2  & 
				\multirowcell{2}{\scriptsize NV,OI, SiIV+OIV], CIV, HeII, CIII] } \\ 		
				& \textbf{\textit{$\kappa$}=5} & -1 & 0.008 & 2$Z_{\odot}$ & 100 & 16.8 \\
				
				\hline
				\multirow{2}{*}{SDSS J080826.02} & M-B & -1 & 0.00276 & 2$Z_{\odot}$ & 10$^4$ & 10.8 & 
				\multirowcell{2}{\scriptsize OI, SiIV+OIV], CIV, HeII, OIII], CIII] } \\ %
				& \textbf{\textit{$\kappa$}=5} & -1 & 0.0035 & 3$Z_{\odot}$ & 100 & 8.2 \\
				
				\hline 
				\multirow{2}{*}{SDSS J081812.72} & M-B & -1 & 0.008 & $Z_{\odot}$ & 10$^4$ & 25.5 & 
				\multirowcell{3}{\scriptsize NV, OI, CIV, HeII, OIII], CIII] } \\ 
				& \textbf{\textit{$\kappa$}=5} & -1 & 0.008 & $Z_{\odot}$ & 10$^4$ & 20.7 \\	
				
				\hline  
				\multirow{2}{*}{SDSS J081452.04} & M-B & -1 & 0.009 & 0.1$Z_{\odot}$ & 100 & 21.2 & 
				\multirowcell{2}{\scriptsize OVI+CII, NV, OI, CIV, CIII] } \\ 
				& \textbf{\textit{$\kappa$}=5} & -1 & 0.009 & 0.1$Z_{\odot}$ & 100 & 15.6 \\
				
				\hline
				\multirow{2}{*}{SDSS J082550.58} & M-B & -1 & 0.0136 & 0.5$Z_{\odot}$ & 10$^4$ & 28.6 & 
				\multirowcell{2}{\scriptsize OVI+CII, NV, OI, SiIV+OIV], CIV, HeII, CIII] } \\ 
				& \textbf{\textit{$\kappa$}=5} & -1 & 0.015 & $Z_{\odot}$ & 10$^4$ & 27.1 \\
				
				\hline
				\multirow{2}{*}{SDSS J083246.92} & M-B & -1 & 0.0035 & 0.1$Z_{\odot}$ & 10$^6$ & 16.01 & 
				\multirowcell{2}{\scriptsize NV, OI, CIV, [NeIV]$\lambda$1602, CIII], [NeIV]$\lambda$2422 } \\
				& \textbf{\textit{$\kappa$}=5} & -1 & 0.0035 & 0.1$Z_{\odot}$ & 10$^6$ & 9.4 \\
				
				\hline										
				\multirow{2}{*}{SDSS J083851.81} & M-B & -1 & 0.008 & 0.5$Z_{\odot}$ & 10$^4$ & 25.8 & 
				\multirowcell{2}{\scriptsize OVI+CII, NV, OI, CIV, HeII, CIII] } \\ 
				& \textbf{\textit{$\kappa$}=5} & -1 & 0.008 & $Z_{\odot}$ & 10$^4$ & 22.97 \\
				
				\hline
				\multirow{2}{*}{SDSS J094308.14} & M-B & -1 & 0.0058 & $Z_{\odot}$ & 100 & 35.02  & 
				\multirowcell{2}{\scriptsize NV, OI, SiIV+OIV], CIV, HeII, OIII], CIII] } \\ 
& \textbf{\textit{$\kappa$}=5} & -1 & 0.0058 & $Z_{\odot}$ & 100 & 25.5 \\

\hline
			\end{tabular}
		\end{center}
	\end{table*}

	\begin{table*}
		\caption{ Continuation of Table \ref{table: Best fit model parameters QSOs 1}. }
		\vspace{-0.2cm}
		\label{table: Best fit model parameters QSOs 2}     
		\begin{center}                         
			\begin{tabular}{c c c c c c c c}        
				Object   & EED &  $\alpha$ & $U$ & $Z$ & $n_H$ & $\chi^2$ & EL \\
				\hline \hline

				\multirow{2}{*}{SDSS J094826.45} & M-B & -1 & 0.0058 & $Z_{\odot}$ & 10$^4$ & 20.9 & 
				\multirowcell{2}{\scriptsize NV, OI, SiIV+OIV], CIV, HeII, OIII], CIII], MgII } \\ 
				& \textbf{\textit{$\kappa$}=5} & -1 & 0.005 & 3$Z_{\odot}$ & 10$^4$ & 11.9 \\
				
				\hline
				\multirow{2}{*}{SDSS J100210.53} & M-B & -1 & 0.015 & $Z_{\odot}$ & 100 & 39.6 & 
				\multirowcell{2}{\scriptsize OVI+CII, NV, OI, SiIV+OIV], CIV, HeII, OIII], CIII] } \\
				& \textbf{\textit{$\kappa$}=5} & -1 & 0.015 & 2$Z_{\odot}$ & 100 & 28.98 \\	
				
				\hline										
				\multirow{2}{*}{SDSS J084949.57} & M-B & -1 & 0.009 & $Z_{\odot}$ & 100 & 26.4 & 
				\multirowcell{2}{\scriptsize OVI+CII, NV, OI, CIV, HeII, OIII] } \\ 
				& \textbf{\textit{$\kappa$}=5} & -1 & 0.004 & 3$Z_{\odot}$ & 10$^4$ & 13.2 \\
				
				\hline 
				\multirow{2}{*}{SDSS J113351.03} & M-B & -1 & 0.005 & 0.5$Z_{\odot}$ & 10$^4$ & 11.4 & 
				\multirowcell{2}{\scriptsize OVI+CII, NV, OI, CIV, HeII, OIII], SiIII], CIII] } \\ 
				& \textbf{\textit{$\kappa$}=5} & -1 & 0.0035 & 3$Z_{\odot}$ & 10$^4$ & 8.5 \\
				\hline
				\multirow{2}{*}{SDSS J104133.36} & M-B & -1 & 0.008 & $Z_{\odot}$ & 10$^4$ & 20.99 & 
				\multirowcell{2}{\scriptsize NV, OI, SiIV+OIV], CIV, HeII, CIII], MgII } \\ 
				& \textbf{\textit{$\kappa$}=5} & -1 & 0.00736 & 3$Z_{\odot}$ & 10$^4$ & 16.1 \\
				
				\hline
				\multirow{2}{*}{SDSS J105324.11} & M-B & -1 & 0.0136 & $Z_{\odot}$ & 100 & 19.5 & 
				\multirowcell{2}{\scriptsize NV, OI, SiIV+OIV], CIV, HeII, 	OIII], CIII], MgII }\\ 
				& \textbf{\textit{$\kappa $}=5} & -1 & 0.012 & 2$Z_{\odot}$ & 100 & 14.1 \\	
				
				\hline 	
				\multirow{2}{*}{SDSS J115335.78} & M-B & -1 & 0.0106 & $Z_{\odot}$ & 100 & 13.5 & 
				\multirowcell{2}{\scriptsize OVI+CII, NV, OI, SiIV+OIV], NIV], CIV, HeII, OIII], CIII]} \\ 
				& \textit{$\kappa$}=5 & -1 & 0.0106 & $Z_{\odot}$ & 100 & 11.3 \\ 
				
				\hline 
				\multirow{2}{*}{SDSS J115411.95} & M-B & -1 & 0.009 & $Z_{\odot}$ & 10$^4$ & 23.1 & 
				\multirowcell{2}{\scriptsize OVI+CII, NV, OI, CIV, HeII, OIII], CIII] } \\
				& \textbf{\textit{$\kappa$}=5} & -1 & 0.009 & 2$Z_{\odot}$ & 10$^4$ & 19.1 \\	
				
				\hline 
				\multirow{2}{*}{SDSS J114856.13} & M-B & -1 & 0.0065 & $Z_{\odot}$ & 10$^4$ & 21.03 & 
				\multirowcell{2}{\scriptsize NV, OI, CIV, HeII, CIII], MgII } \\
				& \textbf{\textit{$\kappa$}=5} & -1 & 0.0058 & 3$Z_{\odot}$ & 10$^4$ & 16.5 \\
				
				\hline
				\multirow{2}{*}{SDSS J115510.34} & M-B & -1 & 0.009 & 0.5$Z_{\odot}$ & 10$^4$ & 20.1 & 
				\multirowcell{2}{\scriptsize NV, OI, CIV, HeII, SiIII],	CIII], MgII } \\ 
				& \textbf{\textit{$\kappa$}=5} & -1 & 0.0065 & 3$Z_{\odot}$ & 10$^4$ & 18.4 \\
				
				\hline	
				\multirow{2}{*}{SDSS J140220.18} & M-B & -1 & 0.004 & 0.1$Z_{\odot}$ & 100 & 8.7 &  
				\multirowcell{2}{\scriptsize NV, OI, NIV], CIV,	SiIII], CIII] } \\ 
				& \textbf{\textit{$\kappa$}=5} & -1 & 0.004 & 0.1$Z_{\odot}$ & 100 & 4.7 \\
				
				\hline
				\multirow{2}{*}{SDSS J140625.75} & M-B & -1 & 0.012 & $Z_{\odot}$ & 100 & 31.03 &  
				\multirowcell{2}{\scriptsize OVI+CII, NV, OI, CIV, HeII, OIII], CIII] } \\ 
				& \textbf{\textit{$\kappa$}=5} & -1 & 0.014 & 2$Z_{\odot}$ & 100 & 23.7 \\
				
				\hline
				\multirow{2}{*}{SDSS J141114.21} & M-B & -1 & 0.007 & 0.1$Z_{\odot}$ & 100 & 17.7 &  
				\multirowcell{2}{\scriptsize OVI+CII, NV, OI, CIV, SiIII, CIII] } \\ 
				& \textbf{\textit{$\kappa$}=5} & -1 & 0.007 & 0.1$Z_{\odot}$ & 100 & 13.04 \\		
				
				\hline
				\multirow{2}{*}{SDSS J141853.65} & M-B & -1 & 0.002 & $Z_{\odot}$ & 10$^4$ & 8.2 &  
				\multirowcell{2}{\scriptsize OI, CIV, HeII, OIII], SiIII], CIII] } \\ 
				& \textbf{\textit{$\kappa$}=5} & -1 & 0.002 & 2$Z_{\odot}$ & 10$^4$ & 6.3 \\
				
				\hline 
				\multirow{2}{*}{SDSS J144441.05} & M-B & -1 & 0.12 & 3$Z_{\odot}$ & 10$^4$ & 2.5 &  
				
				\multirowcell{2}{\scriptsize NV, CIV, HeII, CIII] } \\ 
				& \textit{$\kappa$}=5 & -1 & 1.2 & 2$Z_{\odot}$ & 10$^6$ & 2.4 \\
				
				\hline
				\multirow{2}{*}{SDSS J151747.00} & M-B & -1 & 0.007 & 0.5$Z_{\odot}$ & 10$^4$ & 10.7 &  
				\multirowcell{2}{\scriptsize OVI+CII, NV, OI, CIV, HeII, OIII], CIII] } \\ 
				& \textbf{\textit{$\kappa$}=5} & -1 & 0.0065 & $Z_{\odot}$ & 10$^4$ & 8.2 \\ 
				
				\hline
				\multirow{2}{*}{SDSS J151815.55} & M-B & -1 & 0.0106 & $Z_{\odot}$ & 10$^4$ & 50.3 &  
				\multirowcell{2}{\scriptsize OVI+CII, NV, OI, CIV, [NeIV]$lambda$1602, HeII, OIII], CIII] } \\ 
				& \textbf{\textit{$\kappa$}=5} & -1 & 0.0106 & 2$Z_{\odot}$ & 10$^4$ & 44.07 \\		
				
				\hline
				\multirow{2}{*}{SDSS J152051.00} & M-B & -1 & 0.008 & $Z_{\odot}$ & 100 & 12.1 &  
				\multirowcell{2}{\scriptsize OVI+CII, NV, OI, SiIV+OIV], NIV], CIV, HeII, OIII], CIII] } \\
				& \textbf{\textit{$\kappa$}=5} & -1 & 0.008 & $Z_{\odot}$ & 100 & 9.6 \\
				
				\hline
				\multirow{2}{*}{SDSS J152105.83} & M-B & -1 & 0.0045 & $Z_{\odot}$ & 10$^4$ & 6.9 &  
				\multirowcell{2}{\scriptsize NV, OI, SiIV+OIV], NIV], CIV, HeII, OIII], SiIII], CIII] } \\ 
				& \textbf{\textit{$\kappa$}=5} & -1 & 0.004 & 2$Z_{\odot}$ & 10$^4$ & 5.7 \\
				
				\hline 
				\multirow{2}{*}{SDSS J100345.59} & M-B & -1 & 0.41 & 2$Z_{\odot}$ & 10$^4$ & 0.008 & 
				\multirowcell{2}{\scriptsize NV, CIV, CIII] } \\
				& \textit{$\kappa$}=5 & -1 & 0.37 & 3$Z_{\odot}$ & 100 & 0.0001 \\
				
				\hline
				\multirow{2}{*}{SDSS J161343.40} & M-B & -1 & 0.009 & $Z_{\odot}$ & 10$^4$ & 19.3 & 
				\multirowcell{2}{\scriptsize OVI+CII, NV, OI, CIV, HeII, OIII], CIII] } \\ 
				& \textit{$\kappa$}=5 & -1 & 0.009 & $Z_{\odot}$ & 10$^4$ & 15.5 \\
				
				\hline
				\multirow{2}{*}{SDSS J162327.66} & M-B & -1 & 0.012 & $Z_{\odot}$ & 10$^4$ & 23.1 & 
				\multirowcell{2}{\scriptsize NV, OI, SiIV+OIV], CIV, [NeIV]$\lambda$1602, HeII, OIII], SiIII], CIII], [NeIV]$\lambda$2422 } \\ 
				& \textit{$\kappa$}=5 & -1 & 0.012 & $Z_{\odot}$ & 10$^4$ & 22.2 \\			
				
				\hline					
				\multirow{2}{*}{SDSS J165525.54} & M-B & -1 & 0.009 & $Z_{\odot}$ & 10$^4$ & 32.2 & 
				\multirowcell{2}{\scriptsize OVI+CII, NV, OI, SiIV+OIV], CIV, HeII, OIII], CIII] } \\ 
				& \textit{$\kappa$}=5 & -1 & 0.0106 & 2$Z_{\odot}$ & 10$^4$ & 21.2 \\	
				
				\hline
				\multirow{2}{*}{SDSS J170558.64} & M-B & -1 & 0.007 & $Z_{\odot}$ & 10$^4$ & 14.1 & 
				\multirowcell{2}{\scriptsize NV, OI, SiIV+OIV], CIV, HeII, NIII], SiIII], CIII], MgII } \\ 
				& \textit{$\kappa$}=5 & -1 & 0.0065 & 3$Z_{\odot}$ & 10$^4$ & 12.4 \\
				
				\hline
				\multirow{2}{*}{SDSS J150451.51} & M-B & -1 & 0.007 & $Z_{\odot}$ & 10$^4$ & 25.4 &  
				\multirowcell{2}{\scriptsize OVI+CII, NV, OI, SiIV+OIV], CIV, HeII, CIII] } \\ 
				& \textbf{\textit{$\kappa$}=5} & -1 & 0.005 & 3$Z_{\odot}$ & 10$^4$ & 19.6 \\ 
				
				\hline
				\multirow{2}{*}{SDSS J230451.68} & M-B & -1 & 0.012 & $Z_{\odot}$ & 10$^4$ & 33.1 & 
				\multirowcell{2}{\scriptsize OVI+CII, NV, OI, SiIV+OIV], CIV, HeII, NIII] } \\ 
				& \textbf{\textit{$\kappa$}=5} & -1 & 0.012 & 3$Z_{\odot}$ & 10$^4$ & 23.2 \\ 
				
				\hline
				\multirow{2}{*}{SDSS J234612.64} & M-B & -1 & 0.005 & 0.1$Z_{\odot}$ & 100 & 11.4 & 
				\multirowcell{2}{\scriptsize NV, OI, CIV, SiIII], CIII], MgII } \\ 
				& \textbf{\textit{$\kappa$}=5} & -1 & 0.005 & 0.1$Z_{\odot}$ & 100 & 5.7 \\ 
				
				\hline
				\multirow{2}{*}{SDSS J214609.96} & M-B & -1 & 0.0058 & 0.1$Z_{\odot}$ & 100 & 57.9 & 
				\multirowcell{2}{\scriptsize OVI+CII, OI, CIV, CIII] } \\ 
				& \textit{$\kappa$}=5 & -1 & 0.0058 & 0.1$Z_{\odot}$ & 100 & 33.9 \\
				
				\hline
				\multirow{2}{*}{SDSS J220126.11} & M-B & -1 & 0.0065 & 0.1$Z_{\odot}$ & 100 & 18.4 & 
				\multirowcell{2}{\scriptsize NV, OI, CIV, CIII] } \\ 
				& \textbf{\textit{$\kappa$}=5} & -1 & 0.0065 & 0.1$Z_{\odot}$ & 100 & 12.7 \\
				
				\hline
				\multirow{2}{*}{SDSS J155200.53} & M-B & -1 & 0.002 & 2$Z_{\odot}$ & 10$^4$ & 16.9 & 
				\multirowcell{2}{\scriptsize OI, CIV, HeII, OIII], SiIII], CIII] } \\ 
				& \textbf{\textit{$\kappa$}=5} & -1 & 0.002 & 3$Z_{\odot}$ & 10$^4$ & 11.9 \\
				
				\hline 	
								
				\multirow{2}{*}{SDSS J160103.85} & M-B & -1 & 0.0058 & $Z_{\odot}$ & 10$^4$ & 16.6 & 
				\multirowcell{2}{\scriptsize NV, OI, CIV, HeII, SiIII],	CIII], MgII } \\ 
				& \textbf{\textit{$\kappa$}=5} & -1 & 0.005 & 3$Z_{\odot}$ & 10$^4$ & 12.6 \\
				
				\hline
				\multirow{2}{*}{SDSS J223348.09} & M-B & -1 & 0.0065 & 0.1$Z_{\odot}$ & 10$^6$ & 19.4 &
				\multirowcell{2}{\scriptsize NV, OI, SiIV+OIV], CIV, [NeIV]$\lambda$1602, SiIII], CIII], [NeIV]$\lambda$2422  } \\ 
				& \textbf{\textit{$\kappa$}=5} & -1 & 0.0065 & 0.1$Z_{\odot}$ & 10$^6$ & 15.4 \\
				
				\hline
					
			\end{tabular}
		\end{center}
	\end{table*}

	\begin{table*}
		\caption{ Continuation of Table \ref{table: Best fit model parameters QSOs 1}. }
		\vspace{-0.2cm}
		\label{table: Best fit model parameters QSO2s 3}     
		\begin{center}                         
			\begin{tabular}{c c c c c c c c }        
				Object   & EED &  $\alpha$ & $U$ & $Z$ & $n_H$ & $\chi^2$ & EL \\
				\hline \hline	
				\multirow{2}{*}{SDSS J212951.40} & M-B & -1 & 0.0065 & 0.1$Z_{\odot}$ & 100 & 17.6 & 
				\multirowcell{2}{\scriptsize OVI+CII, NV, OI, CIV, NIII], SiIII], CIII] } \\ 
				& \textit{$\kappa$}=5 & -1 & 0.0065 & 0.1$Z_{\odot}$ & 100 & 12.6 \\
				
				\hline 
				\multirow{2}{*}{SDSS J120835.95} & M-B & -1 & 0.4 & 3$Z_{\odot}$ & 10$^6$ & 0.05 & 
				\multirowcell{2}{\scriptsize NV, CIV, CIII] } \\
				& \textit{$\kappa$}=5 & -1.5 & 1.8 & 3$Z_{\odot}$ & 100 & 0.004 \\	
				
				\hline 
				\multirow{2}{*}{SDSS J074251.43} & M-B & -1.5 & 0.05 & $Z_{\odot}$ & 100 & 0.02 & 
				\multirowcell{2}{\scriptsize OVI+CII, NV, CIV, CIII] } \\ 
				& \textit{$\kappa$}=5 & -1.5 & 0.096 & 2$Z_{\odot}$ & 10$^4$ & 0.004 \\ 
				
				\hline
				\multirow{2}{*}{SDSS J082530.67} & M-B & -1 & 0.25 & 2$Z_{\odot}$ & 10$^6$ & 0.09 & 
				\multirowcell{2}{\scriptsize NV, CIV, HeII, CIII] } \\ 
				& \textit{$\kappa$}=10 & -1 & 0.37 & 3$Z_{\odot}$ & 10$^4$ & 0.1 \\
				
				\hline 
				\multirow{2}{*}{SDSS J100133.85} & M-B & -1 & 0.2 & 3$Z_{\odot}$ & 10$^6$ & 0.04 & 
				\multirowcell{2}{\scriptsize NV, CIV, HeII, CIII] } \\ 
				& \textit{$\kappa$}=10 & -1 & 0.176 & 3$Z_{\odot}$ & 10$^6$ & 0.008 \\			
				
				\hline
				\multirow{2}{*}{SDSS J133927.80} & M-B & -1 & 0.0065 & 0.5$Z_{\odot}$ & 10$^4$ & 11.1 & 
				\multirowcell{2}{\scriptsize NV, OI, NIV], CIV, [NeIV]$\lambda$1602, HeII, OIII], SiIII], CIII], [NeIV]$\lambda$2422 } \\ 
				& \textit{$\kappa$}=10 & -1 & 0.0065 & 0.5$Z_{\odot}$ & 10$^4$ & 11.1 \\
				
				\hline
				\multirow{2}{*}{SDSS J150145.46} & M-B & -1 & 0.29 & 3$Z_{\odot}$ & 10$^6$ & 0.2 &  
				\multirowcell{2}{\scriptsize NV, CIV, HeII, CIII] } \\ 
				& \textit{$\kappa$}=10 & -1 & 0.29 & 3$Z_{\odot}$ & 10$^4$ & 0.2 \\
				
				\hline
				\multirow{2}{*}{SDSS J123742.37} & M-B & -1 & 0.25 & 2$Z_{\odot}$ & 10$^6$ & 0.0002  & 
				\multirowcell{2}{\scriptsize NV, CIV, CIII] } \\ 
				& \textit{$\kappa$}=10 & -1 & 0.37 & 3$Z_{\odot}$ & 10$^4$ & 8E-6 \\	
				
				\hline  
				\multirow{2}{*}{SDSS J114542.07} & M-B & -1 & 0.066 & 3$Z_{\odot}$ & 10$^6$ & 2.6  & 
				\multirowcell{2}{\scriptsize OVI+CII, NV, SiIV+OIV], NIV], CIV, HeII, OIII], NIII], CIII] } \\
				& \textit{$\kappa$}=10 & -1 & 0.059 & 3$Z_{\odot}$ & 10$^4$ & 2.5 \\
				
				\hline
				\multirow{2}{*}{SDSS J131659.84} & M-B & -1.5 & 0.22 & 2$Z_{\odot}$ & 10$^4$ & 0.006 &  
				\multirowcell{2}{\scriptsize OVI+CII, NV, CIV, CIII] } \\ 
				& \textit{$\kappa$}=10 & -1.5 & 0.138 & 2$Z_{\odot}$ & 100 & 0.002 \\
				
				\hline
				\multirow{2}{*}{SDSS J213843.08} & M-B & -1.5 & 0.14 & $Z_{\odot}$ & 10$^4$ & 0.2 & 
				\multirowcell{2}{\scriptsize OVI+CII, NV, CIV, HeII, OIII], CIII] } \\ 
				& \textit{$\kappa$}=10 & -1 & 0.05 & $Z_{\odot}$ & 10$^4$ & 0.2 \\
				
				\hline
				\multirow{2}{*}{SDSS J104722.99} & M-B & -1 & 0.08 & 3$Z_{\odot}$ & 10$^6$ & 0.56 & 
				\multirowcell{2}{\scriptsize OVI+CII, NV, CIV, CIII] } \\ 
				& \textit{$\kappa$}=10 & -1.5 & 0.2 & 3$Z_{\odot}$ & 10$^4$ & 0.57 \\ 
				
				\hline
				\multirow{2}{*}{SDSS J161058.48} & M-B & -1.5 & 0.3 & 3$Z_{\odot}$ & 10$^4$ & 0.003 & 
				\multirowcell{2}{\scriptsize NV, CIV, CIII] } \\ 
				& \textit{$\kappa$}=10 & -1 & 0.2 & 2$Z_{\odot}$ & 100 & 0.0006 \\ 
				
				\hline
				\multirow{2}{*}{SDSS J004423.20} & M-B & -1 & 0.022 & 2$Z_{\odot}$ & 10$^6$ & 0.4 & 
				\multirowcell{2}{\scriptsize NV, CIV, HeII, OIII], CIII] } \\ 
				& \textit{$\kappa$}=10 & -1.5 & 0.86 & 2$Z_{\odot}$ & 10$^6$ & 0.3 \\
				
				\hline
				\multirow{2}{*}{SDSS J221601.21} & M-B & -1 & 0.3 & 2$Z_{\odot}$ & 10$^4$ & 0.0003 & 
				\multirowcell{2}{\scriptsize NV, CIV, CIII] } \\ 
				& \textit{$\kappa$}=20 & -1 & 1.1 & 3$Z_{\odot}$ & 10$^4$ & 0.0001 \\
				
				\hline
				\multirow{2}{*}{SDSS J163343.85} & M-B & -1 & 0.0065 & $Z_{\odot}$ & 10$^4$ & 16.7 & 
				\multirowcell{2}{\scriptsize OVI+CII, NV, OI, CIV, [NeIV]$\lambda$1602, HeII, SiIII], CIII], [NeIV]$\lambda$2422} \\ 
				& \textit{$\kappa$}=20 & -1 & 0.00736 & $Z_{\odot}$ & 10$^4$ & 16.2 \\ 
				
				\hline 	
				\multirow{2}{*}{SDSS J115947.86} & M-B & -1 & 0.25 & 3$Z_{\odot}$ & 10$^4$ & 0.2 & 
				\multirowcell{2}{\scriptsize NV, CIV, HeII, CIII] } \\ 
				& \textit{$\kappa$}=20 & -1 & 0.156 & 3$Z_{\odot}$ & 10$^4$ & 0.1 \\
				
				\hline  
				\multirow{2}{*}{SDSS J103249.55} & M-B & -1 & 0.0065 & $Z_{\odot}$ & 10$^4$ & 18.8 & 
				\multirowcell{2}{\scriptsize NV, OI, SiIV+OIV], CIV, [NeIV]$\lambda$1602, HeII,	SiIII], CIII], [NeIV]$\lambda$2422, MgII } \\ 
				& \textit{$\kappa$}=20 & -1 & 0.00736 & $Z_{\odot}$ & 10$^4$ & 18.1 \\
				
				\hline	
				\multirow{2}{*}{SDSS J091357.87} & M-B & -1 & 0.2 & 3$Z_{\odot}$ & 10$^4$ & 0.35  & 
				\multirowcell{2}{\scriptsize OVI+CII, NV, CIV, HeII, CIII] } \\ 
				& \textit{$\kappa$}=20 & -1 & 0.156 & 3$Z_{\odot}$ & 10$^4$ & 0.3 \\ 			
				
				\hline
				\multirow{2}{*}{SDSS J151544.01} & M-B & -1 & 0.25 & 2$Z_{\odot}$ & 10$^6$ & 0.56 &  
				\multirowcell{2}{\scriptsize NV, SiIV+OIV], NIV], CIV, HeII, OIII], CIII] } \\ 
				& \textit{$\kappa$}=20 & -1 & 0.096 & 2$Z_{\odot}$ & 100 & 0.6 \\			
				
				\hline
				\multirow{2}{*}{SDSS J161447.97} & M-B & -1 & 0.046 & $Z_{\odot}$ & 10$^4$ & 0.27 & 
				\multirowcell{2}{\scriptsize NV, CIV, HeII, OIII], CIII] } \\ 
				& \textit{$\kappa$}=20 & -1 & 0.05 & $Z_{\odot}$ & 10$^4$ & 0.26 \\
				
				\hline
				\multirow{2}{*}{SDSS J162500.57} & M-B & -1 & 0.4 & 3$Z_{\odot}$ & 10$^4$ & 0.0008 & 
				\multirowcell{2}{\scriptsize NV, CIV, HeII } \\ 
				& \textit{$\kappa$}=20 & -1 & 0.29 & 3$Z_{\odot}$ & 10$^4$ & 0.0002 \\ 
				
				\hline
				\multirow{2}{*}{SDSS J122214.45} & M-B & -1.5 & 0.156 & 2$Z_{\odot}$ & 100 & 0.05  & 
				\multirowcell{2}{\scriptsize OVI+CII, NV, CIV, HeII, CIII] } \\ 
				& \textit{$\kappa$}=20 & -1.5 & 0.02 & 3$Z_{\odot}$ & 100 & 0.07 \\
				
				\hline 
				\multirow{2}{*}{SDSS J105344.18} & M-B & -1 & 0.075 & 3$Z_{\odot}$ & 10$^4$ & 0.4 & 
				\multirowcell{2}{\scriptsize OVI+CII, NV, CIV, HeII, CIII] } \\  
				& \textit{$\kappa$}=20 & -1.5 & 0.176 & 3$Z_{\odot}$ & 10$^4$ & 0.38 \\
				
				\hline
				\multirow{2}{*}{SDSS J081257.15} & M-B & -1 & 0.03 & $Z_{\odot}$ & 100 & 0.25 & 
				\multirowcell{2}{\scriptsize NV, CIV, HeII, CIII] } \\ 
				& \textit{$\kappa$}=20 & -1.5 & 1.8 & 2$Z_{\odot}$ & 10$^4$ & 0.05 \\			
				
				\hline
				\multirow{2}{*}{SDSS J162651.76} & M-B & -1 & 0.032 & $Z_{\odot}$ & 10$^6$ & 0.02 & 
				\multirowcell{2}{\scriptsize NV, CIV, HeII, CIII] } \\ 
				& \textit{$\kappa$}=20 & -1.5 & 0.2 & 2$Z_{\odot}$ & 10$^4$ & 0.003 \\ 
				
				\hline
				\multirow{2}{*}{SDSS J212055.57} & M-B & -1 & 0.0058 & $Z_{\odot}$ & 10$^4$ & 13.5 & 
				\multirowcell{2}{\scriptsize NV, OI, SiIV+OIV], CIV, [NeIV]$\lambda$1602, HeII, SiIII], CIII], [NeIV]$\lambda$2422, MgII } \\ 
				& \textit{$\kappa$}=20 & -1 & 0.0065 & $Z_{\odot}$ & 10$^4$ & 13.1 \\	
								
				\hline 
				\multirow{2}{*}{SDSS J023519.66} & M-B & -1 & 0.14 & 3$Z_{\odot}$ & 100 & 1.9 & 
				\multirowcell{2}{\scriptsize OVI+CII, NV, CIV } \\ 
				& \textit{$\kappa$}=40 & -1 & 0.12 & 3$Z_{\odot}$ & 100 & 1.98 \\ 
				
				\hline
				\multirow{2}{*}{SDSS J024525.95} & M-B & -1 & 0.05 & 2$Z_{\odot}$ & 10$^4$ & 0.68 & 
				\multirowcell{2}{\scriptsize OVI+CII, NV, CIV, HeII, OIII], CIII] } \\ 
				& \textit{$\kappa$}=40 & -1 & 0.059 & 2$Z_{\odot}$ & 10$^4$ & 0.7 \\
				
				\hline 
				\multirow{2}{*}{SDSS J073851.85} & M-B & -1 & 0.138 & 2$Z_{\odot}$ & 10$^4$ & 0.04 & 
				\multirowcell{2}{\scriptsize NV, CIV, HeII, OIII], CIII] } \\ 
				& \textit{$\kappa$}=40 & -1 & 0.075 & 2$Z_{\odot}$ & 10$^4$ & 0.048 \\ 
				
				\hline				 
				\multirow{2}{*}{SDSS J025339.00} & M-B & -1 & 0.096 & 2$Z_{\odot}$ & 10$^6$ & 0.9 & 
				\multirowcell{2}{\scriptsize OVI+CII, NV, SiIV+OIV], CIV, HeII, CIII] } \\ 
				& \textit{$\kappa$}=40 & -1 & 0.108 & 2$Z_{\odot}$ & 10$^6$ & 0.9 \\

				\hline
				\multirow{2}{*}{SDSS J082920.43} & M-B & -1 & 0.08 & 2$Z_{\odot}$ & 10$^6$ & 0.2 & 
				\multirowcell{2}{\scriptsize NV, SiIV+OIV], CIV, CIII] } \\ 
				& \textit{$\kappa$}=40 & -1 & 0.22 & 2$Z_{\odot}$ & 10$^6$ & 0.3 \\
				
				\hline
			\end{tabular}
		\end{center}
	\end{table*}

	\begin{table*}
		\caption{ Continuation of Table \ref{table: Best fit model parameters QSOs 1}. }
		\vspace{-0.2cm}
		\label{table: Best fit model parameters QSO2s 4}     
		\begin{center}                         
			\begin{tabular}{c c c c c c c c}        
				Object   & EED &  $\alpha$ & $U$ & $Z$ & $n_H$ & $\chi^2$ & EL \\
				\hline \hline
				 
				\multirow{2}{*}{SDSS J095819.35} & M-B & -1 & 0.008 & $Z_{\odot}$ & 10$^6$ & 8.4 & 
				\multirowcell{2}{\scriptsize OVI+CII, NV, CIV, HeII, OIII], SiIII],	CIII] } \\ 
				& \textit{$\kappa$}=40 & -1 & 0.015 & $Z_{\odot}$ & 10$^6$ & 8.9 \\
				
				\hline 
				\multirow{2}{*}{SDSS J101448.79} & M-B & -1 & 0.2 & 2$Z_{\odot}$ & 10$^6$ & 0.3 & 
				\multirowcell{2}{\scriptsize NV, NIV], CIV, HeII, CIII] } \\ 
				& \textit{$\kappa$}=40 & -1 & 0.29 & 2$Z_{\odot}$ & 10$^6$ & 0.4 \\
				
				\hline
				\multirow{2}{*}{SDSS J112343.18} & M-B & -1 & 0.08 & 0.1$Z_{\odot}$ & 10$^4$ & 1.98 & 
				\multirowcell{2}{\scriptsize NV, CIV, CIII], MgII } \\ 
				& \textit{$\kappa$}=40 & -1 & 0.07 & 0.1$Z_{\odot}$ & 100 & 2.4 \\
				
				\hline			
				\multirow{2}{*}{SDSS J114753.29} & M-B & -1 & 0.01 & 0.1$Z_{\odot}$ & 10$^6$ & 0.99 & 
				\multirowcell{2}{\scriptsize NV, CIV, CIII], MgII } \\ 
				& \textit{$\kappa$}=40 & -1 & 0.06 & 0.1$Z_{\odot}$ & 10$^6$ & 1.5 \\
				
				\hline
				\multirow{2}{*}{SDSS J125733.12} & M-B & -1 & 0.0065 & 0.1$Z_{\odot}$ & 10$^4$ & 2.9 &  
				\multirowcell{2}{\scriptsize OVI+CII, NV, CIV, MgII } \\ 
				& \textit{$\kappa$}=40 & -1 & 0.012 & 0.1$Z_{\odot}$ & 10$^4$ & 3.05 \\
				
				\hline		 
				\multirow{2}{*}{SDSS J124302.62} & M-B & -1 & 0.00736 & 0.5$Z_{\odot}$ & 10$^6$ & 4.5 &  
				\multirowcell{2}{\scriptsize NV, SIV+OIV], CIV, HeII, OIII], SiIII], CIII], MgII } \\
				& \textit{$\kappa$}=40 & -1 & 0.0136 & 0.5$Z_{\odot}$ & 10$^6$ & 4.7 \\
				
				\hline 
				\multirow{2}{*}{SDSS J133059.32} & M-B & -1 & 0.25 & 3$Z_{\odot}$ & 10$^4$ & 0.4  &  
				\multirowcell{2}{\scriptsize NV, CIV, HeII, CIII] } \\ 
				& \textit{$\kappa$}=40 & -1 & 0.2 & 3$Z_{\odot}$ & 10$^4$ & 0.35 \\
				
				\hline
				\multirow{2}{*}{SDSS J133417.04} & M-B & -1 & 0.066 & 2$Z_{\odot}$ & 10$^6$ & 0.3 &  
				\multirowcell{2}{\scriptsize NV, SiIV+OIV], NIV], CIV, HeII, OIII], CIII] } \\ 
				& \textit{$\kappa$}=40 & -1 & 0.1 & 2$Z_{\odot}$ & 10$^6$ & 0.34 \\
				
				\hline
				\multirow{2}{*}{SDSS J155108.96} & M-B & -1 & 0.009 & 0.5$Z_{\odot}$ & 10$^6$ & 6.0 & 
				\multirowcell{2}{\scriptsize NV, NIV], CIV, [NeIV]$\lambda$1602, HeII, SiIII], CIII], [NeIV]$\lambda$2422 } \\ 
				& \textit{$\kappa$}=40 & -1 & 0.017 & 0.5$Z_{\odot}$ & 10$^6$ & 6.5 \\ 	
				
				\hline			
				\multirow{2}{*}{SDSS J155725.27} & M-B & -1 & 0.025 & $Z_{\odot}$ & 10$^6$ & 5.5 & 
				\multirowcell{2}{\scriptsize OVI+CII, NV, SiIV+OIV], CIV, [NeIV]$\lambda$1602, HeII, CIII], [NeIV]$\lambda$2422 } \\ 
				& \textit{$\kappa$}=40 & -1 & 0.04 & $Z_{\odot}$ & 10$^6$ & 5.8 \\
				
				\hline		
				\multirow{2}{*}{SDSS J160158.53} & M-B & -1 & 0.1 & 3$Z_{\odot}$ & 10$^6$ & 0.19 & 
				\multirowcell{2}{\scriptsize OVI+CII, NV, CIV, HeII, CIII] } \\ 
				& \textit{$\kappa$}=40 & -1 & 0.1 & 3$Z_{\odot}$ & 10$^6$ & 0.22 \\
				
				\hline
				\multirow{2}{*}{SDSS J020728.19} & M-B & -1 & 0.008 & $Z_{\odot}$ & 10$^4$ & 12.6 & 
				\multirowcell{2}{\scriptsize OVI+CII, NV, OI, SiIV+OIV], CIV, HeII, OIII], NIII], SiIII], CIII] } \\ 
				& \textit{$\kappa$}=40 & -1 & 0.012 & $Z_{\odot}$ & 10$^4$ & 12.5 \\ 
				
				\hline
				\multirow{2}{*}{SDSS J083031.86} & M-B & -1 & 0.0065 & 0.1$Z_{\odot}$ & 10$^6$ & 1.9 & 
				\multirowcell{2}{\scriptsize OVI+CII, NV, CIV, [NeIV]$\lambda$1602, SiIII], CIII], [NeIV]$\lambda$2422 } \\ 
				& \textit{$\kappa$}=40 & -1 & 0.01 & 0.1$Z_{\odot}$ & 10$^6$ & 2.06 \\
				
				\hline
				\multirow{2}{*}{SDSS J162025.94} & M-B & -1 & 0.0136 & 0.5$Z_{\odot}$ & 10$^6$ & 4.9 & 
				\multirowcell{2}{\scriptsize NV, CIV, [NeIV]$\lambda$1602, HeII, OIII], SiIII], CIII], [NeIV]$\lambda$2422, MgII } \\ 
				& \textit{$\kappa$}=40 & -1 & 0.02 & 0.5$Z_{\odot}$ & 10$^6$ & 5.3 \\
				
				\hline
				\multirow{2}{*}{SDSS J162806.01} & M-B & -1 & 0.009 & 0.5$Z_{\odot}$ & 10$^6$ & 8.3 & 
				\multirowcell{2}{\scriptsize NV, CIV, HeII, SiIII], CIII] } \\ 
				& \textit{$\kappa$}=40 & -1 & 0.02 & 0.5$Z_{\odot}$ & 10$^6$ & 9.1 \\
				
				\hline
				\multirow{2}{*}{SDSS J170110.12} & M-B & -1 & 0.04 & 3$Z_{\odot}$ & 10$^6$ & 2.6 & 
				\multirowcell{2}{\scriptsize OVI+CII, NV, SiIV+OIV], CIV, HeII, OIII], NIII], CIII] } \\
				& \textit{$\kappa$}=40 & -1 & 0.05 & 3$Z_{\odot}$ & 10$^6$ & 2.7 \\	
				
				\hline
				\multirow{2}{*}{SDSS J171908.90} & M-B & -1 & 0.02 & $Z_{\odot}$ & 10$^6$ & 0.009 & 
				\multirowcell{2}{\scriptsize NV, CIV, HeII, OIII] CIII] } \\ 
				& \textit{$\kappa$}=40 & -1 & 0.03 & $Z_{\odot}$ & 10$^6$ & 0.008 \\

				\hline
				\multirow{2}{*}{SDSS J225607.63} & M-B & -1 & 0.05 & 3$Z_{\odot}$ & 10$^4$ & 0.7 & 
				\multirowcell{2}{\scriptsize OVI+CII, NV, CIV, HeII } \\ 
				& \textit{$\kappa$}=40 & -1 & 0.05 & 3$Z_{\odot}$ & 10$^4$ & 0.7 \\	
				
				\hline
				\multirow{2}{*}{SDSS J215341.33} & M-B & -1 & 0.2 & 2$Z_{\odot}$ & 100 & 0.5 & 
				\multirowcell{2}{\scriptsize NV, SiIV+OIV], CIV, HeII, CIII] } \\ 
				& \textit{$\kappa$}=40 & -1 & 0.176 & 2$Z_{\odot}$ & 100 & 0.6 \\
				
				\hline
				\multirow{2}{*}{SDSS J213557.35} & M-B & -1 & 0.036 & 0.5$Z_{\odot}$ & 10$^6$ & 2.4  & 
				\multirowcell{2}{\scriptsize NV, CIV, [NeIV]$\lambda$1602, HeII, CIII], [NeIV]$\lambda$2422 } \\ 
				& \textit{$\kappa$}=40 & -1 & 0.08 & 0.5$Z_{\odot}$ & 10$^6$ & 2.5 \\
				
				\hline
				\multirow{2}{*}{SDSS J114703.82} & M-B & -1.5 & 0.04 & 2$Z_{\odot}$ & 10$^6$ & 1.4 & 
				\multirowcell{3}{\scriptsize NV, CIV, HeII, CIII], MgII } \\ 
				& \textit{$\kappa$}=40 & -1.5 & 0.5 & 2$Z_{\odot}$ & 10$^4$ & 1.6 \\
				
				\hline										
				\multirow{2}{*}{SDSS J090612.64} & M-B & -1.5 & 0.096 & 2$Z_{\odot}$ & 10$^4$ & 2.0 & 
				\multirowcell{2}{\scriptsize NV, CIV, HeII, CIII], MgII } \\ 
				& \textit{$\kappa$}=40 & -1.5 & 0.29 & 3$Z_{\odot}$ & 10$^4$ & 2.4 \\
				
				\hline		
				\multirow{2}{*}{SDSS J073637.54} & M-B & -1.5 & 0.036 & 2$Z_{\odot}$ & 10$^6$ & 1.2 & 
				\multirowcell{2}{\scriptsize NV, CIV, HeII, CIII], MgII } \\ 
				& \textit{$\kappa$}=40 & -1.5 & 0.176 & 2$Z_{\odot}$ & 10$^6$ & 1.6 \\					
				
				\hline			
				\multirow{2}{*}{SDSS J003605.26} & M-B & -1.5 & 0.53 & 2$Z_{\odot}$ & 10$^4$ & 0.2 & 
				\multirowcell{2}{\scriptsize NV, CIV, CIII] } \\ 
				& \textit{$\kappa$}=40 & -1.5 & 1.8 & 3$Z_{\odot}$ & 10$^4$ & 0.1 \\
				
				\hline	
								
				\multirow{2}{*}{SDSS J004600.48} & M-B & -1.5 & 0.04 & 2$Z_{\odot}$ & 10$^6$ & 2.01 & 
				\multirowcell{2}{\scriptsize NV, SiIV+OIV], CIV, HeII, CIII], MgII } \\ 
				& \textit{$\kappa$}=40 & -1.5 & 0.096 & 2$Z_{\odot}$ & 10$^6$ & 2.1 \\	
				
				\hline 
				\multirow{2}{*}{SDSS J004728.77} & M-B & -1.5 & 0.37 & 3$Z_{\odot}$ & 10$^4$ & 0.002 & 
				\multirowcell{2}{\scriptsize OVI+CII, NV, CIV } \\ 
				& \textit{$\kappa$}=40 & -1 & 0.12 & 3$Z_{\odot}$ & 10$^4$ & 0.0003 \\ 
				
				\hline
				\multirow{2}{*}{SDSS J224532.92} & M-B & -1.5 & 0.2 & 2$Z_{\odot}$ & 100 & 0.12 & 
				\multirowcell{2}{\scriptsize OVI+CII, NV, CIV, CIII] } \\ 
				& \textit{$\kappa$}=40 & -1.5 & 0.08 & 2$Z_{\odot}$ & 100 & 0.16 \\ 
				
				\hline		
				\multirow{2}{*}{SDSS J161353.27} & M-B & -1.5 & 0.05 & 2$Z_{\odot}$ & 10$^4$ & 5.5 & 
				\multirowcell{2}{\scriptsize OVI+CII, CIV, HeII, CIII], CII] } \\ 
				& \textit{$\kappa$}=40 & -1.5 & 0.156 & 3$Z_{\odot}$ & 10$^4$ & 6 \\ 
				
				\hline
				\multirow{2}{*}{SDSS J125154.02} & M-B & -1.5 & 0.156 & 3$Z_{\odot}$ & 100 & 0.26 &  
				\multirowcell{2}{\scriptsize OVI+CII, NV, CIV } \\
				& \textit{$\kappa$}=40 & -1.5 & 0.176 & 3$Z_{\odot}$ & 100 & 0.28 \\
				
				\hline
			\multirow{2}{*}{SDSS J023210.88} & M-B & -1.5 & 0.015 & $Z_{\odot}$ & 100 & 8.2 & 
				\multirowcell{2}{\scriptsize NV, CIV, CIII], MgII } \\ 
				& \textit{$\kappa$}=40 & -1 & 0.066 & 0.1$Z_{\odot}$ & 10$^6$ & 11.8 \\
				
				\hline
				\multirow{2}{*}{SDSS J075656.49} &\textbf{ M-B} & -1 & 0.007 & 0.5$Z_{\odot}$ & 10$^4$ & 21.95 &
				\multirowcell{2}{\scriptsize OVI+CII, NV, OI, SiIV+OIV], CIV, [NeIV]$\lambda$1602, \\
					\scriptsize	HeII, NIII], SiIII], CIII], [NeIV]$\lambda$2422 } \\
				& \textit{$\kappa$}=40 & -1 & 0.0106 & 0.5$Z_{\odot}$ & 10$^4$ & 22.8 \\  
				
				\hline
	\multirow{2}{*}{SDSS J112230.35} & M-B & -1 & 0.108 & $Z_{\odot}$ & 10$^6$ & 1.6 & 
			\multirowcell{2}{\scriptsize NV, SiIV+OIV], CIV, [NeIV]$\lambda$1602, HeII, CIII], [NeIV]2422 } \\ 
		& \textit{$\kappa$}=40 & -1 & 0.06 & 0.5$Z_{\odot}$ & 10$^6$ & 1.6 \\
				
				\hline
					
			\end{tabular}
		\end{center}
	\end{table*}

	\begin{table*}
		\caption{ Continuation of Table \ref{table: Best fit model parameters QSOs 1}. }
		\vspace{-0.2cm}
		\label{table: Best fit model parameters QSO2s 5}     
		\begin{center}                         
			\begin{tabular}{c c c c c c c c}        
				Object   & EED &  $\alpha$ & $U$ & $Z$ & $n_H$ & $\chi^2$ & ELs \\
				\hline \hline 
		
				\multirow{2}{*}{SDSS J153306.06} & M-B & -1 & 0.012 & $Z_{\odot}$ & 10$^4$ & 18.9 & 
				\multirowcell{2}{\scriptsize OVI+CII, NV, OI, SiIV+OIV], CIV, HeII, OIII], NIII], SiIII], CIII] } \\ 
				& \textit{$\kappa$}=40 & -1 & 0.015 & $Z_{\odot}$ & 10$^4$ & 19.2 \\
				\hline	
				\multirow{2}{*}{SDSS J222946.61} & M-B & -1 & 0.008 & $Z_{\odot}$ & 10$^6$ & 8.2 & 
				\multirowcell{2}{\scriptsize NV, CIV, HeII, SiIII], CIII] } \\ 
				& \textit{$\kappa$}=40 & -1 & 0.0136 & $Z_{\odot}$ & 10$^6$ & 9.01 \\
				
				\hline
				\multirow{2}{*}{SDSS J001040.82} & \textbf{M-B} & -1 & 0.0106 & 0.5$Z_{\odot}$ & 10$^6$ & 4.98 & 
				\multirowcell{2}{\scriptsize OVI+CII, NV, SiIV+OIV], CIV, [NeIV]$\lambda$1602, HeII, SiIII], CIII], [NeIV]$\lambda$2422 } \\
				& \textit{$\kappa$}=40 & -1 & 0.0173 & 0.5$Z_{\odot}$ & 10$^6$ & 5.4 \\    
				
				\hline
				\multirow{2}{*}{SDSS J001814.72} & M-B & -1 & 0.0196 & $Z_{\odot}$ & 10$^4$ & 0.13 & 
				\multirowcell{2}{\scriptsize OVI+CII, CIV, HeII, CIII] } \\ 
				& \textit{$\kappa$}=40 & -1 & 0.028 & $Z_{\odot}$ & 10$^4$ & 0.15 \\
				
				\hline
				\multirow{2}{*}{SDSS J012552.08} & M-B & -1 & 0.066 & 2$Z_{\odot}$ & 10$^6$ & 2.0 & 
				\multirowcell{2}{\scriptsize OVI+CII, NV, SiIV+OIV], CIV, HeII, CIII] } \\ 
				& \textit{$\kappa$}=40 & -1 & 0.1 & 2$Z_{\odot}$ & 10$^6$ & 2.0 \\
				
				\hline 
			\end{tabular}
		\end{center}
	\end{table*}

	The QSO2s are generally better fitted by \textit{$\kappa$} distributions (87 in 126 objects, see Fig. \ref{fig:count_EED_SDSS_HzRGs}). Of these, 65 (52\%) sources are best fitted by a \textit{$\kappa$} distribution with \textit{$\kappa$}=5. There are 38 objects (31\%) whose $\chi^2$ is lower when the M-B distribution is used. 
	
\begin{figure*}
		\centering
		\includegraphics[width=0.96\linewidth]{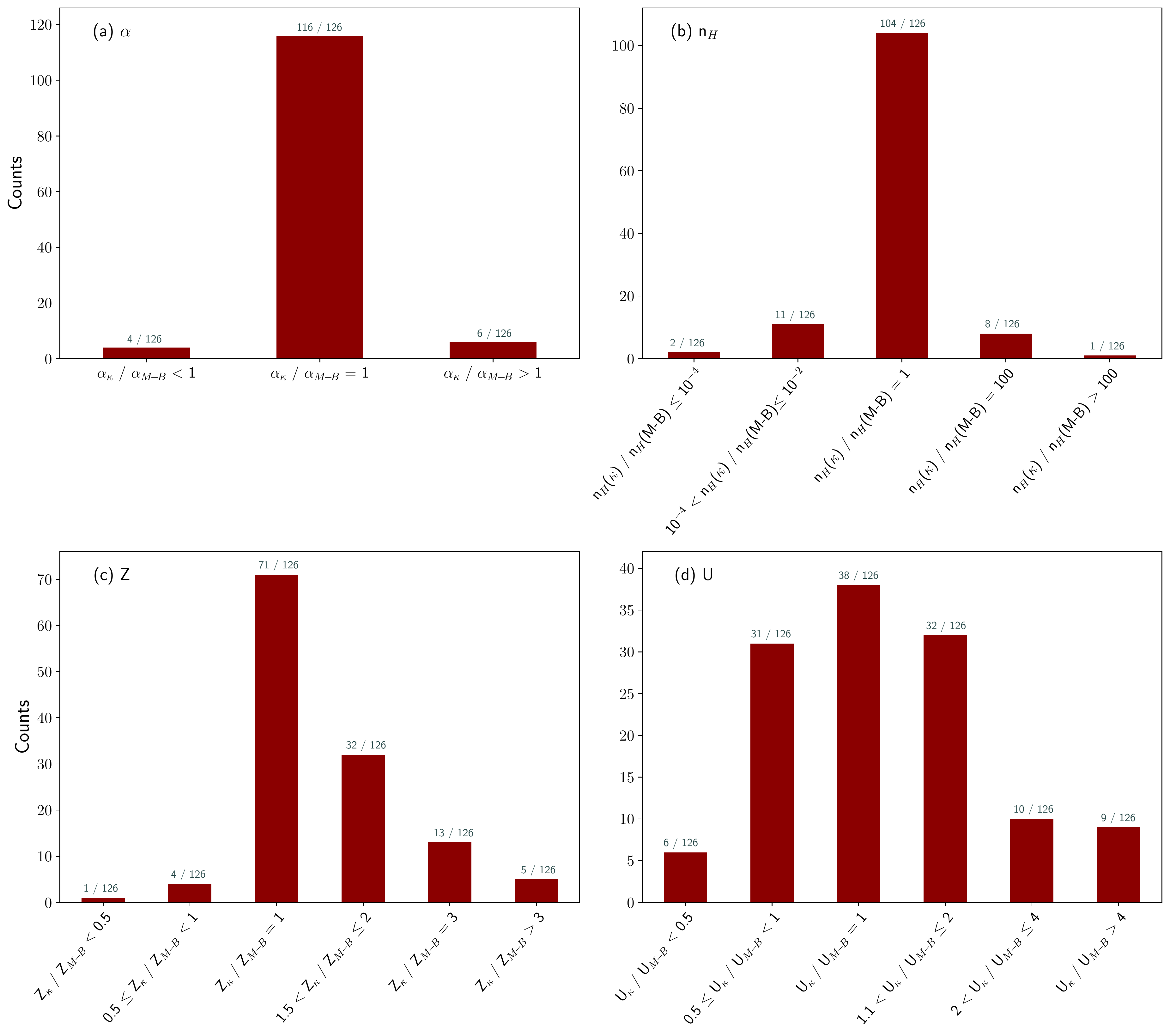}
		\caption{QSO2s binned according to discrepancy in (a) $\alpha$, (b) n$_H$, (c) Z, and (d) U between \textit{$\kappa$} and M-B distributions.}
		\label{fig:qso2s}
	\end{figure*}

	Fig. \ref{fig:qso2s} presents the QSO2s binned according to discrepancy in $\alpha$ (panel (a)), n$_H$ (panel (b)), Z (panel (c)) and U (panel (d)).
	Similarly to the HzRGs and the Seyfert galaxies, generally a spectral index of -1 gives the best fit to the emission line ratios. 
	The assumption of a \textit{$\kappa$} or M-B distribution has no impact on the inferred spectral index, $\alpha$, in most QSO2s (116/126 or 92\%).  
	
	Panel (b) of Fig. \ref{fig:qso2s} presents the ratio n$_H$(\textit{$\kappa$}) / n$_H$(M-B). 
	Assuming a \textit{$\kappa$} distribution instead of the M-B does not cause significant changes to the derived n$_H$. For 104 (83\%) QSO2s the value of n$_H$ is identical in \textit{$\kappa$} and M-B models. Of these 104 objects, 24 are best-fitted by a model with n$_H$=100 cm$^{-3}$, 55 by a model with n$_H$=10$^4$ cm$^{-3}$ and the remaining 25 are best-fitted by a model with n$_H$=10$^6$ cm$^{-3}$. 
	Deviations in n$_H$ by a factor as large as $\geq$ 100 are identified in the remaining 17\% of QSO2.	
	
	In panel (c) of Fig. \ref{fig:qso2s} the ratio Z$_{\kappa}$ / Z$_{M-B}$ is presented. \textit{$\kappa$} models produce identical or higher (Z$_{\kappa}$ $\geq$ Z$_{M-B}$) $Z$ than M-B models in 96\% (121/126) of cases. 
	The predicted metallicities are consistent (Z$_\textit{$\kappa$}$ / Z$_{M\!-\!B}$ = 1) in 56\% of QSO2s. 
	For 25\% of objects \textit{$\kappa$} models predict metallicities a factor of 1.5 to 2 times higher than M-B models.  
	Large discrepancies (Z$_\textit{$\kappa$}$ / Z$_{M\!-\!B}$ \textless 0.5 or Z$_\textit{$\kappa$}$ / Z$_{M\!-\!B}$ $\geq$ 3) in the predicted metallicities occur for 19 QSO2s (15\%).
	
	Panel (d) of Fig. \ref{fig:qso2s} presents the results for the ionization parameter. 
	The values of $U$ predicted by \textit{$\kappa$} and M-B models differ in 70\% (88/26) of QSO2s. The values predicted by \textit{$\kappa$} models are a factor of 1.1 to 2 times higher than U$_{M\!-\!B}$ in 25\% (32/126) of cases. 
	For 37 QSO2s (29\%) \textit{$\kappa$} models predict a lower ionization parameter than the M-B models. In 25\% of these cases, \textit{$\kappa$} models predict an ionization parameter that is a factor of 1.1 to 2 times lower than the value predicted by the M-B. Large discrepancies (U$_\textit{$\kappa$}$ / U$_{M\!-\!B}$ \textless 0.5 or U$_\textit{$\kappa$}$ / U$_{M\!-\!B}$ \textgreater 2) in the predicted ionization parameter occur for 25 QSO2s (20\%).

	\section{Discussion} \label{sec:discussion}
	
	Line ratios of infrared, optical and ultraviolet emission lines have been extensively used to study the properties of the gas in the narrow line regions of AGNs. Historically, photoionization models generally adopt a M-B distribution for their electron energies. There are, however, some discrepancies between the results of these models and the observed emission line flux ratios. These discrepancies can sometimes be resolved using \textit{$\kappa$} distributions of electron energies \citep[e.g.][]{Nicholls2012, Humphrey2014, Zhang2016}. 
	
	\subsection{Impact on NLR chemical abundance estimates}
	
	Super-solar metallicities have generally been invoked to explain the emission in the NLR of AGN \citep[e.g.][]{Nagao2006, Dors2014}. However, lower values have been found by \citet{Mignoli2019} who estimated a subsolar or close to solar metallicity in 89\% of their objects. They argued that the discrepancy between their results and the ones from other studies are due to the fact that they measured a larger number of emission lines and used a wider range of free parameters in their model grid, which allows them to explain the enhanced NV emission observed in AGN spectra at z$\sim$2.
	
	Following \citet{Silva2020} we assumed secondary production of carbon in addition to nitrogen. There are some objects whose best fitting model has sub-solar metallicity (SDSS J081257.15+181916.8, 0850-206, SDSS J001814.72+023258.8, SDSS J020245.82+000848.4), but generally our models predict solar or super-solar metallicities. This is not due to the assumption of secondary production of C; in the case where only primary C production was adopted, the predicted gas metallicities of the objects would be even higher. 
	
	The assumption of a \textit{$\kappa$} distribution instead of the M-B distribution alters the inferred metallicity in a significant fraction of objects. In 47\% of HzRGs and 44\% of QSO2s the \textit{$\kappa$} and M-B models predict different metallicities. In most of these objects, \textit{$\kappa$} models imply 1.5 to 2 times higher metallicities than M-B models. More extreme devitations, by a factor of $\geq$ 3 are inferred in 15\% of the total QSO2 sample.

	\subsection{Do \textit{$\kappa$}-distributions exist in the NLR?}
	
	The physical properties of gaseous nebulae are usually inferred under the assumption that the electrons are in an equilibrium M-B distribution. This is assumed because the energy redistribution through elastic collisions of electrons occurs faster than any other process \citep[e.g.][]{Draine2018}. However, if energetic electrons are injected into the gas continually and sufficiently quickly that the particle distribution does not have time to relax to a classical equilibrium distribution, then \textit{$\kappa$} distributions may arise. 
	
	\citet{Nicholls2012} proposed several mechanisms that would be capable of creating and maintaining \textit{$\kappa$} distributions in photoionized regions. These include the injection of a population of energetic electrons by acceleration mechanisms such as magnetic reconnection and development of inertial Alfvén waves, shocks, and the injection of high-energy electrons through the photoionization process. Energetic electrons can be produced by the photoionization of dust \citep{Dopita2000}, and X-ray or EUV photoionization \citep{Shull1985, Petrini1997}. AGNs are regions where shocks, winds, and turbulence are known to be present \citep[e.g.][]{Baring1991, Humphrey2008a, Humphrey2010, Zhang2013, Silva2018}, and these processes may also be sufficient to accelerate particles creating \textit{$\kappa$} distributions of electron energies. Several other processes, which are plausubly associated with AGN activity, may also be responsible for deviations from equilibrium, such as ionization by cosmic rays \citep{Giammanco2005}, and dissipative turbulence internal to the gas clouds \citep{Mignoli2019}.
	
	In this work we have found tantalizing evidence for the possible presence of \textit{$\kappa$}-distributions in the NLR some AGN, based on the fact that their emission line ratios are better reproduced by photoionization models that use \textit{$\kappa$}-distributed EEDs. However, more work is needed to identify and understand physical processes that could establish \textit{$\kappa$}-distributed EEDs in the NLR.

	\section{Conclusions} \label{sec:conclusion}
	
	In this work we study how the presence of \textit{$\kappa$}-distributed electron energies affects the emission line ratios of photoionized nebulae associated with AGN. We also provide best fitting parameters ($Z$, n$_H$, $\alpha$ and $U$) for both \textit{$\kappa$} and M-B models for 143 type 2 AGN: 15 HzRG, 126 QSO2s and 2 Seyfert galaxies.  
	
	We find that the choice of EED can have a large impact on some UV and optical emission lines, and that the impact of adopting a \textit{$\kappa$} distribution is strongly dependent on gas metallicity and ionization parameter. 
	The fluxes of UV lines with high excitation energies (e.g. CIII $\lambda$977, NV $\lambda$1240) are greatly enhanced by \textit{$\kappa$} distributions. The enhancement is stronger when the deviation from thermal equilibrium is large, i.e. for lower values of \textit{$\kappa$}. \\
	In the optical, some line fluxes are enhanced when \textit{$\kappa$} distributions are adopted (e.g. [OIII]$\lambda$4363), but in most cases the line fluxes are fainter in the \textit{$\kappa$} models compared to the M-B models (\textit{$\kappa$} / M-B \textless 1). \\
	Using \textit{$\kappa$} distributions results, generally, in a decrease of the predicted IR line fluxes. With increasing metallicity and ionization parameter the differences between ratios obtained using M-B distributions and \textit{$\kappa$} distributions becomes smaller. 
	
	We have compared our models against the rest-frame UV to optical emission line ratios of 143 type 2 active galaxies. We found that for a subset of the sources (one of the two studied Seyfert galaxies, 67\% of HzRGs and 69\% of QSO2s) a photoionization model using \textit{$\kappa$}-distributed electron energies provides a significantly better fit to the data, which suggests that \textit{$\kappa$}-distributed electron energy distributions may indeed be present in the NLR of some type 2 AGN.  
	
	The assumption of a \textit{$\kappa$} or M-B distribution has no impact on the inferred spectral index, $\alpha$ in most HzRGs (14/15 or 93\%) QSO2s (116/126 or 92\%). 
	
	The assumption of a \textit{$\kappa$} distribution instead of the M-B distribution has no impact on the derived gas density in most HzRG (11/15 or 73\%) and QSO2s (104/126 or 83\%). Deviations in n$_H$ by a factor as large as $\geq$ 100 are identified in the remaining 17\% of QSO2s.
	
	Regarding the metallicity, the general trend in both HzRGs and QSO2s is that \textit{$\kappa$} models produce identical (roughly in half of both samples) or higher $Z$ than M-B models. More specifically, the predicted metallicities are consistent (Z$\kappa$ = Z$_{M-B}$) in 53\% of HzRGs (8/15) and 56\% of QSO2s (71/126). 
	For the rest of both samples, but for a few exceptions, \textit{$\kappa$} models produce 1.5 to 2 times higher $Z$ than M-B models in 40\% HzRG (6/15) and 25\% (32/126) QSO2s. Large deviations (a factor \textgreater 2) are found in a small fraction of HzRGs (1/15) and QSO2s (19/126).
	
	The predicted $U$ values for the \textit{$\kappa$} and M-B models differ in most objects (13/15 or 87\% of HzRG and 88/126 or 70\% of QSO2). HzRGs and QSO2s behave differently in the sense that in general \textit{$\kappa$} models produce higher $U$ values (1.1 \textless U$\textit{$\kappa$}$ / U$_{M-B}$ \textless 2), while the relative factor U$\textit{$\kappa$}$ / U$_{M-B}$ is more evenly spread in QSO2s at both sides of U$\textit{$\kappa$}$ / U$_{M-B}$ = 1. Large discrepancies occur for 25 QSO2s (20\%, all of which have U$\textit{$\kappa$}$ / U$_{M-B}$ \textgreater 2) and one HzRG (with U$\textit{$\kappa$}$ / U$_{M-B}$ \textless 0.5).

	\section*{Acknowledgements}
	
	SGM acknowledges support from the Fundação para a Ciência e a Tecnologia (FCT) through the Fellowship PD/BD/135228/2017 (PhD::SPACE Doctoral Network PD/00040/2012) and POCH/FSE (EC).
	SGM and AH were supported by Fundação para a Ciência e a Tecnologia (FCT) through the research grants UIDB/04434/2020 and UIDP/04434/2020, and an FCT-CAPES Transnational Cooperation Project "Parceria Estratégica em Astrofísica Portugal-Brasil". 
	MVM acknowledges support from grant PGC2018-094671-BI00 (MCIU/AEI/FEDER,UE). Her work was done under project No. MDM-2017-0737 Unidad de Excelencia “Mar\'\i a de Maeztu” Centro de Astrobiolog\'\i a (CSIC-INTA).
	
	
	\section*{Data Availability}

	The data underlying this article will be shared on reasonable request to the corresponding author.

	
	\bibliographystyle{mnras}
	\bibliography{refs_kpaper}

\begin{thebibliography}{}
\makeatletter
\relax
\def\mn@urlcharsother{\let\do\@makeother \do\$\do\&\do\#\do\^\do\_\do\%\do\~}
\def\mn@doi{\begingroup\mn@urlcharsother \@ifnextchar [ {\mn@doi@}
  {\mn@doi@[]}}
\def\mn@doi@[#1]#2{\def\@tempa{#1}\ifx\@tempa\@empty \href
  {http://dx.doi.org/#2} {doi:#2}\else \href {http://dx.doi.org/#2} {#1}\fi
  \endgroup}
\def\mn@eprint#1#2{\mn@eprint@#1:#2::\@nil}
\def\mn@eprint@arXiv#1{\href {http://arxiv.org/abs/#1} {{\tt arXiv:#1}}}
\def\mn@eprint@dblp#1{\href {http://dblp.uni-trier.de/rec/bibtex/#1.xml}
  {dblp:#1}}
\def\mn@eprint@#1:#2:#3:#4\@nil{\def\@tempa {#1}\def\@tempb {#2}\def\@tempc
  {#3}\ifx \@tempc \@empty \let \@tempc \@tempb \let \@tempb \@tempa \fi \ifx
  \@tempb \@empty \def\@tempb {arXiv}\fi \@ifundefined
  {mn@eprint@\@tempb}{\@tempb:\@tempc}{\expandafter \expandafter \csname
  mn@eprint@\@tempb\endcsname \expandafter{\@tempc}}}

\bibitem[\protect\citeauthoryear{{Asplund}, {Grevesse}  \& {Jacques
  Sauval}}{{Asplund} et~al.}{2006}]{Asplund2006}
{Asplund} M.,  {Grevesse} N.,   {Jacques Sauval} A.,  2006, \mn@doi [Nuclear
  Physics A] {10.1016/j.nuclphysa.2005.06.010}, \href
  {http://adsabs.harvard.edu/abs/2006NuPhA.777....1A} {777, 1}

\bibitem[\protect\citeauthoryear{{Baldwin}, {Phillips}  \&
  {Terlevich}}{{Baldwin} et~al.}{1981}]{Baldwin1981}
{Baldwin} J.~A.,  {Phillips} M.~M.,   {Terlevich} R.,  1981, \mn@doi [\pasp]
  {10.1086/130766}, \href {http://adsabs.harvard.edu/abs/1981PASP...93....5B}
  {93, 5}

\bibitem[\protect\citeauthoryear{{Baring}}{{Baring}}{1991}]{Baring1991}
{Baring} M.~G.,  1991, \mn@doi [\mnras] {10.1093/mnras/253.3.388}, \href
  {http://adsabs.harvard.edu/abs/1991MNRAS.253..388B} {253, 388}

\bibitem[\protect\citeauthoryear{{Binette}, {Dopita}  \& {Tuohy}}{{Binette}
  et~al.}{1985}]{Binette1985}
{Binette} L.,  {Dopita} M.~A.,   {Tuohy} I.~R.,  1985, \mn@doi [\apj]
  {10.1086/163544}, \href {http://adsabs.harvard.edu/abs/1985ApJ...297..476B}
  {297, 476}

\bibitem[\protect\citeauthoryear{{Binette}, {Wilson}  \&
  {Storchi-Bergmann}}{{Binette} et~al.}{1996}]{Binette1996}
{Binette} L.,  {Wilson} A.~S.,   {Storchi-Bergmann} T.,  1996, \aap, \href
  {http://adsabs.harvard.edu/abs/1996A%26A...312..365B} {312, 365}

\bibitem[\protect\citeauthoryear{{Binette}, {Matadamas}, {H{\"a}gele},
  {Nicholls}, {Magris C.}, {Pe{\~n}a-Guerrero}, {Morisset}  \&
  {Rodr{\'{\i}}guez-Gonz{\'a}lez}}{{Binette} et~al.}{2012}]{Binette2012}
{Binette} L.,  {Matadamas} R.,  {H{\"a}gele} G.~F.,  {Nicholls} D.~C.,  {Magris
  C.} G.,  {Pe{\~n}a-Guerrero} M.~{\'A}.,  {Morisset} C.,
  {Rodr{\'{\i}}guez-Gonz{\'a}lez} A.,  2012, \mn@doi [\aap]
  {10.1051/0004-6361/201219515}, \href
  {http://adsabs.harvard.edu/abs/2012A\%26A...547A..29B} {547, A29}

\bibitem[\protect\citeauthoryear{{Binsack}}{{Binsack}}{1966}]{Binsack1966}
{Binsack} J.~H.,  1966, PhD thesis, MIT

\bibitem[\protect\citeauthoryear{{Bohm} \& {Aller}}{{Bohm} \&
  {Aller}}{1947}]{Bohm1947}
{Bohm} D.,  {Aller} L.~H.,  1947, \mn@doi [\apj] {10.1086/144890}, \href
  {http://adsabs.harvard.edu/abs/1947ApJ...105..131B} {105, 131}

\bibitem[\protect\citeauthoryear{{Bradshaw} \& {Raymond}}{{Bradshaw} \&
  {Raymond}}{2013}]{Bradshaw2013}
{Bradshaw} S.~J.,  {Raymond} J.,  2013, \mn@doi [Space Sci. Rev.]
  {10.1007/s11214-013-9970-0}, \href
  {http://adsabs.harvard.edu/abs/2013SSRv..178..271B} {178, 271}

\bibitem[\protect\citeauthoryear{{Cano-D{\'{\i}}az}, {Maiolino}, {Marconi},
  {Netzer}, {Shemmer}  \& {Cresci}}{{Cano-D{\'{\i}}az} et~al.}{2012}]{Cano2012}
{Cano-D{\'{\i}}az} M.,  {Maiolino} R.,  {Marconi} A.,  {Netzer} H.,  {Shemmer}
  O.,   {Cresci} G.,  2012, \mn@doi [A\&A] {10.1051/0004-6361/201118358}, \href
  {http://adsabs.harvard.edu/abs/2012A%26A...537L...8C} {537, L8}

\bibitem[\protect\citeauthoryear{{Collins}, {Kraemer}, {Crenshaw}, {Ruiz},
  {Deo}  \& {Bruhweiler}}{{Collins} et~al.}{2005}]{Collins2005}
{Collins} N.~R.,  {Kraemer} S.~B.,  {Crenshaw} D.~M.,  {Ruiz} J.,  {Deo} R.,
  {Bruhweiler} F.~C.,  2005, \mn@doi [\apj] {10.1086/426314}, \href
  {http://adsabs.harvard.edu/abs/2005ApJ...619..116C} {619, 116}

\bibitem[\protect\citeauthoryear{{Collins}, {Kraemer}, {Crenshaw}, {Bruhweiler}
   \& {Mel{\'e}ndez}}{{Collins} et~al.}{2009}]{Collins2009}
{Collins} N.~R.,  {Kraemer} S.~B.,  {Crenshaw} D.~M.,  {Bruhweiler} F.~C.,
  {Mel{\'e}ndez} M.,  2009, \mn@doi [\apj] {10.1088/0004-637X/694/2/765}, \href
  {http://adsabs.harvard.edu/abs/2009ApJ...694..765C} {694, 765}

\bibitem[\protect\citeauthoryear{{Croton} et~al.,}{{Croton}
  et~al.}{2006}]{Croton2006}
{Croton} D.~J.,  et~al., 2006, \mn@doi [\mnras]
  {10.1111/j.1365-2966.2005.09675.x}, \href
  {http://adsabs.harvard.edu/abs/2006MNRAS.365...11C} {365, 11}

\bibitem[\protect\citeauthoryear{{Decker} \& {Krimigis}}{{Decker} \&
  {Krimigis}}{2003}]{Decker2003}
{Decker} R.~B.,  {Krimigis} S.~M.,  2003, \mn@doi [Advances in Space Research]
  {10.1016/S0273-1177(03)00356-9}, \href
  {http://adsabs.harvard.edu/abs/2003AdSpR..32..597D} {32, 597}

\bibitem[\protect\citeauthoryear{{Diaz}, {Prieto}  \& {Wamsteker}}{{Diaz}
  et~al.}{1988}]{Diaz1988}
{Diaz} A.~I.,  {Prieto} M.~A.,   {Wamsteker} W.,  1988, \aap, \href
  {http://adsabs.harvard.edu/abs/1988A%26A...195...53D} {195, 53}

\bibitem[\protect\citeauthoryear{{Dopita} \& {Sutherland}}{{Dopita} \&
  {Sutherland}}{2000}]{Dopita2000}
{Dopita} M.~A.,  {Sutherland} R.~S.,  2000, \mn@doi [\apj] {10.1086/309241},
  \href {http://adsabs.harvard.edu/abs/2000ApJ...539..742D} {539, 742}

\bibitem[\protect\citeauthoryear{{Dopita}, {Sutherland}, {Nicholls}, {Kewley}
  \& {Vogt}}{{Dopita} et~al.}{2013}]{Dopita2013}
{Dopita} M.~A.,  {Sutherland} R.~S.,  {Nicholls} D.~C.,  {Kewley} L.~J.,
  {Vogt} F.~P.~A.,  2013, \mn@doi [\apjs] {10.1088/0067-0049/208/1/10}, \href
  {http://adsabs.harvard.edu/abs/2013ApJS..208...10D} {208, 10}

\bibitem[\protect\citeauthoryear{{Dopita} et~al.,}{{Dopita}
  et~al.}{2014}]{Dopita2014}
{Dopita} M.~A.,  et~al., 2014, \mn@doi [\aap] {10.1051/0004-6361/201423467},
  \href {https://ui.adsabs.harvard.edu/abs/2014A%26A...566A..41D} {566, A41}

\bibitem[\protect\citeauthoryear{{Dors}, {Cardaci}, {H{\"a}gele}  \&
  {Krabbe}}{{Dors} et~al.}{2014}]{Dors2014}
{Dors} O.~L.,  {Cardaci} M.~V.,  {H{\"a}gele} G.~F.,   {Krabbe} {\^A}.~C.,
  2014, \mn@doi [\mnras] {10.1093/mnras/stu1218}, \href
  {https://ui.adsabs.harvard.edu/abs/2014MNRAS.443.1291D} {443, 1291}

\bibitem[\protect\citeauthoryear{{Draine} \& {Kreisch}}{{Draine} \&
  {Kreisch}}{2018}]{Draine2018}
{Draine} B.~T.,  {Kreisch} C.~D.,  2018, \mn@doi [\apj]
  {10.3847/1538-4357/aac891}, \href
  {http://adsabs.harvard.edu/abs/2018ApJ...862...30D} {862, 30}

\bibitem[\protect\citeauthoryear{{Feltre}, {Charlot}  \& {Gutkin}}{{Feltre}
  et~al.}{2016}]{Feltre2016}
{Feltre} A.,  {Charlot} S.,   {Gutkin} J.,  2016, \mn@doi [\mnras]
  {10.1093/mnras/stv2794}, \href
  {http://adsabs.harvard.edu/abs/2016MNRAS.456.3354F} {456, 3354}

\bibitem[\protect\citeauthoryear{{Ferland}, {Henney}, {O'Dell}  \&
  {Peimbert}}{{Ferland} et~al.}{2016}]{Ferland2016}
{Ferland} G.~J.,  {Henney} W.~J.,  {O'Dell} C.~R.,   {Peimbert} M.,  2016,
  \rmxaa, \href {http://adsabs.harvard.edu/abs/2016RMxAA..52..261F} {52, 261}

\bibitem[\protect\citeauthoryear{{Ferruit}, {Binette}, {Sutherland}  \&
  {Pecontal}}{{Ferruit} et~al.}{1997}]{Ferruit1997}
{Ferruit} P.,  {Binette} L.,  {Sutherland} R.~S.,   {Pecontal} E.,  1997, \aap,
  \href {http://adsabs.harvard.edu/abs/1997A%26A...322...73F} {322, 73}

\bibitem[\protect\citeauthoryear{{Giammanco} \& {Beckman}}{{Giammanco} \&
  {Beckman}}{2005}]{Giammanco2005}
{Giammanco} C.,  {Beckman} J.~E.,  2005, \mn@doi [\aap]
  {10.1051/0004-6361:200500129}, \href
  {https://ui.adsabs.harvard.edu/abs/2005A&A...437L..11G} {437, L11}

\bibitem[\protect\citeauthoryear{{Gloeckler} \& {Hamilton}}{{Gloeckler} \&
  {Hamilton}}{1987}]{Gloeckler1987}
{Gloeckler} G.,  {Hamilton} D.~C.,  1987, \mn@doi [Physica Scripta Volume T]
  {10.1088/0031-8949/1987/T18/009}, \href
  {http://adsabs.harvard.edu/abs/1987PhST...18...73G} {18, 73}

\bibitem[\protect\citeauthoryear{{Gloeckler} et~al.,}{{Gloeckler}
  et~al.}{1992}]{Gloeckler1992}
{Gloeckler} G.,  et~al., 1992, \aaps, \href
  {http://adsabs.harvard.edu/abs/1992A%26AS...92..267G} {92, 267}

\bibitem[\protect\citeauthoryear{{Groves}, {Dopita}  \& {Sutherland}}{{Groves}
  et~al.}{2004}]{Groves2004}
{Groves} B.~A.,  {Dopita} M.~A.,   {Sutherland} R.~S.,  2004, \mn@doi [\apjs]
  {10.1086/421113}, \href {http://adsabs.harvard.edu/abs/2004ApJS..153....9G}
  {153, 9}

\bibitem[\protect\citeauthoryear{{Hagihara}}{{Hagihara}}{1939a}]{Hagihara1939a}
{Hagihara} Y.,  1939a, Japanese Journal of Astronomy and Geophysics, \href
  {http://adsabs.harvard.edu/abs/1939JaJAG..17..199H} {17, 199}

\bibitem[\protect\citeauthoryear{{Hagihara}}{{Hagihara}}{1939b}]{Hagihara1939b}
{Hagihara} Y.,  1939b, Japanese Journal of Astronomy and Geophysics, \href
  {http://adsabs.harvard.edu/abs/1939JaJAG..17..417H} {17, 417}

\bibitem[\protect\citeauthoryear{{Hamann} \& {Ferland}}{{Hamann} \&
  {Ferland}}{1992}]{Hamann1992}
{Hamann} F.,  {Ferland} G.,  1992, \mn@doi [\apjl] {10.1086/186397}, \href
  {https://ui.adsabs.harvard.edu/abs/1992ApJ...391L..53H} {391, L53}

\bibitem[\protect\citeauthoryear{{Hamann} \& {Ferland}}{{Hamann} \&
  {Ferland}}{1993}]{Hamann1993}
{Hamann} F.,  {Ferland} G.,  1993, \mn@doi [\apj] {10.1086/173366}, \href
  {https://ui.adsabs.harvard.edu/abs/1993ApJ...418...11H} {418, 11}

\bibitem[\protect\citeauthoryear{{Hebb} \& {Menzel}}{{Hebb} \&
  {Menzel}}{1940}]{Hebb1940}
{Hebb} M.~H.,  {Menzel} D.~H.,  1940, \mn@doi [ApJ] {10.1086/144230}, \href
  {http://adsabs.harvard.edu/abs/1940ApJ....92..408H} {92, 408}

\bibitem[\protect\citeauthoryear{{Heerikhuisen}, {Pogorelov}, {Florinski},
  {Zank}  \& {le Roux}}{{Heerikhuisen} et~al.}{2008}]{Heerikhuisen2008}
{Heerikhuisen} J.,  {Pogorelov} N.~V.,  {Florinski} V.,  {Zank} G.~P.,   {le
  Roux} J.~A.,  2008, \mn@doi [ApJ] {10.1086/588248}, \href
  {http://adsabs.harvard.edu/abs/2008ApJ...682..679H} {682, 679}

\bibitem[\protect\citeauthoryear{{Henry}, {Edmunds}  \& {K{\"o}ppen}}{{Henry}
  et~al.}{2000}]{Henry2000}
{Henry} R.~B.~C.,  {Edmunds} M.~G.,   {K{\"o}ppen} J.,  2000, \mn@doi [\apj]
  {10.1086/309471}, \href {http://adsabs.harvard.edu/abs/2000ApJ...541..660H}
  {541, 660}

\bibitem[\protect\citeauthoryear{{Humphrey}}{{Humphrey}}{2019}]{Humphrey2019b}
{Humphrey} A.,  2019, \mn@doi [\mnras] {10.1093/mnras/stz687}, \href
  {http://adsabs.harvard.edu/abs/2019MNRAS.tmp..790H} {}

\bibitem[\protect\citeauthoryear{{Humphrey} \& {Binette}}{{Humphrey} \&
  {Binette}}{2014}]{Humphrey2014}
{Humphrey} A.,  {Binette} L.,  2014, \mn@doi [\mnras] {10.1093/mnras/stu723},
  \href {http://adsabs.harvard.edu/abs/2014MNRAS.442..753H} {442, 753}

\bibitem[\protect\citeauthoryear{{Humphrey}, {Villar-Mart{\'{\i}}n}, {Vernet},
  {Fosbury}, {di Serego Alighieri}  \& {Binette}}{{Humphrey}
  et~al.}{2008}]{Humphrey2008a}
{Humphrey} A.,  {Villar-Mart{\'{\i}}n} M.,  {Vernet} J.,  {Fosbury} R.,  {di
  Serego Alighieri} S.,   {Binette} L.,  2008, \mn@doi [\mnras]
  {10.1111/j.1365-2966.2007.12506.x}, \href
  {http://adsabs.harvard.edu/abs/2008MNRAS.383...11H} {383, 11}

\bibitem[\protect\citeauthoryear{{Humphrey}, {Villar-Mart{\'{\i}}n},
  {S{\'a}nchez}, {Mart{\'{\i}}nez-Sansigre}, {Delgado}, {P{\'e}rez},
  {Tadhunter}  \& {P{\'e}rez-Torres}}{{Humphrey} et~al.}{2010}]{Humphrey2010}
{Humphrey} A.,  {Villar-Mart{\'{\i}}n} M.,  {S{\'a}nchez} S.~F.,
  {Mart{\'{\i}}nez-Sansigre} A.,  {Delgado} R.~G.,  {P{\'e}rez} E.,
  {Tadhunter} C.,   {P{\'e}rez-Torres} M.~A.,  2010, \mn@doi [\mnras]
  {10.1111/j.1745-3933.2010.00906.x}, \href
  {http://adsabs.harvard.edu/abs/2010MNRAS.408L...1H} {408, L1}

\bibitem[\protect\citeauthoryear{{Jones} \& {Lambourne}}{{Jones} \&
  {Lambourne}}{2004}]{Jones2004}
{Jones} M.~H.,  {Lambourne} R.~J.~A.,  2004, {An Introduction to Galaxies and
  Cosmology}

\bibitem[\protect\citeauthoryear{{Jurac}, {McGrath}, {Johnson}, {Richardson},
  {Vasyliunas}  \& {Eviatar}}{{Jurac} et~al.}{2002}]{Jurac2002}
{Jurac} S.,  {McGrath} M.~A.,  {Johnson} R.~E.,  {Richardson} J.~D.,
  {Vasyliunas} V.~M.,   {Eviatar} A.,  2002, \mn@doi [Geophysical Research
  Letters] {10.1029/2002GL015855}, \href
  {http://adsabs.harvard.edu/abs/2002GeoRL..29.2172J} {29, 2172}

\bibitem[\protect\citeauthoryear{{Kraemer}, {Ruiz}  \& {Crenshaw}}{{Kraemer}
  et~al.}{1998}]{Kraemer1998}
{Kraemer} S.~B.,  {Ruiz} J.~R.,   {Crenshaw} D.~M.,  1998, \mn@doi [\apj]
  {10.1086/306389}, \href
  {https://ui.adsabs.harvard.edu/abs/1998ApJ...508..232K} {508, 232}

\bibitem[\protect\citeauthoryear{{Krimigis}, {Carbary}, {Keath}, {Armstrong},
  {Lanzerotti}  \& {Gloeckler}}{{Krimigis} et~al.}{1983}]{Krimigis1983}
{Krimigis} S.~M.,  {Carbary} J.~F.,  {Keath} E.~P.,  {Armstrong} T.~P.,
  {Lanzerotti} L.~J.,   {Gloeckler} G.,  1983, \mn@doi [\jgr]
  {10.1029/JA088iA11p08871}, \href
  {http://adsabs.harvard.edu/abs/1983JGR....88.8871K} {88, 8871}

\bibitem[\protect\citeauthoryear{{Krimigis}, {Armstrong}, {Axford}, {Cheng}  \&
  {Gloeckler}}{{Krimigis} et~al.}{1986}]{Krimigis1986}
{Krimigis} S.~M.,  {Armstrong} T.~P.,  {Axford} W.~I.,  {Cheng} A.~F.,
  {Gloeckler} G.,  1986, \mn@doi [Science] {10.1126/science.233.4759.97}, \href
  {http://adsabs.harvard.edu/abs/1986Sci...233...97K} {233, 97}

\bibitem[\protect\citeauthoryear{{Kriss}, {Davidsen}, {Blair}, {Ferguson}  \&
  {Long}}{{Kriss} et~al.}{1992}]{Kriss1992}
{Kriss} G.~A.,  {Davidsen} A.~F.,  {Blair} W.~P.,  {Ferguson} H.~C.,   {Long}
  K.~S.,  1992, \mn@doi [\apjl] {10.1086/186467}, \href
  {http://adsabs.harvard.edu/abs/1992ApJ...394L..37K} {394, L37}

\bibitem[\protect\citeauthoryear{{Leubner}}{{Leubner}}{2002}]{Leubner2002}
{Leubner} M.~P.,  2002, \mn@doi [\apss] {10.1023/A:1020990413487}, \href
  {http://adsabs.harvard.edu/abs/2002Ap%26SS.282..573L} {282, 573}

\bibitem[\protect\citeauthoryear{{Livadiotis} \& {McComas}}{{Livadiotis} \&
  {McComas}}{2009}]{Livadiotis2009}
{Livadiotis} G.,  {McComas} D.~J.,  2009, \mn@doi [Journal of Geophysical
  Research (Space Physics)] {10.1029/2009JA014352}, \href
  {http://adsabs.harvard.edu/abs/2009JGRA..11411105L} {114, A11105}

\bibitem[\protect\citeauthoryear{{Maksimovic}, {Pierrard}  \&
  {Lemaire}}{{Maksimovic} et~al.}{1997}]{Maksimovic1997}
{Maksimovic} M.,  {Pierrard} V.,   {Lemaire} J.~F.,  1997, \aap, \href
  {http://adsabs.harvard.edu/abs/1997A%26A...324..725M} {324, 725}

\bibitem[\protect\citeauthoryear{{Mathews} \& {Ferland}}{{Mathews} \&
  {Ferland}}{1987}]{Mathews1987}
{Mathews} W.~G.,  {Ferland} G.~J.,  1987, \mn@doi [\apj] {10.1086/165843},
  \href {https://ui.adsabs.harvard.edu/abs/1987ApJ...323..456M} {323, 456}

\bibitem[\protect\citeauthoryear{{Matsuoka}, {Nagao}, {Maiolino}, {Marconi}  \&
  {Taniguchi}}{{Matsuoka} et~al.}{2009}]{Matsuoka2009}
{Matsuoka} K.,  {Nagao} T.,  {Maiolino} R.,  {Marconi} A.,   {Taniguchi} Y.,
  2009, \mn@doi [\aap] {10.1051/0004-6361/200811478}, \href
  {http://adsabs.harvard.edu/abs/2009A%26A...503..721M} {503, 721}

\bibitem[\protect\citeauthoryear{{McCarthy}, {Spinrad}, {Dickinson}, {van
  Breugel}, {Liebert}, {Djorgovski}  \& {Eisenhardt}}{{McCarthy}
  et~al.}{1990}]{McCarthy1990}
{McCarthy} P.~J.,  {Spinrad} H.,  {Dickinson} M.,  {van Breugel} W.,  {Liebert}
  J.,  {Djorgovski} S.,   {Eisenhardt} P.,  1990, \mn@doi [\apj]
  {10.1086/169503}, \href {http://adsabs.harvard.edu/abs/1990ApJ...365..487M}
  {365, 487}

\bibitem[\protect\citeauthoryear{{Mendoza} \& {Bautista}}{{Mendoza} \&
  {Bautista}}{2014}]{Mendoza2014}
{Mendoza} C.,  {Bautista} M.~A.,  2014, \mn@doi [\apj]
  {10.1088/0004-637X/785/2/91}, \href
  {http://adsabs.harvard.edu/abs/2014ApJ...785...91M} {785, 91}

\bibitem[\protect\citeauthoryear{{Meyer-Vernet}, {Moncuquet}  \&
  {Hoang}}{{Meyer-Vernet} et~al.}{1995}]{Meyer-Vernet1995}
{Meyer-Vernet} N.,  {Moncuquet} M.,   {Hoang} S.,  1995, \mn@doi [\icarus]
  {10.1006/icar.1995.1121}, \href
  {https://ui.adsabs.harvard.edu/abs/1995Icar..116..202M} {116, 202}

\bibitem[\protect\citeauthoryear{{Mignoli} et~al.,}{{Mignoli}
  et~al.}{2019}]{Mignoli2019}
{Mignoli} M.,  et~al., 2019, \mn@doi [\aap] {10.1051/0004-6361/201935062},
  \href {https://ui.adsabs.harvard.edu/abs/2019A&A...626A...9M} {626, A9}

\bibitem[\protect\citeauthoryear{{Moe}, {Arav}, {Bautista}  \& {Korista}}{{Moe}
  et~al.}{2009}]{Moe2009}
{Moe} M.,  {Arav} N.,  {Bautista} M.~A.,   {Korista} K.~T.,  2009, \mn@doi
  [ApJ] {10.1088/0004-637X/706/1/525}, \href
  {http://adsabs.harvard.edu/abs/2009ApJ...706..525M} {706, 525}

\bibitem[\protect\citeauthoryear{{Moncuquet}, {Bagenal}  \&
  {Meyer-Vernet}}{{Moncuquet} et~al.}{2002}]{Moncuquet2002}
{Moncuquet} M.,  {Bagenal} F.,   {Meyer-Vernet} N.,  2002, \mn@doi [J. Geophys.
  Res.] {10.1029/2001JA900124}, \href
  {http://adsabs.harvard.edu/abs/2002JGRA..107.1260M} {107, 1260}

\bibitem[\protect\citeauthoryear{{Morganti}, {Oosterloo}, {Tadhunter}, {van
  Moorsel}  \& {Emonts}}{{Morganti} et~al.}{2005}]{Morganti2005}
{Morganti} R.,  {Oosterloo} T.~A.,  {Tadhunter} C.~N.,  {van Moorsel} G.,
  {Emonts} B.,  2005, \mn@doi [\aap] {10.1051/0004-6361:20053175}, \href
  {http://adsabs.harvard.edu/abs/2005A%26A...439..521M} {439, 521}

\bibitem[\protect\citeauthoryear{{Moy} \& {Rocca-Volmerange}}{{Moy} \&
  {Rocca-Volmerange}}{2002}]{Moy2002}
{Moy} E.,  {Rocca-Volmerange} B.,  2002, \mn@doi [A\&A]
  {10.1051/0004-6361:20011727}, \href
  {http://adsabs.harvard.edu/abs/2002A%26A...383...46M} {383, 46}

\bibitem[\protect\citeauthoryear{{Nagao}, {Maiolino}  \& {Marconi}}{{Nagao}
  et~al.}{2006}]{Nagao2006}
{Nagao} T.,  {Maiolino} R.,   {Marconi} A.,  2006, \mn@doi [\aap]
  {10.1051/0004-6361:20054127}, \href
  {http://adsabs.harvard.edu/abs/2006A%26A...447..863N} {447, 863}

\bibitem[\protect\citeauthoryear{{Netzer}}{{Netzer}}{1997}]{Netzer1997}
{Netzer} H.,  1997, \mn@doi [\apss] {10.1023/A:1000509024526}, \href
  {http://adsabs.harvard.edu/abs/1997Ap%26SS.248..127N} {248, 127}

\bibitem[\protect\citeauthoryear{{Nicholls}, {Dopita}  \&
  {Sutherland}}{{Nicholls} et~al.}{2012}]{Nicholls2012}
{Nicholls} D.~C.,  {Dopita} M.~A.,   {Sutherland} R.~S.,  2012, \mn@doi [\apj]
  {10.1088/0004-637X/752/2/148}, \href
  {http://adsabs.harvard.edu/abs/2012ApJ...752..148N} {752, 148}

\bibitem[\protect\citeauthoryear{{Nicholls}, {Dopita}, {Sutherland}, {Kewley}
  \& {Palay}}{{Nicholls} et~al.}{2013}]{Nicholls2013}
{Nicholls} D.~C.,  {Dopita} M.~A.,  {Sutherland} R.~S.,  {Kewley} L.~J.,
  {Palay} E.,  2013, \mn@doi [\apjs] {10.1088/0067-0049/207/2/21}, \href
  {http://adsabs.harvard.edu/abs/2013ApJS..207...21N} {207, 21}

\bibitem[\protect\citeauthoryear{{Olbert}}{{Olbert}}{1968}]{Olbert1968}
{Olbert} S.,  1968, in {Carovillano} R.~D.~L.,  {McClay} J.~F.,  eds,
  Astrophysics and Space Science Library Vol. 10, Physics of the Magnetosphere.
  p.~641, \mn@doi{10.1007/978-94-010-3467-8_23}

\bibitem[\protect\citeauthoryear{{Owocki} \& {Scudder}}{{Owocki} \&
  {Scudder}}{1983}]{Owocki1983}
{Owocki} S.~P.,  {Scudder} J.~D.,  1983, \mn@doi [\apj] {10.1086/161167}, \href
  {http://adsabs.harvard.edu/abs/1983ApJ...270..758O} {270, 758}

\bibitem[\protect\citeauthoryear{{Petrini} \& {da Silva}}{{Petrini} \& {da
  Silva}}{1997}]{Petrini1997}
{Petrini} D.,  {da Silva} E.~P.,  1997, \aap, \href
  {https://ui.adsabs.harvard.edu/abs/1997A&A...317..262P} {317, 262}

\bibitem[\protect\citeauthoryear{{Pierrard} \& {Lazar}}{{Pierrard} \&
  {Lazar}}{2010}]{Pierrard2010}
{Pierrard} V.,  {Lazar} M.,  2010, \mn@doi [\solphys]
  {10.1007/s11207-010-9640-2}, \href
  {http://adsabs.harvard.edu/abs/2010SoPh..267..153P} {267, 153}

\bibitem[\protect\citeauthoryear{{Radovich}, {Hasinger}  \&
  {Rafanelli}}{{Radovich} et~al.}{1998}]{Radovich1998}
{Radovich} M.,  {Hasinger} G.,   {Rafanelli} P.,  1998, \mn@doi [Astronomische
  Nachrichten] {10.1002/asna.2123190602}, \href
  {http://adsabs.harvard.edu/abs/1998AN....319..325R} {319, 325}

\bibitem[\protect\citeauthoryear{{Richardson}, {Allen}, {Baldwin}, {Hewett}  \&
  {Ferland}}{{Richardson} et~al.}{2014}]{Richardson2014}
{Richardson} C.~T.,  {Allen} J.~T.,  {Baldwin} J.~A.,  {Hewett} P.~C.,
  {Ferland} G.~J.,  2014, \mn@doi [\mnras] {10.1093/mnras/stt2056}, \href
  {http://adsabs.harvard.edu/abs/2014MNRAS.437.2376R} {437, 2376}

\bibitem[\protect\citeauthoryear{{Robinson}, {Binette}, {Fosbury}  \&
  {Tadhunter}}{{Robinson} et~al.}{1987}]{Robinson1987}
{Robinson} A.,  {Binette} L.,  {Fosbury} R.~A.~E.,   {Tadhunter} C.~N.,  1987,
  \mnras, \href {http://adsabs.harvard.edu/abs/1987MNRAS.227...97R} {227, 97}

\bibitem[\protect\citeauthoryear{{Shull} \& {van Steenberg}}{{Shull} \& {van
  Steenberg}}{1985}]{Shull1985}
{Shull} J.~M.,  {van Steenberg} M.~E.,  1985, \mn@doi [\apj] {10.1086/163605},
  \href {https://ui.adsabs.harvard.edu/abs/1985ApJ...298..268S} {298, 268}

\bibitem[\protect\citeauthoryear{{Silva} et~al.,}{{Silva}
  et~al.}{2018}]{Silva2018}
{Silva} M.,  et~al., 2018, \mn@doi [\mnras] {10.1093/mnras/stx3019}, \href
  {http://adsabs.harvard.edu/abs/2018MNRAS.474.3649S} {474, 3649}

\bibitem[\protect\citeauthoryear{{Silva}, {Humphrey}, {Lagos}  \&
  {Morais}}{{Silva} et~al.}{2020}]{Silva2020}
{Silva} M.,  {Humphrey} A.,  {Lagos} P.,   {Morais} S.~G.,  2020, \mn@doi
  [\mnras] {10.1093/mnras/staa1409}, \href
  {https://ui.adsabs.harvard.edu/abs/2020MNRAS.495.4707S} {495, 4707}

\bibitem[\protect\citeauthoryear{{Sol{\'o}rzano-I{\~n}arrea}, {Best},
  {R{\"o}ttgering}  \& {Cimatti}}{{Sol{\'o}rzano-I{\~n}arrea}
  et~al.}{2004}]{Solorzano2004}
{Sol{\'o}rzano-I{\~n}arrea} C.,  {Best} P.~N.,  {R{\"o}ttgering} H.~J.~A.,
  {Cimatti} A.,  2004, \mn@doi [\mnras] {10.1111/j.1365-2966.2004.07842.x},
  \href {https://ui.adsabs.harvard.edu/abs/2004MNRAS.351..997S} {351, 997}

\bibitem[\protect\citeauthoryear{{Spitzer}}{{Spitzer}}{1962}]{Spitzer1962}
{Spitzer} L.,  1962, {Physics of Fully Ionized Gases}

\bibitem[\protect\citeauthoryear{{Springel} et~al.,}{{Springel}
  et~al.}{2005}]{Springel2005}
{Springel} V.,  et~al., 2005, \mn@doi [\nat] {10.1038/nature03597}, \href
  {http://adsabs.harvard.edu/abs/2005Natur.435..629S} {435, 629}

\bibitem[\protect\citeauthoryear{{Storey} \& {Sochi}}{{Storey} \&
  {Sochi}}{2013}]{Storey2013}
{Storey} P.~J.,  {Sochi} T.,  2013, \mn@doi [\mnras] {10.1093/mnras/sts660},
  \href {http://adsabs.harvard.edu/abs/2013MNRAS.430..599S} {430, 599}

\bibitem[\protect\citeauthoryear{{Storey} \& {Sochi}}{{Storey} \&
  {Sochi}}{2014}]{Storey2014a}
{Storey} P.~J.,  {Sochi} T.,  2014, \mn@doi [\mnras] {10.1093/mnras/stu477},
  \href {http://adsabs.harvard.edu/abs/2014MNRAS.440.2581S} {440, 2581}

\bibitem[\protect\citeauthoryear{{Tadhunter}, {Robinson}  \&
  {Morganti}}{{Tadhunter} et~al.}{1989}]{Tadhunter1989}
{Tadhunter} C.~N.,  {Robinson} A.,   {Morganti} R.,  1989, in {Meurs} E.~J.~A.,
   {Fosbury} R.~A.~E.,  eds,  European Southern Observatory Conference and
  Workshop Proceedings Vol. 32, European Southern Observatory Conference and
  Workshop Proceedings. p.~293

\bibitem[\protect\citeauthoryear{{Tsallis}, {Levy}, {Souza}  \&
  {Maynard}}{{Tsallis} et~al.}{1995}]{Tsallis1995}
{Tsallis} C.,  {Levy} S.~V.~F.,  {Souza} A.~M.~C.,   {Maynard} R.,  1995,
  \mn@doi [Physical Review Letters] {10.1103/PhysRevLett.75.3589}, \href
  {http://adsabs.harvard.edu/abs/1995PhRvL..75.3589T} {75, 3589}

\bibitem[\protect\citeauthoryear{{Vasyliunas}}{{Vasyliunas}}{1968}]{Vasyliunas1968}
{Vasyliunas} V.~M.,  1968, \mn@doi [\jgr] {10.1029/JA073i009p02839}, \href
  {http://adsabs.harvard.edu/abs/1968JGR....73.2839V} {73, 2839}

\bibitem[\protect\citeauthoryear{{Vernet}, {Fosbury}, {Villar-Mart{\'{\i}}n},
  {Cohen}, {Cimatti}, {di Serego Alighieri}  \& {Goodrich}}{{Vernet}
  et~al.}{2001}]{Vernet2001}
{Vernet} J.,  {Fosbury} R.~A.~E.,  {Villar-Mart{\'{\i}}n} M.,  {Cohen} M.~H.,
  {Cimatti} A.,  {di Serego Alighieri} S.,   {Goodrich} R.~W.,  2001, \mn@doi
  [\aap] {10.1051/0004-6361:20000076}, \href
  {http://adsabs.harvard.edu/abs/2001A%26A...366....7V} {366, 7}

\bibitem[\protect\citeauthoryear{{V{\'e}ron-Cetty} \&
  {V{\'e}ron}}{{V{\'e}ron-Cetty} \& {V{\'e}ron}}{2000}]{Veron2000}
{V{\'e}ron-Cetty} M.~P.,  {V{\'e}ron} P.,  2000, \mn@doi [\aapr]
  {10.1007/s001590000006}, \href
  {http://adsabs.harvard.edu/abs/2000A%26ARv..10...81V} {10, 81}

\bibitem[\protect\citeauthoryear{{Villar-Mart{\'{\i}}n}, {Binette}  \&
  {Fosbury}}{{Villar-Mart{\'{\i}}n} et~al.}{1996}]{VM1996}
{Villar-Mart{\'{\i}}n} M.,  {Binette} L.,   {Fosbury} R.~A.~E.,  1996, \aap,
  \href {http://adsabs.harvard.edu/abs/1996A%26A...312..751V} {312, 751}

\bibitem[\protect\citeauthoryear{{Villar-Mart{\'{\i}}n}, {Tadhunter}  \&
  {Clark}}{{Villar-Mart{\'{\i}}n} et~al.}{1997}]{VM1997}
{Villar-Mart{\'{\i}}n} M.,  {Tadhunter} C.,   {Clark} N.,  1997, \aap, \href
  {http://adsabs.harvard.edu/abs/1997A%26A...323...21V} {323, 21}

\bibitem[\protect\citeauthoryear{{Villar-Mart{\'{\i}}n}, {Tadhunter},
  {Morganti}, {Axon}  \& {Koekemoer}}{{Villar-Mart{\'{\i}}n}
  et~al.}{1999a}]{VM1999b}
{Villar-Mart{\'{\i}}n} M.,  {Tadhunter} C.,  {Morganti} R.,  {Axon} D.,
  {Koekemoer} A.,  1999a, \mn@doi [\mnras] {10.1046/j.1365-8711.1999.02603.x},
  \href {http://adsabs.harvard.edu/abs/1999MNRAS.307...24V} {307, 24}

\bibitem[\protect\citeauthoryear{{Villar-Mart{\'{\i}}n}, {Fosbury}, {Binette},
  {Tadhunter}  \& {Rocca-Volmerange}}{{Villar-Mart{\'{\i}}n}
  et~al.}{1999b}]{VM1999c}
{Villar-Mart{\'{\i}}n} M.,  {Fosbury} R.~A.~E.,  {Binette} L.,  {Tadhunter}
  C.~N.,   {Rocca-Volmerange} B.,  1999b, \aap, \href
  {http://adsabs.harvard.edu/abs/1999A%26A...351...47V} {351, 47}

\bibitem[\protect\citeauthoryear{{Villar-Mart{\'{\i}}n}, {Fosbury}, {Vernet},
  {Cohen}, {Cimatti}  \& {di Serego Alighieri}}{{Villar-Mart{\'{\i}}n}
  et~al.}{2001}]{VM2001}
{Villar-Mart{\'{\i}}n} M.,  {Fosbury} R.,  {Vernet} J.,  {Cohen} M.,  {Cimatti}
  A.,   {di Serego Alighieri} S.,  2001, \mn@doi [Astrophysics and Space
  Science Supplement] {10.1023/A:1012764309451}, \href
  {http://adsabs.harvard.edu/abs/2001ApSSS.277..571V} {277, 571}

\bibitem[\protect\citeauthoryear{{Villar-Mart{\'{\i}}n}, {Humphrey}, {De
  Breuck}, {Fosbury}, {Binette}  \& {Vernet}}{{Villar-Mart{\'{\i}}n}
  et~al.}{2007}]{VM2007a}
{Villar-Mart{\'{\i}}n} M.,  {Humphrey} A.,  {De Breuck} C.,  {Fosbury} R.,
  {Binette} L.,   {Vernet} J.,  2007, \mn@doi [\mnras]
  {10.1111/j.1365-2966.2006.11371.x}, \href
  {http://adsabs.harvard.edu/abs/2007MNRAS.375.1299V} {375, 1299}

\bibitem[\protect\citeauthoryear{{Wyse}}{{Wyse}}{1942}]{Wyse1942}
{Wyse} A.~B.,  1942, \mn@doi [\apj] {10.1086/144409}, \href
  {http://adsabs.harvard.edu/abs/1942ApJ....95..356W} {95, 356}

\bibitem[\protect\citeauthoryear{{Zhang}, {Liang}  \& {Hammer}}{{Zhang}
  et~al.}{2013}]{Zhang2013}
{Zhang} Z.~T.,  {Liang} Y.~C.,   {Hammer} F.,  2013, \mn@doi [\mnras]
  {10.1093/mnras/sts713}, \href
  {https://ui.adsabs.harvard.edu/abs/2013MNRAS.430.2605Z} {430, 2605}

\bibitem[\protect\citeauthoryear{{Zhang}, {Zhang}  \& {Liu}}{{Zhang}
  et~al.}{2016}]{Zhang2016}
{Zhang} Y.,  {Zhang} B.,   {Liu} X.-W.,  2016, \mn@doi [\apj]
  {10.3847/0004-637X/817/1/68}, \href
  {http://adsabs.harvard.edu/abs/2016ApJ...817...68Z} {817, 68}

\bibitem[\protect\citeauthoryear{{Zheng}, {Kriss}, {Telfer}, {Grimes}  \&
  {Davidsen}}{{Zheng} et~al.}{1997}]{Zheng1997}
{Zheng} W.,  {Kriss} G.~A.,  {Telfer} R.~C.,  {Grimes} J.~P.,   {Davidsen}
  A.~F.,  1997, \mn@doi [\apj] {10.1086/303560}, \href
  {http://adsabs.harvard.edu/abs/1997ApJ...475..469Z} {475, 469}

\makeatother
\end{thebibliography}

	
	

	\bsp	
	\label{lastpage}
\end{document}